\documentclass[fleqn,usenatbib]{mnras}

\usepackage{newtxtext,newtxmath}
\usepackage[T1]{fontenc}

\DeclareRobustCommand{\VAN}[3]{#2}
\let\VANthebibliography\thebibliography
\def\thebibliography{\DeclareRobustCommand{\VAN}[3]{##3}\VANthebibliography}

\usepackage{graphicx}	% Including figure files
\usepackage{amsmath}	% Advanced maths commands
\usepackage{adjustbox}
\usepackage{natbib}
\title[Study of HgMn and Metallic A Stars]{A Homogeneous Spectroscopic and Evolutionary Study of cool HgMn and hot Metallic A Stars}

\author[F. Kahraman Ali\c{c}avu\c{s} et. al.]{F. Kahraman Ali\c{c}avu\c{s}$^{1,2}$\thanks{E-mail: filizkahraman01@gmail.com},  E. \c{C}elik$^{3}$, E. Niemczura$^{4}$\\%, F. Ali\c{c}avu\c{s}$^{1,2}$
\\
$^{1}$\c{C}anakkale Onsekiz Mart University, Faculty of Sciences, Physics Department, 17100, \c{C}anakkale, T\"{u}rkiye\\
$^{2}$\c{C}anakkale Onsekiz Mart University, Astrophysics Research Center and Ulup{\i}nar Observatory, TR-17100, \c{C}anakkale, T\"{u}rkiye\\
$^{3}$\c{C}anakkale Onsekiz Mart University, School of Graduate Studies, Physics Department, TR-17100, Çanakkale, T\"{u}rkiye\\
$^{4}$Instytut Astronomiczny, Uniwersytet Wroc\l{}awski, ul. Kopernika 11, 51-622 Wroc\l{}aw, Poland\\%$^{6}$Dipartimento di Fisica e Astronomia, Sezione Astrofisica, Universit´a di Catania, Via S. Sofia 78, I-95123 Catania, Italy\\
}

\date{Accepted XXX. Received YYY; in original form ZZZ}

\pubyear{\the\year{}}

\begin{document}
\label{firstpage}
\pagerange{\pageref{firstpage}--\pageref{lastpage}}
\maketitle

\begin{abstract}
We present a homogeneous spectroscopic and evolutionary analysis of chemically peculiar stars, focusing on HgMn and metallic A (Am) subclasses. Particular emphasis was placed on the coolest HgMn stars and the hottest Am stars, motivated by the long-standing hypothesis that these objects may represent consecutive evolutionary stages of a similar stellar population. High-resolution spectra of 13 HgMn and 14 Am targets were analyzed to derive atmospheric parameters and detailed chemical abundances.  While all HgMn stars were confirmed to retain their chemically peculiar nature, only six of the fourteen candidate Am stars exhibit abundance patterns consistent with genuine Am characteristics, highlighting the importance of homogeneous spectroscopic classification. We investigate the temperature dependence of elemental abundances, finding a positive correlation between Mn abundance and $T_{\rm eff}$ and a negative correlation for Ni, supporting the operation of atomic diffusion. Stellar evolutionary models were computed with the MESA code using atmospheric parameters derived from the spectroscopic analysis. The locations of the confirmed HgMn and Am stars in the Hertzsprung--Russell diagram support the possibility that the cool HgMn stars and the hot Am stars may occupy overlapping evolutionary sequences. Although only a limited number of candidate pairs were identified, further analyses using evolutionary models that include atomic diffusion are needed to test the possible connection between the two subclasses.
\end{abstract}

\begin{keywords}
stars: chemically peculiar -- stars: abundances -- stars: evolution
\end{keywords}

%%%%%%%%%%%%%%%%%%%%%%%%%%%%%%%%%%%%%%%%%%%%%%%%%%

%%%%%%%%%%%%%%%%% BODY OF PAPER %%%%%%%%%%%%%%%%%%

\section{Introduction}

Stars are the basic components of the Universe and play an important role in galactic evolution through their life cycles, internal structure, and surface chemical composition. Among stars, a significant fraction exhibits chemical abundance patterns that differ from the solar chemical composition \citep{2009ARA&A..47..481A}. These objects, known as chemically peculiar (CP) stars, display atmospheric over- or underabundances of specific elements relative to stars of similar spectral type.

The origin of these chemical anomalies is primarily attributed to atomic diffusion processes operating in radiative stellar envelopes \citep{1970ApJ...160..641M, 1982ARA&A..20...37V, 2000ApJ...529..338R}. In slowly rotating stars with relatively stable atmospheres, the competition between gravitational settling and radiative acceleration leads to vertical abundance stratification. Elements with numerous ultraviolet absorption lines (e.g., Fe-peak and heavy elements) may be pushed toward the surface by radiative levitation, while others (e.g., He, Mg, Ca) tend to sink under gravity \citep{2009ssc..book.....G}. For diffusion to operate efficiently, rotational velocities must generally remain below $\sim 120$~km~s$^{-1}$, ensuring limited turbulent mixing \citep{2000ApJ...544..933A}.

Chemically peculiar stars are traditionally divided into magnetic and non-magnetic groups \citep{1974ARA&A..12..257P}. While Ap/Bp stars host strong magnetic fields, weak or undetectable magnetic fields characterize other CP subgroups, including the Mercury--Manganese (HgMn) stars and the metallic A (Am/Fm) stars. The present study focuses on these two weak-field subclasses, which share several physical and observational properties.

HgMn stars are typically found in the spectral range B8--A0 and are characterized by strong Hg~\textsc{ii} and Mn~\textsc{ii} lines, along with enhancements in heavy elements such as P, Ga, Sr, Y, Xe, and Pt \citep{1996Ap&SS.237...77S, 2016MNRAS.460.1912G}. They generally exhibit slow rotation and a high incidence of binarity \citep{1995ComAp..18..167H, 2000ApJ...544..933A}. Observational studies have shown that the number of HgMn stars decreases with increasing rotational velocity \citep{1974ApJ...194...65W}, reinforcing the importance of rotational braking in enabling diffusion-driven anomalies.

Similarly, Am stars exhibit Ca and Sc underabundances, along with overabundances of iron-peak and heavy elements such as Fe, Zn, Sr, and Ba \citep{1974ARA&A..12..257P, 2009ssc..book.....G}. These stars occupy the spectral range A0--F4 and are also predominantly slow rotators, frequently found in spectroscopic binary systems \citep{1995ApJS...99..135A, 2017MNRAS.465.2662S}. Tidal interactions in close binaries are believed to play a crucial role in reducing rotational velocities, thereby facilitating atomic diffusion.

The similarities between HgMn and Am stars, slow rotation, binarity fraction, weak magnetic fields, and diffusion-dominated atmospheres have long motivated the hypothesis of a possible evolutionary connection between the two groups. Evolutionary investigations indicate that HgMn stars tend to occupy the hotter, generally less-evolved region of the main sequence. In contrast, Am stars are found at slightly cooler temperatures and often at more advanced evolutionary stages \citep{2003A&A...397..267A}. Based on this distribution, it has been suggested that the coolest HgMn stars may evolve into the hottest Am stars.

Recent spectroscopic surveys have expanded the known populations of these stars and provided new insights into their abundance patterns. Large-sample analyses using LAMOST DR4 data identified dozens of new HgMn stars and confirmed systematic heavy-element enhancements \citep{2021A&A...645A..34P}. Infrared spectroscopic analyses from APOGEE have revealed numerous HgMn candidates exhibiting strong signatures of Mn, Hg, and other heavy elements \citep{2020MNRAS.496..832C}. Detailed abundance studies of individual objects further demonstrate the diversity of chemical patterns within the HgMn class \citep{2021MNRAS.502.3670G}.

For Am stars, modern high-resolution abundance analyses confirm characteristic Ca and Sc deficiencies, iron-group enrichments, and, in some cases, rare-earth overabundances \citep{2021MNRAS.500.2451W}. Moreover, detailed investigations of specific systems, such as V772~Cas, have shown that surface chemical inhomogeneities may be present even in weak-magnetic-field CP stars, suggesting a more complex interplay among diffusion, rotation, and stellar structure \citep{2021MNRAS.500.2577K}.

Particularly compelling evidence for a possible evolutionary link was presented by \citet{2018ApJ...852..116K}, who identified abundance characteristics consistent with an evolved HgMn nature in specific systems. \citet{1982ARA&A..20...37V, 2003A&A...397..267A} also demonstrated that the coolest HgMn stars and the hottest Am stars occupy overlapping regions in the Hertzsprung--Russell (HR) diagram and may follow similar evolutionary tracks. %These findings strongly support the notion that the chemical and physical differences between the two classes could represent different evolutionary stages of stars with comparable initial masses and rotational histories.

Despite the studies on HgMn and Am stars, a comprehensive and homogeneous spectroscopic and evolutionary comparison of HgMn and Am stars remains lacking. Many previous studies rely on heterogeneous atmospheric parameters compiled from the literature or adopt generic evolutionary tracks rather than models tailored to individually determined stellar parameters.

In this study, we performed a homogeneous high-resolution spectroscopic analysis of 13 cool HgMn and 14 hot Am targets to determine precise atmospheric parameters and detailed chemical abundances. Using these results as input constraints, we computed evolutionary models with the MESA code to investigate whether cool HgMn stars can evolve into hot Am stars. We aim to provide observational and theoretical evidence addressing the long-standing question of whether cool HgMn stars may serve as progenitors of hot Am stars.

\section{Target Selection and Spectroscopic Data}

The stars analyzed in this study consist of chemically peculiar HgMn and Am targets. The initial sample was selected from the catalog of CP stars compiled by \citet{2009A&A...498..961R}. The targets were selected based not only on their chemical classification but also on their suitability for investigating the proposed evolutionary connection between cool HgMn (B8--A0) and hot Am (A0--A2) stars. This choice allows us to investigate the possible transition between the two groups along the effective temperature ($T_{\rm eff}$) axis.

The main goal of this work is to test whether cool HgMn stars can evolve into hot Am stars. For this purpose, reliable evolutionary modeling requires accurate atmospheric parameters such as $T_{\rm eff}$, surface gravity ($\log g$), metallicity (Z), and projected rotational velocity ($v \sin i$). Therefore, only stars with available high-resolution spectra in public archives were included in the final sample.

We restricted our search to spectrographs with resolving power $R \gtrsim 30~000$. The ESO Science Archive Facility was extensively explored, together with the ELODIE and SOPHIE archives. In particular, data obtained with HARPS ($R \sim 115~000$; \citealt{2011arXiv1109.2497M}), UVES ($R \sim 40~000$; \citealt{2000SPIE.4008..534D}), FEROS ($R \sim 48~000$; \citealt{1999Msngr..95....8K}), and GIRAFFE ($R \sim 55~000$--$65~000$; \citealt{2004A&A...420L..31F}) were considered. High-resolution spectra from ELODIE ($R \sim 45~000$) and SOPHIE ($R \sim 75~000/39~000$; \citealt{2004PASP..116..693M}) were also included.

Among the stars listed in \citet{2009A&A...498..961R}, those with suitable archival spectra were selected. As a result, the final sample consists of 14 hot Am stars and 13 cool HgMn stars. Some of these stars have not been studied with high-resolution spectroscopy, while others have been analyzed only at medium or low resolution. Including such objects allows us to update their atmospheric parameters and derive more accurate chemical abundances. Basic information about the selected targets and the corresponding spectroscopic data is given in Table~\ref{tab:targets}. Additional references for stars previously analyzed in the literature are provided in Table~\ref{tab:A1} in the Appendix.

Although the archival spectra collected for this study had already been processed for bias correction, flat-fielding, and wavelength calibration, they were not always normalized to the continuum level. Since accurate continuum normalization is essential for reliable line-profile analysis and abundance determination, all spectra were therefore re-normalized before further analysis. The normalization was performed using the SUPPNet (Suppression Network) software developed by \citet{2022A&A...659A.199R}. SUPPNet is based on a neural network algorithm that automatically and consistently separates spectral lines from the continuum. This approach is particularly useful for CP stars, where line blending is common and classical polynomial fitting methods may fail. The program takes one-dimensional spectra (wavelength--flux) as input and provides normalized spectra as output. The resulting spectra were visually inspected to ensure that no artificial structures or over-corrections were introduced. When necessary, minor manual adjustments were applied.

In addition, the spectra were examined for possible signatures of binarity. Radial velocity shifts were investigated by cross-correlating with synthetic template spectra, and secondary spectral features were carefully examined. The synthetic spectra used as templates were generated using the ATLAS9 model atmospheres \citep{1993sssp.book.....K} and the SYNTHE code \citep{1981SAOSR.391.....K}. The $T_{\rm eff}$ values of the templates were selected according to the values listed in Table~\ref{tab:A1}. Specifically, synthetic spectra with $T_{\mathrm{eff}} = 7000$, $8000$, and $9000$ K were computed for Am stars, while models with $T_{\mathrm{eff}} = 11000$, $12000$, and $13000$ K were generated for HgMn stars. The $\log g$ was fixed at $\log g = 4.0$ for all models. $v \sin i$ values were adopted based on the average values given in Table~\ref{tab:A1} and applied to the synthetic spectra to ensure consistency with the observed line broadening. These tailored templates were then used in the cross-correlation analysis to search for radial velocity variations and to identify possible secondary components.

For two Am (BD+52~1733, CD-49~4801) and one HgMn (BD+40~407) targets, evidence of a secondary component was detected. Since the majority of the targets were single stars and to maintain a homogeneous sample, these stars were replaced with alternative objects selected using the same criteria described above. The final list of targets analyzed in this work is presented in Table~\ref{tab:targets}.
%In addition, the spectra were examined for possible binarity signatures. Radial velocity shifts were assessed by cross-correlation with synthetic spectra, and secondary spectral features were carefully inspected. For two metallic A stars and one HgMn star, evidence of a secondary component was detected. Since the majority of the targets were single stars, and to keep the sample homogeneous, these stars were replaced with alternative objects selected using the same criteria described above. The final list of targets analysed in this work is presented in Table~\ref{tab:targets}.

\begin{table*}
\centering
\small
\caption{Selected hot Am and cool HgMn targets analyzed in this study. S/N denotes the average signal-to-noise ratio of the spectra measured at around 550~nm. The spectral types were taken from Renson et. al. (2009).}
\label{tab:targets}
\resizebox{\textwidth}{!}{
\begin{tabular}{llllllll}
\hline
\hline
Name & RA (deg) & DEC (deg) & Spectral Type & V (mag) & Spectrograph & Number of Spectra & S/N \\
\hline
\multicolumn{8}{c}{\textbf{Am Targets}} \\
\hline
BD+20 527  & 48.725  & 21.044  & A1V       & 4.87 & UVES    & 1  & 344 \\
BD+25 641  & 58.394  & 25.683  & A0        & 6.35 & ELODIE  & 1  & 225 \\
BD+3 1437  & 102.914 & 3.042   & A1        & 6.37 & HARPS   & 16 & 130 \\
BD+50 1460 & 116.017 & 50.434  & A0        & 5.36 & SOPHIE  & 1  & 320 \\
BD-1 2074  & 128.507 & -2.152  & A0        & 5.79 & HARPS   & 26 & 145 \\
BD+19 2097 & 131.444 & 19.049  & A0        & 8.00 & SOPHIE  & 2  & 153 \\
BD+3 2280  & 148.051 & 2.454   & A1        & 6.02 & ELODIE  & 2  & 100 \\
BD+25 2498 & 185.545 & 24.774  & A1        & 6.16 & ELODIE  & 1  & 254 \\
BD-8 3372  & 188.445 & -9.452  & A0        & 5.48 & UVES    & 5  & 309 \\
BD-5 3535  & 189.197 & -5.832  & A2        & 5.88 & UVES    & 4  & 257 \\
CD-47 7893 & 192.581 & -48.459 & A0        & 6.26 & HARPS   & 26 & 134 \\
BD+69 850  & 246.996 & 68.768  & B9        & 4.95 & ELODIE  & 1  & 356 \\
BD+25 3246 & 26.041  & 25.537  & A2        & 5.36 & SOPHIE  & 1  & 465 \\
BD+50 2468 & 267.268 & 50.781  & A1        & 5.00 & ELODIE  & 1  & 180 \\
\hline
\multicolumn{8}{c}{\textbf{HgMn Star Targets}} \\
\hline
BD+28 4     & 2.097   & 29.090  & B8        & 2.06 & UVES    & 5  & 479 \\
BD+15 177   & 18.532  & 16.133  & B9        & 5.96 & ELODIE  & 2  & 139 \\
BD+21 535   & 57.479  & 22.244  & B9        & 6.07 & SOPHIE  & 1  & 385 \\
BD+20 733   & 64.859  & 21.142  & B9        & 5.48 & ELODIE  & 1  & 142 \\
BD+14 787   & 73.959  & 15.040  & B8        & 5.78 & FEROS   & 1  & 248 \\
BD-16 1072  & 78.233  & -16.205 & B9        & 3.29 & HARPS   & 16 & 323 \\
BD-20 1544  & 101.409 & -20.512 & B9        & 9.36 & GIRAFFE & 2  & 194 \\
CD-36 14166 & 307.195 & -35.596 & B9        & 6.08 & FEROS   & 1  & 254 \\
BD+40 5068  & 350.145 & 41.113  & B9        & 6.72 & ELODIE  & 1  & 146 \\
BD+24 4778  & 351.918 & 25.167  & A0        & 5.98 & ELODIE  & 1  & 148 \\
CD-38 15527 & 353.243 & -37.818 & B9        & 4.37 & UVES    & 3  & 290 \\
BD+15 4033  & 300.875 & 16.031  & B9        & 5.66 & ELODIE  & 2  & 240 \\
BD+5 3223   & 248.149 & 5.521   & B9        & 5.63 & ELODIE  & 1  & 90 \\
\hline
\hline
\end{tabular}
}
\end{table*}

\section{Analysis and parameter determination}
\subsection{Spectral Classification}

Spectral classification provides essential information about the chemical peculiarity status of a star and offers initial estimates of its atmospheric parameters. The procedure is based on comparing the observed spectra of the targets with standard stars whose spectral types (e.g., A0, F2) and luminosity classes (e.g., IV, V) are well established. %In this comparison, particular attention is given to hydrogen Balmer lines and selected metal lines.

Since most of the stars analyzed in this study fall within the B8--A2 spectral range, standard spectra of B- and A-type stars published by \citet{2009ssc..book.....G} were used as references. For Am targets, the classification was performed by examining hydrogen and metal features simultaneously. The basic spectral type was first determined from the H$\gamma$ and H$\delta$ lines. This was followed by an inspection of metal lines such as Fe, Ca, and Mg, including their relative strengths compared to the Balmer lines, which provide information about chemical peculiarities. In addition, the Ca~\textsc{ii} K line was used as an indicator. %for stars earlier than F3, while for later F-type stars, the G-band served as an auxiliary classification criterion.
In normal (non-peculiar) A-type stars, the spectral types derived from these three criteria are expected to be consistent. However, in Am stars, discrepancies between classifications based on hydrogen and metal lines are observed \citep{2009ssc..book.....G}. These differences arise from abnormal surface abundances and therefore require a careful and multi-criteria evaluation.

The luminosity class of A-type stars was determined using additional spectral indicators. In particular, the blended profiles of ionized iron (Fe~\textsc{ii}) and ionized titanium (Ti~\textsc{ii}) lines around 4500~\AA\ were used as primary diagnostics \citep{2009ssc..book.....G}. For A and early F-type stars, the shape and width of the Balmer lines also provide information on luminosity class, while for later F-type stars, the G-band serves as a useful indicator. The final luminosity class was assigned based on the combined evaluation of these features.

For B-type HgMn targets, the classification was primarily based on hydrogen Balmer lines and metal lines. When the spectral coverage allowed, the presence of Hg at 3984~\AA\ and Mn lines at 4136, 4206, and 4252~\AA\ was also examined as additional confirmation of the HgMn nature.

This multi-criteria classification approach allowed us to reliably determine both the spectral type and the luminosity class of the analyzed stars, while also providing preliminary insight into their chemical peculiarity. The updated spectral classifications of the targets are presented in Table~\ref{tab:phot_params}.

\subsection{Atmospheric Parameter Determination}

Before analyzing the high-resolution spectra in detail, initial atmospheric parameters were estimated using photometric indicators. Johnson $B\!-\!V$ color and Strömgren $uvby\beta$ indices were adopted to derive preliminary values of $T_{\rm eff}$ and $\log g$. Since photometric measurements are sensitive to interstellar reddening, the color excess $E(B-V)$ was determined before the parameter estimation.

The $E(B-V)$ values were obtained from three-dimensional dust maps using the \texttt{dustmaps} Python package \citep{2018JOSS....3..695G}. In particular, the Bayestar 2019 map \citep{2019ApJ...887...93G} was queried using the Galactic coordinates and distances of the targets. Gaia DR3 parallaxes \citep{2023A&A...674A...1G} were adopted when available; otherwise, Hipparcos distances \citep{1997A&A...323L..49P} were used. The derived $E(B-V)$ values were applied to correct the photometric colours.

Strömgren indices $(b-y)$, $m_1$, $c_1$, and $\beta$ were used to calculate $T_{\rm eff}$ and $\log g$ following the calibration of \citet{1985MNRAS.217..305M}, which is suitable for early-type stars. Additionally, $T_{\rm eff}$ values were independently derived from the Johnson $B\!-\!V$ color using the calibration of \citet{2000AJ....120.1072S}, assuming $\log g = 4.0$ and solar metallicity as initial conditions. In most cases, the $T_{\rm eff}$ derived from the two systems was consistent within uncertainties. The determined photometric parameters and $E(B-V)$ values are listed in Table~\ref{tab:phot_params}. Uncertainties were estimated by considering photometric errors, an assumed reddening uncertainty of 0.02~mag, and possible variations in metallicity and $\log g$.

Refined atmospheric parameters were then obtained from spectroscopic analysis. The $T_{\rm eff}$ and $\log g$ were first constrained using the Balmer lines (H$\beta$/~H$\gamma$/~H$\delta$) through spectrum synthesis. Observed profiles were compared with synthetic spectra computed using ATLAS9 model atmospheres \citep{1993sssp.book.....K} and the SYNTHE code \citep{1981SAOSR.391.....K}. The models assume plane-parallel geometry, hydrostatic equilibrium, and local thermodynamic equilibrium (LTE).

\begin{table*}
\centering
\small
\caption{Photometric $T_{\rm eff}$ and $\log g$ values, updated spectral classifications, and $E(B-V)$ values of the analyzed stars. * indicates spectra where the Ca~\textsc{ii} K line is not available. ** stars whose spectra do not cover the Hg line. Mn lines are found to be overabundant in all HgMn stars.}
\label{tab:phot_params}
\resizebox{\textwidth}{!}{
\begin{tabular}{llllllll}
\hline
\hline
Name & Spectral Type & $E(B-V)$ & $T_{\rm eff}^{uvby}$ (K) & $\log g_{uvby}$ & $T_{\rm eff}^{BV}$ (K) & $T_{\rm eff}^{TIC}$ (K) & $\log g^{TIC}$ \\
\hline
\multicolumn{8}{c}{\textbf{Am Targets}} \\
\hline
BD+20 527  & kA1hB9mA2V  & 0.000 & $9710 \pm 390$  & $3.86 \pm 0.12$ & $9540 \pm 445$  & $9528 \pm 117$  & $3.88 \pm 0.07$ \\
BD+25 641  & A2IV*       & 0.130 & --               & --              & $8435 \pm 360$  & $8413 \pm 267$  & $3.49 \pm 0.07$ \\
BD+3 1437  & A3V         & 0.000 & $9140 \pm 414$  & $4.16 \pm 0.18$ & $9080 \pm 420$  & $8954 \pm 134$  & $4.10 \pm 0.07$ \\
BD+50 1460 & A0III       & 0.000 & $9880 \pm 206$  & $3.50 \pm 0.02$ & $9660 \pm 500$  & $9654 \pm 198$  & $3.56 \pm 0.09$ \\
BD-1 2074  & kA1hA0mA2V  & 0.000 & $9749 \pm 80$   & $4.05 \pm 0.19$ & $9550 \pm 340$  & $9618 \pm 148$  & $4.06 \pm 0.06$ \\
BD+19 2097 & kA2hA1mA3V  & 0.000 & --               & --              & $7650 \pm 230$  & $7737 \pm 130$  & $4.01 \pm 0.08$ \\
BD+3 2280  & hA0mA3III*  & 0.000 & $10200 \pm 216$ & $3.77 \pm 0.24$ & $9930 \pm 910$  & $10071 \pm 131$ & $3.55 \pm 0.06$ \\
BD+25 2498 & hA0mA2V     & 0.000 & $9750 \pm 520$  & $4.10 \pm 0.08$ & $9570 \pm 490$  & $9625 \pm 113$  & $4.15 \pm 0.06$ \\
BD-8 3372  & B9V*        & 0.010 & $10410 \pm 226$ & $4.13 \pm 0.18$ & $10080 \pm 850$ & $10209 \pm 123$ & $4.12 \pm 0.07$ \\
BD-5 3535  & kA2hA1mA3V  & 0.000 & $8920 \pm 300$  & $4.20 \pm 0.06$ & $8910 \pm 460$  & $8770 \pm 100$  & $4.22 \pm 0.07$ \\
CD-47 7893 & A1IV        & 0.157 & $9410 \pm 240$  & $3.95 \pm 0.08$ & $9230 \pm 540$  & $9333 \pm 163$  & $3.72 \pm 0.07$ \\
BD+69 850  & hB9mA1IV*   & 0.000 & $10130 \pm 320$ & $3.59 \pm 0.02$ & $9880 \pm 510$  & $9994 \pm 177$  & $3.47 \pm 0.07$ \\
BD+25 3246 & kA0hA1mA3V  & 0.000 & $8675 \pm 240$  & $3.50 \pm 0.04$ & $9000 \pm 530$  & $9038 \pm 134$  & $3.50 \pm 0.07$ \\
BD+50 2468 & A1V*        & 0.000 & $9293 \pm 180$  & $4.24 \pm 0.09$ & $9200 \pm 420$  & $9099 \pm 105$  & -- \\
\hline
\multicolumn{8}{c}{\textbf{HgMn Star Targets}} \\
\hline
BD+28 4     & B8V*       & 0.000 & $13940 \pm 460$ & $4.09 \pm 0.10$ & $10210 \pm 520$ & --              & -- \\
BD+15 177**   & B8IV/III   & 0.000 & $12900 \pm 880$ & $3.99 \pm 0.12$ & $10380 \pm 610$ & $12605 \pm 80$  & -- \\
BD+21 535   & B8IV       & 0.078 & $13120 \pm 90$  & $3.98 \pm 0.10$ & $11340 \pm 3860$& $11706 \pm 603$ & $4.00 \pm 0.10$ \\
BD+20 733**   & B9IV       & 0.000 & $12000 \pm 240$ & $4.21 \pm 0.06$ & $10730 \pm 3590$& $11376 \pm 135$ & $4.30 \pm 0.06$ \\
BD+14 787   & B8V        & 0.000 & $14000 \pm 530$ & $4.02 \pm 0.10$ & $10360 \pm 630$ & $13456 \pm 80$  & -- \\
BD-16 1072  & B8IV/V     & 0.000 & $12740 \pm 350$ & $3.69 \pm 0.08$ & $10670 \pm 650$ & $12274 \pm 85$  & -- \\
BD-20 1544**  & --         & 0.088 & --              & --              & $11160 \pm 650$ & $13082 \pm 128$ & -- \\
CD-36 14166 & B8III      & 0.049 & $13820 \pm 440$ & $3.60 \pm 0.10$ & $11060 \pm 1840$& $12601 \pm 170$ & -- \\
BD+40 5068**  & B8IV/V     & 0.120 & $12160 \pm 280$ & $3.35 \pm 0.09$ & $10280 \pm 3430$& $10192 \pm 124$ & -- \\
BD+24 4778**  & B9IV/V     & 0.030 & $10980 \pm 230$ & $4.06 \pm 0.06$ & $10500 \pm 1140$& $11237 \pm 256$ & $4.21 \pm 0.06$ \\
CD-38 15527** & B8IV/V     & 0.016 & $12670 \pm 640$ & $4.20 \pm 0.05$ & $10680 \pm 880$ & $12106 \pm 84$  & -- \\
BD+15 4033**  & B8IV*      & 0.000 & $13220 \pm 560$ & $3.51 \pm 0.07$ & $10445 \pm 350$ & $12823 \pm 148$ & -- \\
BD+5 3223** & B9IV*      & 0.000 & $10920 \pm 450$ & $3.85 \pm 0.05$ & $9912 \pm 280$  & $10623 \pm 270$ & $3.94 \pm 0.06$ \\
\hline
\hline
\end{tabular}
}
\end{table*}

\begin{figure}
 \centering
 \begin{minipage}[t]{0.5\textwidth}
  \centering
  \includegraphics[width=\linewidth]{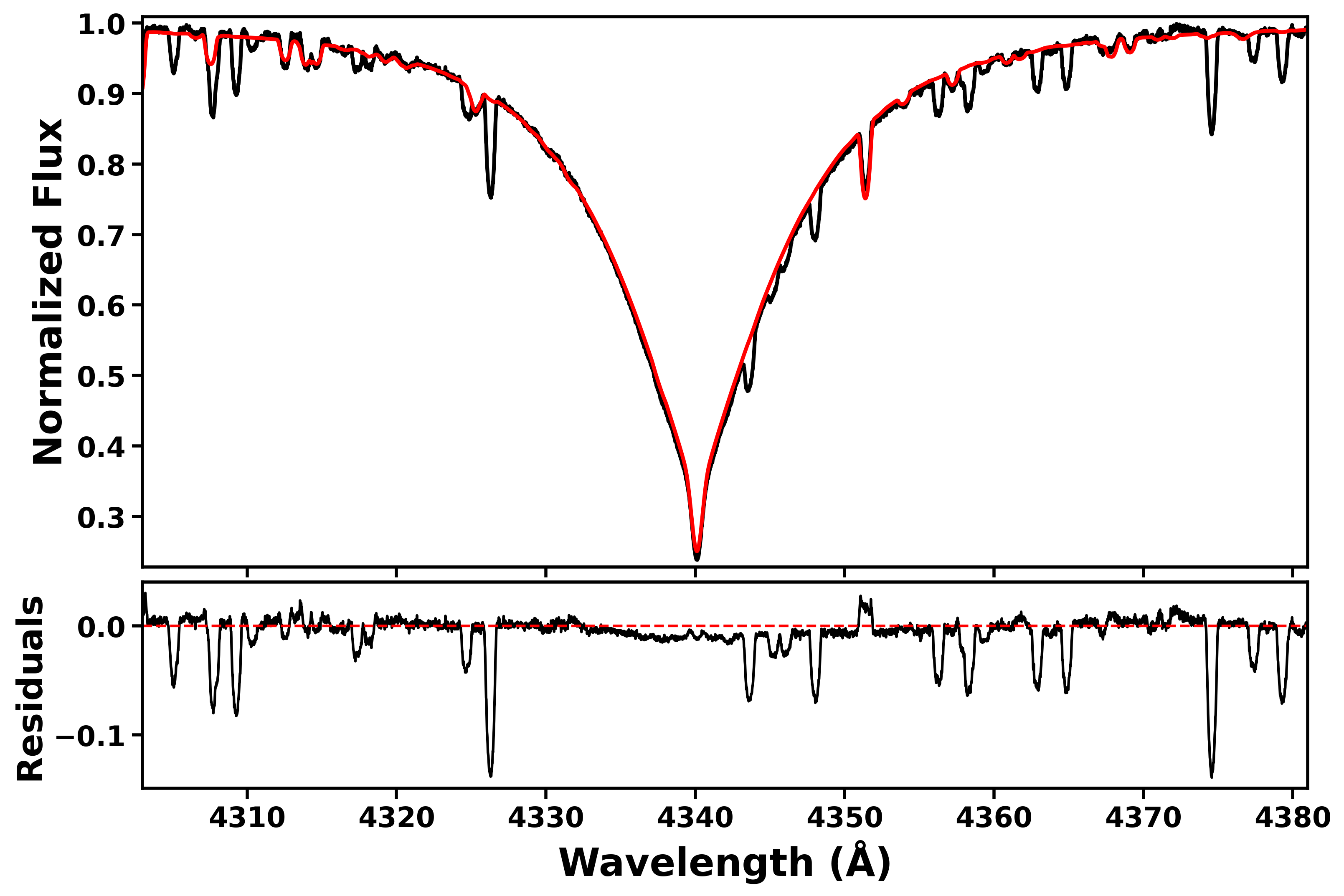}
 \end{minipage}\hfill
 \caption{Comparison between the observed (black) and synthetic (red) H$\gamma$ spectra of CD$-$38~1552. The lower panel shows the residuals (observed minus synthetic). The synthetic spectrum was computed using the adopted stellar atmospheric parameters and solar chemical composition. Therefore, only the H$\gamma$ line profile is expected to be well reproduced.}
 \label{fig:hlinefit}
\end{figure}

Balmer lines are particularly sensitive to both $T_{\rm eff}$ and $\log g$ in B- and early A-type stars. Therefore, starting from the photometric estimates, synthetic spectra were generated within a grid with steps of 100~K in $T_{\rm eff}$ and 0.1~dex in $\log g$ assuming the solar composition. The best-fitting parameters were determined using a minimum-difference approach following \citet{2004A&A...425..641C}. Projected rotational velocities ($v \sin i$) were adopted from the literature when available or initially estimated from metal lines and kept fixed during the Balmer fitting to account for rotational broadening. The resulting parameters are given in Table~\ref{tab:final_params}, and an example Balmer line fit is shown in Fig.~\ref{fig:hlinefit}.

Final atmospheric parameters were derived using Fe lines through spectrum synthesis, following the procedure described by \citet{2015MNRAS.450.2764N}. This method is suitable for both slow and fast rotators, especially when line blending is significant.

The analysis relies on the different sensitivities of Fe~\textsc{i} and Fe~\textsc{ii} lines. Fe~\textsc{i} lines are primarily sensitive to $T_{\rm eff}$ and only weakly dependent on $\log g$, whereas Fe~\textsc{ii} lines are more sensitive to $\log g$ through the ionization balance. The sensitivity of spectral lines to microturbulent velocity ($\xi$) depends mainly on line strength and saturation rather than on the ionization stage. Weak lines are only marginally affected by $\xi$, while stronger and partially saturated lines exhibit a greater sensitivity to changes in $\xi$. Therefore, $\xi$ was determined by minimizing any correlation between the derived Fe abundances and line strength (or line depth) \citep{2008oasp.book.....G}.

The parameters were determined iteratively in three steps: (1) $\xi$ was adjusted to remove any correlation between the derived Fe abundances and line strength; 2) $T_{\rm eff}$ was refined by minimizing the trend between Fe~\textsc{i} abundances and excitation potential; (3) $\log g$ was obtained by enforcing ionization equilibrium between Fe~\textsc{i} and Fe~\textsc{ii}. For all stars analyzed in this study, the line selection was performed using the Kurucz line list \citep{1995all..book.....K} and the adopted atmospheric parameters. Only lines predicted to have normalized central fluxes below 0.95 (corresponding to line depths greater than approximately 5\%) were considered in the final analysis. Depending on the target, typically several tens of Fe~\textsc{i} (and some Fe~\textsc{ii}) lines met this criterion and were used to determine atmospheric parameters and abundances. The atmospheric parameters derived from the Fe-line analysis were also found to be in good agreement with those independently constrained from Balmer-line profile fitting. Uncertainties of the Fe-based atmospheric parameters were estimated by evaluating the sensitivity of each parameter to variations in the others. The minimum uncertainties, based on the adopted grid steps, are 100~K in $T_{\rm eff}$, 0.1~dex in $\log g$, and 0.1~km~s$^{-1}$ in $\xi$. The final atmospheric parameters listed in Table~\ref{tab:final_params} were adopted for the subsequent chemical abundance analysis.

\begin{table*}
\centering
\small
\caption{Final atmospheric parameters derived from hydrogen Balmer and Fe lines' analyses, together with $v \sin i$ values. The superscript H represents the values derived from hydrogen lines. *For these targets, there were no hydrogen lines in the spectral coverage.}
\label{tab:final_params}
\resizebox{\textwidth}{!}{
\begin{tabular}{lcccccc}
\hline
\hline
Name & $T_{\rm eff}^{\rm H}$ (K) & $\log g^{\rm H}$ & $T_{\rm eff}$ (K) & $\log g$ & $\xi$ (km~s$^{-1}$) & $v \sin i$ (km~s$^{-1}$) \\
\hline
\multicolumn{7}{c}{\textbf{Am Targets}} \\
\hline
BD+20 527  & $9600 \pm 200$  & $3.80 \pm 0.20$ & $9800 \pm 200$  & $3.9 \pm 0.2$ & $1.5 \pm 0.3$ & $126 \pm 7$ \\
BD+25 641  & $9700 \pm 200$  & $3.90 \pm 0.10$ & $9700 \pm 200$  & $3.9 \pm 0.1$ & $1.3 \pm 0.2$ & $10 \pm 3$ \\
BD+3 1437  & $9200 \pm 100$  & $3.90 \pm 0.10$ & $9200 \pm 100$  & $3.9 \pm 0.1$ & $2.6 \pm 0.2$ & $59 \pm 3$ \\
BD+50 1460 & $10000 \pm 300$ & $3.50 \pm 0.10$ & $10000 \pm 300$ & $3.5 \pm 0.1$ & $2.5 \pm 0.4$ & $194 \pm 9$ \\
BD-1 2074  & $9800 \pm 200$  & $4.00 \pm 0.10$ & $9700 \pm 100$  & $4.0 \pm 0.1$ & $2.6 \pm 0.2$ & $4 \pm 1$ \\
BD+19 2097 & $7500 \pm 200$  & $4.00 \pm 0.10$ & $7500 \pm 100$  & $4.0 \pm 0.1$ & $2.6 \pm 0.1$ & $29 \pm 4$ \\
BD+3 2280  & $10200 \pm 300$ & $3.90 \pm 0.10$ & $10200 \pm 100$ & $3.9 \pm 0.1$ & $1.1 \pm 0.2$ & $29 \pm 2$ \\
BD+25 2498 & $9900 \pm 100$  & $4.20 \pm 0.10$ & $9800 \pm 100$  & $4.1 \pm 0.1$ & $2.1 \pm 0.2$ & $45 \pm 3$ \\
BD-8 3372  & $10400 \pm 300$ & $4.10 \pm 0.10$ & $10400 \pm 300$ & $4.1 \pm 0.1$ & $2.5 \pm 0.5$ & $141 \pm 20$ \\
BD-5 3535  & $8700 \pm 200$  & $4.00 \pm 0.20$ & $9100 \pm 300$  & $4.0 \pm 0.1$ & $1.5 \pm 0.5$ & $150 \pm 6$ \\
CD-47 7893 & $9400 \pm 200$  & $3.80 \pm 0.10$ & $9200 \pm 100$  & $3.8 \pm 0.1$ & $2.0 \pm 0.2$ & $39 \pm 3$ \\
BD+69 850  & $10200 \pm 300$ & $3.50 \pm 0.20$ & $9800 \pm 200$  & $4.0 \pm 0.3$ & $2.0 \pm 0.4$ & $164 \pm 10$ \\
BD+25 3246 & $9100 \pm 200$  & $3.60 \pm 0.20$ & $9100 \pm 200$  & $3.6 \pm 0.1$ & $2.0 \pm 0.1$ & $11 \pm 1$ \\
BD+50 2468 & $9200 \pm 300$  & $4.10 \pm 0.10$ & $9100 \pm 200$  & $4.0 \pm 0.2$ & $1.4 \pm 0.2$ & $118 \pm 4$ \\
\hline
\multicolumn{7}{c}{\textbf{HgMn Star Targets}} \\
\hline
BD+28 4     & $12800 \pm 400$ & $3.80 \pm 0.20$ & $12250 \pm 500$ & $4.0 \pm 0.1$ & $2.7 \pm 0.2$ & $57 \pm 5$ \\
BD+15 177   & $13500 \pm 300$ & $4.10 \pm 0.10$ & $13500 \pm 250$ & $4.2 \pm 0.1$ & $1.9 \pm 0.3$ & $22 \pm 5$ \\
BD+21 535   & $13000 \pm 500$ & $4.00 \pm 0.20$ & $13000 \pm 250$ & $4.0 \pm 0.1$ & $1.0 \pm 0.1$ & $72 \pm 6$ \\
BD+20 733   & $11400 \pm 400$ & $3.90 \pm 0.10$ & $11500 \pm 250$ & $4.3 \pm 0.1$ & $1.2 \pm 0.3$ & $5 \pm 1$ \\
BD+14 787   & $13800 \pm 500$ & $3.80 \pm 0.20$ & $13750 \pm 250$ & $4.0 \pm 0.1$ & $1.3 \pm 0.3$ & $76 \pm 4$ \\
BD-16 1072  & $12800 \pm 400$ & $3.70 \pm 0.10$ & $12500 \pm 500$ & $3.7 \pm 0.1$ & $2.3 \pm 0.2$ & $16 \pm 2$ \\
BD-20 1544*  & --              & --              & $10900 \pm 200$ & $4.0 \pm 0.1$ & $1.8 \pm 0.2$ & $12 \pm 2$ \\
CD-36 14166 & $13000 \pm 500$ & $3.50 \pm 0.10$ & $13750 \pm 750$ & $3.6 \pm 0.1$ & $1.4 \pm 0.2$ & $35 \pm 3$ \\
BD+40 5068  & $13000 \pm 500$ & $3.60 \pm 0.10$ & $12750 \pm 500$ & $3.5 \pm 0.1$ & $1.0 \pm 0.2$ & $26 \pm 2$ \\
BD+24 4778  & $11600 \pm 400$ & $4.20 \pm 0.10$ & $10700 \pm 500$ & $4.0 \pm 0.1$ & $1.0 \pm 0.2$ & $35 \pm 3$ \\
CD-38 15527 & $12000 \pm 750$ & $4.20 \pm 0.10$ & $12000 \pm 750$ & $4.2 \pm 0.1$ & $1.7 \pm 0.3$ & $26 \pm 1$ \\
BD+15 4033  & $13000 \pm 500$ & $3.60 \pm 0.10$ & $13500 \pm 500$ & $3.5 \pm 0.1$ & $0.8 \pm 0.2$ & $14 \pm 3$ \\
BD+5 3223   & $11000 \pm 250$ & $3.80 \pm 0.10$ & $10900 \pm 500$ & $3.9 \pm 0.1$ & $1.3 \pm 0.3$ & $10 \pm 2$ \\
\hline
\hline
\end{tabular}
}
\end{table*}

\subsection{Chemical Abundance Analysis}

After determining the fundamental atmospheric parameters ($T_{\rm eff}$, $\log g$, and $\xi$), we performed a detailed chemical abundance analysis based on high-resolution spectra. Our main aim was to derive reliable relative abundances of chemical elements in the atmospheres of cool HgMn and hot Am stars.

\begin{figure}
 \centering
 \begin{minipage}[t]{0.5\textwidth}
  \centering
  \includegraphics[width=\linewidth]{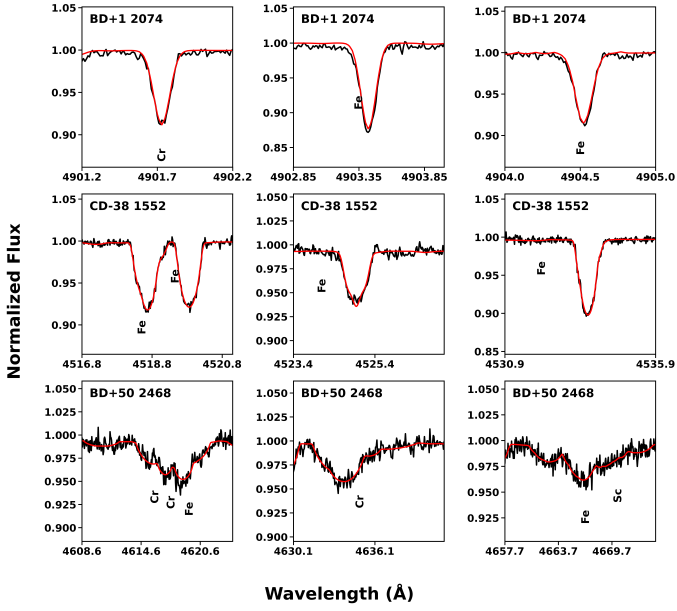}
 \end{minipage}\hfill
 \caption{Comparison between the observed (black) and synthetic (red) spectra of three stars in selected wavelength regions. From top to bottom, the panels correspond to BD$+$1~2074, CD$-$38~1552, and BD$+$50~2468, respectively, representing stars with increasing $v\sin i$. Each column shows a different spectral interval. The spectra were obtained using different instruments, and the synthetic spectra were computed using the adopted stellar atmospheric parameters. The dominant contributing chemical species to the absorption features are indicated by bold labels.}
 \label{fig:abunfit}
\end{figure}

For each target, the observed spectrum was divided into smaller wavelength segments by taking the $v\sin i$ values into account. For slow rotators, narrow intervals (typically 1--5~nm) were selected, which usually contain one or a few isolated or mildly blended lines. For moderate and fast rotators, broader intervals were preferred because rotational broadening increases line blending. Line identification in each segment was carried out using the Kurucz line list \citep{1995all..book.....K}. Since we adopted spectrum synthesis rather than equivalent-width measurements, all potentially contributing lines within a given interval were included. Blended lines were not removed; instead, the contribution of each element was evaluated directly through synthetic spectra. For each star, the line list was refined using its atmospheric parameters, and transitions producing line depths above 5\% in the relevant wavelength range were considered in the final selection.

Abundances were derived using the spectrum synthesis method by iteratively minimizing the difference between observed and synthetic spectra. In practice, synthetic spectra were adjusted until the best match was obtained for each wavelength segment, using a least-squares approach. This procedure allows abundances and $v\sin i$ to be constrained simultaneously. Examples of observed versus synthetic fits for stars observed with different spectrographs are presented in Fig.~\ref{fig:abunfit}.

\begin{figure}
 \centering
 \begin{minipage}[t]{0.47\textwidth}
  \centering
  \includegraphics[width=\linewidth]{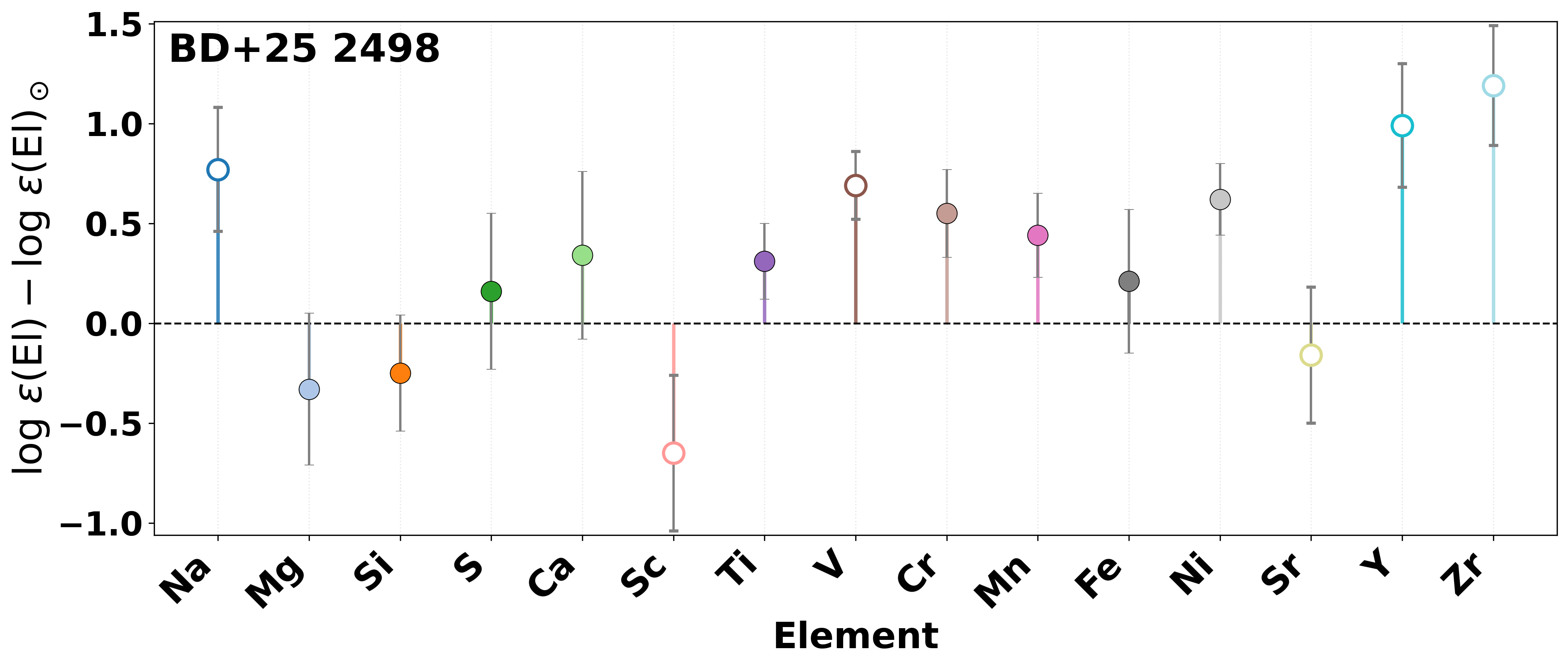}
 \end{minipage}\hfill
\begin{minipage}[t]{0.47\textwidth}
   \centering
  \includegraphics[width=\linewidth]{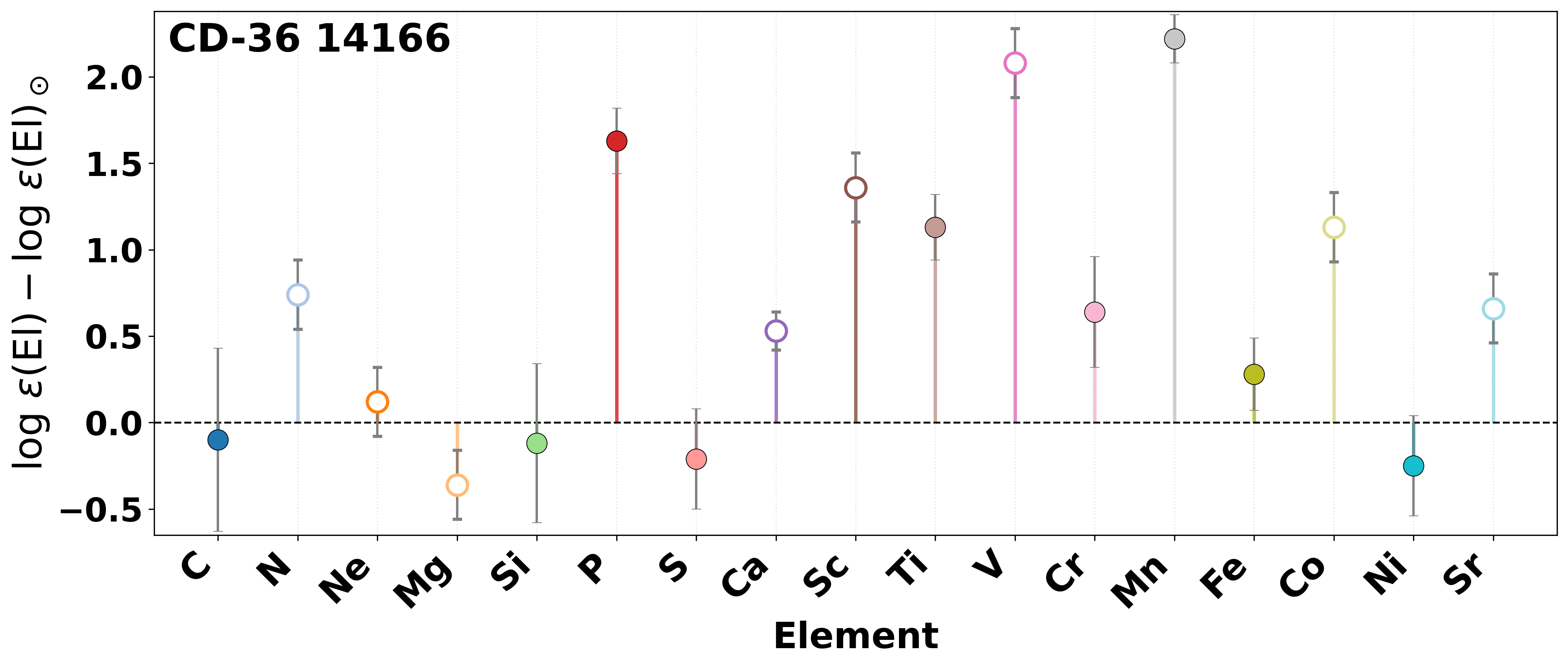}
 \end{minipage}
 \caption{Elemental abundance pattern of BD$+$25 2498 and CD$-$36~14166 relative to the Sun \citep{2009ARA&A..47..481A}, expressed as $\log \epsilon(\mathrm{El}) - \log \epsilon(\mathrm{El})_{\odot}$ as a function of element. Open circles denote abundances derived from fewer than five spectral lines, while filled circles represent abundances based on five or more lines. Error bars indicate the uncertainties of the abundance determinations. The dashed line marks the solar reference level.}
 \label{fig:abundis}
\end{figure}

For each star, results from all analyzed segments were combined to determine mean elemental abundances and an updated $v\sin i$. The abundance patterns relative to the solar composition (adopting \citealt{2009ARA&A..47..481A}) are provided in the Appendix (Fig.~\ref{fig:ap1} and Fig.~\ref{fig:ap2}), and an example distribution is shown in Fig.~\ref{fig:abundis}. The line-to-line scatter (standard deviation) from the analyzed segments was adopted as the statistical uncertainty for each element.

The uncertainties listed in Tables~\ref{tab:table_abundance1} and \ref{tab:table_abundance2} correspond to the line-to-line scatter (standard deviation) of the abundances derived from the analyzed spectral regions and therefore represent the internal statistical uncertainties of the abundance determination. Additional systematic uncertainties may arise from uncertainties in the adopted atmospheric parameters ($T_{\rm eff}$, $\log g$, and $\xi$), $v\sin i$, continuum normalization, S/N, and spectral resolution. Based on previous abundance studies of A- and B-type stars, these effects are expected to contribute abundance uncertainties of the order of 0.2~dex for most elements \citep{2020MNRAS.499.3706M, 2016MNRAS.458.2307K}. %They may reach up to about 0.3~dex in spectra affected by strong line blending or high rotational broadening. }

Furthermore, the present analysis was performed under the LTE assumption. Non-LTE effects were not explicitly considered and may introduce systematic abundance offsets for some elements. Recent studies have shown that NLTE corrections are generally small for many Fe-peak elements, but may reach several tenths of a dex for ions such as Sc~\textsc{ii}, Sr~\textsc{ii}, Zr~\textsc{ii}, and Ba~\textsc {ii} in stars with atmospheric parameters similar to those analyzed here \citep{2013A&A...551A..57H,2020MNRAS.499.3706M,2024MNRAS.527.8234M}. However, the magnitude and even the sign of the NLTE corrections depend on the spectral lines and stellar atmospheric parameters adopted, and therefore no general NLTE correction can be applied to the present sample. The abundances presented in this work should therefore be regarded as homogeneous LTE abundances suitable primarily for relative comparisons within the sample.

%In addition to statistical scatter, systematic uncertainties arise from model assumptions (e.g., LTE, plane-parallel geometry, hydrostatic equilibrium), uncertainties in atmospheric parameters ($T_{\rm eff}$, $\log g$, $\xi$), and atomic data. Following typical estimates in the literature, we assume that these effects can contribute an additional uncertainty of about $\pm 0.1$~dex to the final abundances \citep{2008A&A...478..529M}.

\subsection{Chemical Classification Based on Abundance Analysis}

The targets analyzed in this study were initially selected from the catalog of CP stars compiled by \citet{2009A&A...498..961R}. It should be noted that the classification of HgMn stars in the literature is primarily based on the presence of Hg and Mn lines in their spectra, making their identification relatively reliable. In contrast, the previous classification of Am targets was generally based on Strömgren photometric indices and/or spectral classification, which may not always accurately reflect their true chemical nature.

For instance, \citet{2025PASJ...77.1135K} showed that some stars previously classified as Am stars and exhibiting $\delta$ Scuti-type pulsations were, in fact, chemically normal A-type stars when analyzed using high-resolution spectroscopy. This highlights the importance of detailed spectroscopic analysis in confirming the chemical nature of such stars.

In this study, we reassessed the classifications of the target stars based on the derived chemical abundances. The evaluation was carried out by considering the characteristic abundance patterns of HgMn and Am stars described in previous sections, together with the element abundance distributions relative to the Sun (presented in Appendix~A). The results of this reclassification are summarized in Table~\ref{tab:chem_class}. According to this analysis, several stars initially classified as Am stars were found to exhibit chemically normal abundance patterns. In addition, one A-type star was identified as chemically metal-poor. On the other hand, all HgMn stars analyzed in this study display abundance patterns consistent with their literature classifications.

Overall, our spectroscopic analysis indicates that all 13 HgMn stars retain their CP nature, whereas only 6 out of 14 Am stars show abundance patterns consistent with Am characteristics.

\begin{table*}
\centering
\caption{Revised chemical classification of the analyzed targets based on abundance analysis.}
\label{tab:chem_class}
\small
\begin{tabular}{lll}
\hline
Name & Chemical properties & Classification \\
\hline
\multicolumn{3}{c}{\textbf{Am Targets}} \\
\hline
BD+20 527 & Rapid rotation; Mostly solar abundances & Normal A star \\
BD+25 641 & Cr, Mn, and Fe enhanced relative to the Sun; classification limited due to high rotation & Am star \\
BD+3 1437 & Slightly low Sc abundance, otherwise solar-like composition & Normal A star \\
BD+50 1460 & Fe, Ti, Cr underabundant; Ca and Sc near solar & Normal A star \\
BD-1 2074 & Sc underabundant; heavy elements enhanced & Am star \\
BD+19 2097 & Ca and Sc underabundant; heavy elements enhanced & Am star \\
BD+3 2280 & Heavy elements enhanced despite normal Ca and Sc & Am star \\
BD+25 2498 & Sc underabundant; Ca and heavy elements enhanced & Am star \\
BD-8 3372 & Rapid rotation; spectral classification indicates normal A star & Normal A star \\
BD-5 3535 & Rapid rotation; No clear Fe-peak enhancement & Normal A star \\
CD-47 7893 & Heavy elements underabundant; possible $\lambda$ Bootis nature & Metal-poor star \\
BD+69 850 & Rapid rotation; Heavy-element enhancement without classical Am pattern & Normal A star \\
BD+25 3246 & Ca and Sc underabundant; heavy elements enhanced & Am star \\
BD+50 2468 & Mostly solar or sub-solar abundances & Normal A star \\
\hline
\multicolumn{3}{c}{\textbf{HgMn Star Targets}} \\
\hline
BD+28 4 & Strong Mn and Hg overabundance (Hg from 398.4 nm line) & HgMn star \\
BD+15 177 & Mn overabundant; Hg line not measurable* & HgMn star \\
BD+21 535 & Strong Mn and Hg overabundance & HgMn star \\
BD+20 733 & Strong Mn overabundance; Hg not measurable* & HgMn star \\
BD+14 787 & Strong Mn and Hg overabundance & HgMn star \\
BD-16 1072 & Strong Mn and Hg overabundance & HgMn star \\
BD-20 1544 & Mn overabundant; Hg not measurable* & HgMn star \\
CD-36 14166 & Strong Mn and Hg overabundance & HgMn star \\
BD+40 5068 & Mn overabundant; Hg not measurable* & HgMn star \\
BD+24 4778 & Mn overabundant; Hg not measurable* & HgMn star \\
CD-38 15527 & Mn moderately enhanced; Hg not measurable* & HgMn star \\
BD+15 4033 & Mn moderately enhanced; Hg not measurable* & HgMn star \\
BD+5 3223 & Mn moderately enhanced; Hg not measurable* & HgMn star \\
\hline
\end{tabular}
\begin{itemize}
    \item *due to the limited wavelength range of the spectra
\end{itemize}
\end{table*}

\subsection{Calculation of Astrophysical Parameters}

Using the atmospheric parameters derived from the spectroscopic analysis, we calculated the fundamental astrophysical parameters only for the Am and HgMn stars identified in Table~\ref{tab:chem_class}. The luminosity ($L$), absolute ($M_V$) and bolometric ($M_{\rm bol}$) magnitudes were calculated. The $L$ values of the stars were determined using the Gaia DR3 parallaxes \citep{2023A&A...674A...1G} together with apparent magnitudes ($m_V$). The $M_V$ values were computed using the standard relation:

\begin{equation}
M_V = m_V - 5 \log d + 5 - A_V
\end{equation}

where $d$ is the distance in parsecs derived from Gaia parallaxes, and $A_V$ is the interstellar extinction in the V band. The extinction values were estimated from the color excess $E(B-V)$ assuming $A_V = 3.1 \times E(B-V)$.

Bolometric corrections (BC) were adopted from the calibration of \citet{1996ApJ...469..355F} as a function of $T_{\rm eff}$. The $M_{\rm bol}$ values were then computed as:

\begin{equation}
M_{\rm bol} = M_V + BC
\end{equation}

The $L$ parameters were calculated using:

\begin{equation}
\log \left( \frac{L}{L_\odot} \right) = -0.4 \left( M_{\rm bol} - M_{{\rm bol},\odot} \right)
\end{equation}

where $M_{{\rm bol},\odot} = 4.74$ mag \citep{2010A&ARv..18...67T}.

The derived astrophysical parameters were used to place the stars on the HR diagram and to compare their positions with theoretical evolutionary tracks. These positions provide essential constraints for the evolutionary modeling presented in Sect.~\ref{evo}.

The uncertainties in the derived parameters were estimated by propagating the errors in $T_{\rm eff}$, parallax, photometry, and reddening. Typical uncertainties are dominated by the errors in distance and $T_{\rm eff}$. The final astrophysical parameters for the Am and HgMn stars identified in Table~\ref{tab:chem_class} are listed in Table~\ref{tab:astro_params}.

\begin{table}
\centering
\footnotesize
\setlength{\tabcolsep}{4pt}
\caption{Calculated astrophysical parameters of the Am and HgMn stars identified in Table~\ref{tab:chem_class} of the study.}
\label{tab:astro_params}
\begin{tabular}{lccc}
\hline
\hline
Name &  $M_V$ (mag) & $M_{\rm bol}$ (mag) & $\log (L/L_{\odot})$ \\
\hline
\multicolumn{4}{c}{\textbf{Am Stars}} \\
\hline
%BD+20 527  & $0.246 \pm 0.080$ & $0.077 \pm 0.978$ & $1.865 \pm 0.391$ \\
BD+25 641  & $-0.263 \pm 0.015$ & $-0.451 \pm 0.975$ & $2.076 \pm 0.389$ \\
%BD+3 1437  & Normal A star   & $1.267 \pm 0.012$ & $1.172 \pm 0.975$ & $1.427 \pm 0.389$ \\
%BD+50 1460 & Normal A star   & $-0.244 \pm 0.101$ & $-0.493 \pm 0.980$ & $2.093 \pm 0.392$ \\
BD-1 2074  & $0.832 \pm 0.015$ & $0.644 \pm 0.975$ & $1.638 \pm 0.389$ \\
BD+19 2097 & $1.748 \pm 0.017$ & $1.783 \pm 0.975$ & $1.183 \pm 0.389$ \\
BD+3 2280  & $-0.510 \pm 0.021$ & $-0.801 \pm 0.975$ & $2.217 \pm 0.390$ \\
BD+25 2498 & $1.079 \pm 0.016$ & $0.871 \pm 0.975$ & $1.548 \pm 0.389$ \\
%BD-8 3372  & Normal A star   & $0.787 \pm 0.029$ & $0.453 \pm 0.975$ & $1.715 \pm 0.390$ \\
%BD-5 3535  & $1.621 \pm 0.013$ & $1.543 \pm 0.975$ & $1.279 \pm 0.390$ \\
%CD-47 7893 & Metal-poor star & $-0.438 \pm 0.048$ & $-0.533 \pm 0.976$ & $2.109 \pm 0.390$ \\
%BD+69 850  & $-0.891 \pm 0.026$ & $-1.099 \pm 0.975$ & $2.336 \pm 0.390$ \\
BD+25 3246 & $-0.349 \pm 0.046$ & $-0.412 \pm 0.976$ & $2.061 \pm 0.390$ \\
%BD+50 2468 & Normal A star   & $0.973 \pm 0.039$ & $0.895 \pm 0.976$ & $1.538 \pm 0.390$ \\
\hline
\multicolumn{4}{c}{\textbf{HgMn Stars}} \\
\hline
BD+28 4     & $-0.307 \pm 0.025$ & $-1.041 \pm 0.976$ & $2.313 \pm 0.390$ \\
BD+15 177   & $0.388 \pm 0.090$ & $-0.592 \pm 0.979$ & $2.133 \pm 0.392$ \\
BD+21 535   & $0.180 \pm 0.032$ & $-0.704 \pm 0.101$ & $2.178 \pm 0.390$ \\
BD+20 733   & $0.865 \pm 0.018$ & $0.289 \pm 0.975$ & $1.780 \pm 0.390$ \\
BD+14 787   & $0.032 \pm 0.042$ & $-0.994 \pm 0.976$ & $2.294 \pm 0.390$ \\
BD-16 1072  & $-0.296 \pm 0.038$ & $-1.082 \pm 0.976$ & $2.329 \pm 0.390$ \\
BD-20 1544  & $-0.257 \pm 0.044$ & $-0.701 \pm 0.976$ & $2.176 \pm 0.390$ \\
CD-36 14166 & $-0.861 \pm 0.036$ & $-1.887 \pm 0.977$ & $2.651 \pm 0.391$ \\
BD+40 5068  & $-1.695 \pm 0.053$ & $-2.531 \pm 0.977$ & $2.909 \pm 0.391$ \\
BD+24 4778  & $0.552 \pm 0.015$ & $0.151 \pm 0.975$ & $1.835 \pm 0.390$ \\
CD-38 1552  & $0.573 \pm 0.026$ & $-0.109 \pm 0.976$ & $1.939 \pm 0.390$ \\
BD+15 4033  & $-0.867 \pm 0.026$ & $-1.847 \pm 0.976$ & $2.635 \pm 0.390$ \\
BD+5 3223   & $0.290 \pm 0.036$ & $-0.154 \pm 0.976$ & $1.958 \pm 0.390$ \\
\hline
\hline
\end{tabular}
\end{table}

\section{Evolutionary Modeling}\label{evo}

To investigate a possible evolutionary connection between cool HgMn and hot Am stars, we first constructed a grid of stellar evolutionary models spanning a wide range of input parameters. The primary aim at this stage was to examine the positions of all target stars on the HR diagram and to explore whether cool HgMn and hot Am stars occupy regions consistent with a common evolutionary sequence.

For this purpose, evolutionary tracks were computed using the MESA code \citep{2010ascl.soft10083P, 2015ApJS..220...15P, 2019ApJS..243...10P}, considering initial metallicity ($Z_{\rm int}$), convective core overshooting ($\alpha_{\rm ov}$), mixing-length parameter ($\alpha_{\rm MLT}$), rotation rate ($V/V_{\rm crit}$), and mass-loss rate ($\dot{M}$). These models were then used to construct HR diagrams, where the observed positions of all HgMn and Am stars were overplotted. In this context, models were computed for stellar masses in the range $1.5$--$5.0~M_{\odot}$ with a step size of $0.1~M_{\odot}$.

The choice of model parameters was guided by both theoretical considerations and prior studies. The convective core overshooting parameter was fixed at $\alpha_{\rm ov} = 0.2$, consistent with the values reported for intermediate-mass stars \citep{2016A&A...592A..15C, 2019ApJ...876..134C}. Similarly, the mixing-length parameter was adopted as $\alpha_{\rm MLT} = 2.0$, corresponding to the solar-calibrated value, since stars in this mass range possess radiative envelopes and are not significantly affected by variations in this parameter \citep{2023Galax..11...75J}.

For the initial set of evolutionary models, constructed to assess whether HgMn and Am stars follow similar evolutionary sequences in the HR diagram, a solar metallicity of $Z = 0.013$ \citep{2009ARA&A..47..481A} was adopted as the reference. This assumption was used only at this stage to facilitate a homogeneous comparison between the targets and to determine candidate stars that may be evolutionarily connected.

%old version
%\textbf{Mass-loss rates in CP stars remain poorly constrained. Previous studies have suggested that weak stellar winds with rates of the order of $10^{-14}$--$10^{-13}~M_{\odot}~\mathrm{yr}^{-1}$ may account for the observed abundance patterns of Am and HgMn stars \citep{1983ApJ...269..239M,2010A&A...521A..62V,2019MNRAS.482.4519A,2021ApJ...910...48L}. As a preliminary step, we investigated the sensitivity of our evolutionary models to the adopted mass-loss rate by computing evolutionary tracks over a broad range, from $10^{-16}$ to $10^{-10}~M_{\odot}~\mathrm{yr}^{-1}$, and models without mass loss. We found that the resulting tracks occupy nearly identical positions on the H--R diagram within the observational uncertainties for our targets, yielding essentially the same masses and evolutionary states. Therefore, the choice of mass-loss rate does not significantly affect the derived stellar parameters in the present study. Consequently, we adopted a fixed mass-loss rate of $\dot{M}=10^{-14}~M_{\odot}~\mathrm{yr}^{-1}$ for all final evolutionary calculations. This value lies within the range commonly considered plausible for CP late-B and A-type stars and has been widely adopted in diffusion models as a representative weak-wind prescription \citep{1983ApJ...269..239M,2010A&A...521A..62V}.}

%new version
Mass-loss rates in CP stars remain poorly constrained. Previous studies have suggested that weak stellar winds, with rates of about $10^{-14}$--$10^{-13} M_{\odot} \mathrm{yr}^{-1}$, may help explain the observed abundance patterns of Am and HgMn stars \citep{1983ApJ...269..239M,2010A&A...521A..62V, 2019MNRAS.482.4519A,2021ApJ...910...48L}. In this study, we focus on the effect of mass loss on the evolutionary tracks and on the stellar parameters derived from them. We therefore tested the sensitivity of our models to the adopted mass-loss rate. We computed tracks without mass loss and tracks with constant mass-loss rates from $10^{-16}$ to $10^{-10} M_{\odot} \mathrm{yr}^{-1}$. The resulting tracks occupy almost the same positions in the H--R diagram, within the observational uncertainties of our targets. As a result, they give nearly the same stellar masses and evolutionary states. This means that the adopted mass-loss rate does not significantly affect the stellar parameters derived in this work. For the final evolutionary calculations, we adopted a representative weak-wind value of $\dot{M}=10^{-14} M_{\odot} \mathrm{yr}^{-1}$. This value is consistent with mass-loss rates commonly used in diffusion models of CP late-B and A-type stars \citep{1983ApJ...269..239M, 2010A&A...521A..62V}.

%Mass-loss rates were chosen based on previous studies indicating that CP stars exhibit very low mass loss, typically in the range $10^{-14}$--$10^{-13}~M_{\odot}~\mathrm{yr}^{-1}$ \citep{2014A&A...564A..70K, 2011MNRAS.418..195H}. For the initial evolutionary modeling, a value of $\dot{M} = 10^{-14}~M_{\odot}~\mathrm{yr}^{-1}$ was adopted.

The effects of rotation were taken into account using the dimensionless parameter $V/V_{\rm crit}$, which provides a physically consistent description of stellar rotation across different masses. Since the critical velocity depends on stellar mass, adopting a fixed equatorial velocity would correspond to significantly different rotational regimes for different stars. Therefore, the use of $V/V_{\rm crit}$ ensures a homogeneous treatment of rotation in the model grid. Following \citet{2012A&A...537A.146E}, an initial rotation rate of $V/V_{\rm crit} = 0.4$ was adopted.
%This value corresponds to the peak of the observed rotational velocity distribution of young B-type stars and translates to typical main-sequence equatorial velocities of $\sim110$--$220~\mathrm{km~s^{-1}}$, which are consistent with observational constraints.

Using this grid of models, we compared the observed $\log T_{\rm eff}$ and $\log (L/L_{\odot})$ values of all target stars with the theoretical evolutionary tracks. This comparison allowed us to visually assess whether HgMn and Am stars could be aligned along similar evolutionary paths.

%very old version
%\textbf{As can be seen from Fig.~\ref{fig:evomodel}, several Am and HgMn stars occupy positions along evolutionary tracks corresponding to similar stellar masses, particularly in the range of $2.7$--$3.3 M_{\odot}$. A subset of HgMn stars (BD+20~733, BD$-$20~1544, BD+24~4778, CD$-$38~1552, and BD+5~3223) and Am stars (BD+25~641, BD+3~2280, and BD+25~3246) are found to lie on closely related evolutionary tracks. Their locations in the HR diagram are consistent with the possibility that at least some HgMn stars may evolve into Am stars as they cool and move toward lower $T_{\rm eff}$. Although the present sample does not allow us to establish a direct evolutionary link, the observed distribution supports the hypothesis of an evolutionary connection between some members of the cool HgMn and hot Am groups.}

%old version
%As shown in Fig.~\ref{fig:evomodel}, several Am and HgMn stars are located along evolutionary tracks corresponding to similar stellar masses, especially in the range of $2.7$--$3.3 M_{\odot}$. A subset of HgMn stars (BD+20~733, BD$-$20~1544, BD+24~4778, CD$-$38~1552, and BD+5~3223) and Am stars (BD+25~641, BD+3~2280, and BD+25~3246) follow closely related tracks. Their positions in the H--R diagram are consistent with the possibility that at least some HgMn stars may evolve into Am stars as they cool and move toward lower $T_{\rm eff}$. Although the present sample does not allow us to establish a direct evolutionary link, the observed distribution supports a possible evolutionary connection between some cool HgMn and hot Am stars.

\begin{figure}
 \centering
 \begin{minipage}[t]{0.45\textwidth}
  \centering
  \includegraphics[width=\linewidth]{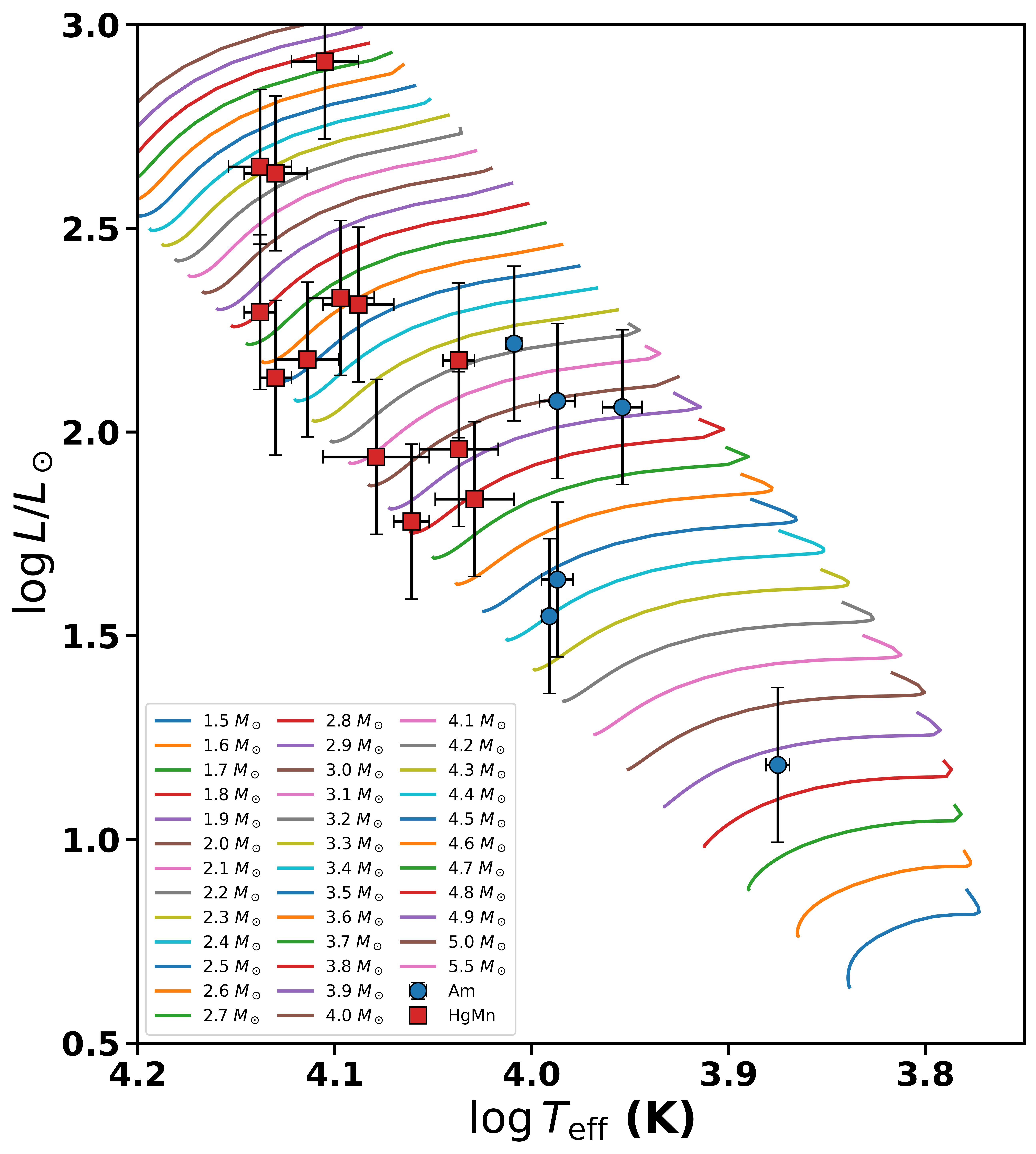}
 \end{minipage}\hfill
 \caption{HR diagram showing the comparison between the observed positions of Am (blue circles) and HgMn (red squares) stars and the theoretical evolutionary tracks computed for a metallicity of $Z = 0.013$, overshooting parameter $\alpha_{\rm ov} = 0.2$, mixing-length parameter $\alpha_{\rm MLT} = 2.0$, and mass-loss rate of $\dot{M} = 10^{-14}~M_{\odot}~\mathrm{yr}^{-1}$. The tracks correspond to stellar masses in the range $1.5$--$5.0~M_{\odot}$. Error bars represent the uncertainties in $\log T_{\rm eff}$ and $\log (L/L_{\odot})$.}
 \label{fig:evomodel}
\end{figure}

%new version
As shown in Fig.~\ref{fig:evomodel}, several Am and HgMn stars are located along evolutionary tracks corresponding to similar stellar masses, especially in the range of $2.7$--$3.3~M_{\odot}$. A subset of HgMn stars (BD+20~733, BD$-$20~1544, BD+24~4778, CD$-$38~1552, and BD+5~3223) and Am stars (BD+25~641, BD+3~2280, and BD+25~3246) follow closely related tracks. These stars were therefore selected for more detailed evolutionary modeling.

To examine this possible evolutionary connection in more detail, dedicated evolutionary models were calculated for the candidate cool HgMn and hot Am stars identified above as following similar evolutionary tracks. For each of these targets, the initial metallicity ($Z_{\rm initial}$\footnote{The $Z_{\rm initial}$ refers to the global composition assumed at the start of the evolutionary models and should be distinguished from the present-day surface metallicity, which may be altered by diffusion.}) was varied over the range $0.010 \leq Z_{\rm initial} \leq 0.030$ with a step of 0.001. A fixed mass-loss rate of $\dot{M}=10^{-14} M_{\odot}~\mathrm{yr}^{-1}$ was adopted for all models, since our tests showed that varying the mass-loss rate over several orders of magnitude produces no significant changes in the evolutionary tracks within the observational uncertainties for the targets. The fixed mass-loss rate allowed us to examine the effect of $Z_{\rm initial}$ separately. %This approach allowed us to investigate the influence of the initial metallicity on the evolutionary status of these candidate stars without introducing uncertainties from poorly constrained mass-loss rates.

\begin{figure}
 \centering
 \begin{minipage}[t]{0.45\textwidth}
  \centering
  \includegraphics[width=\linewidth]{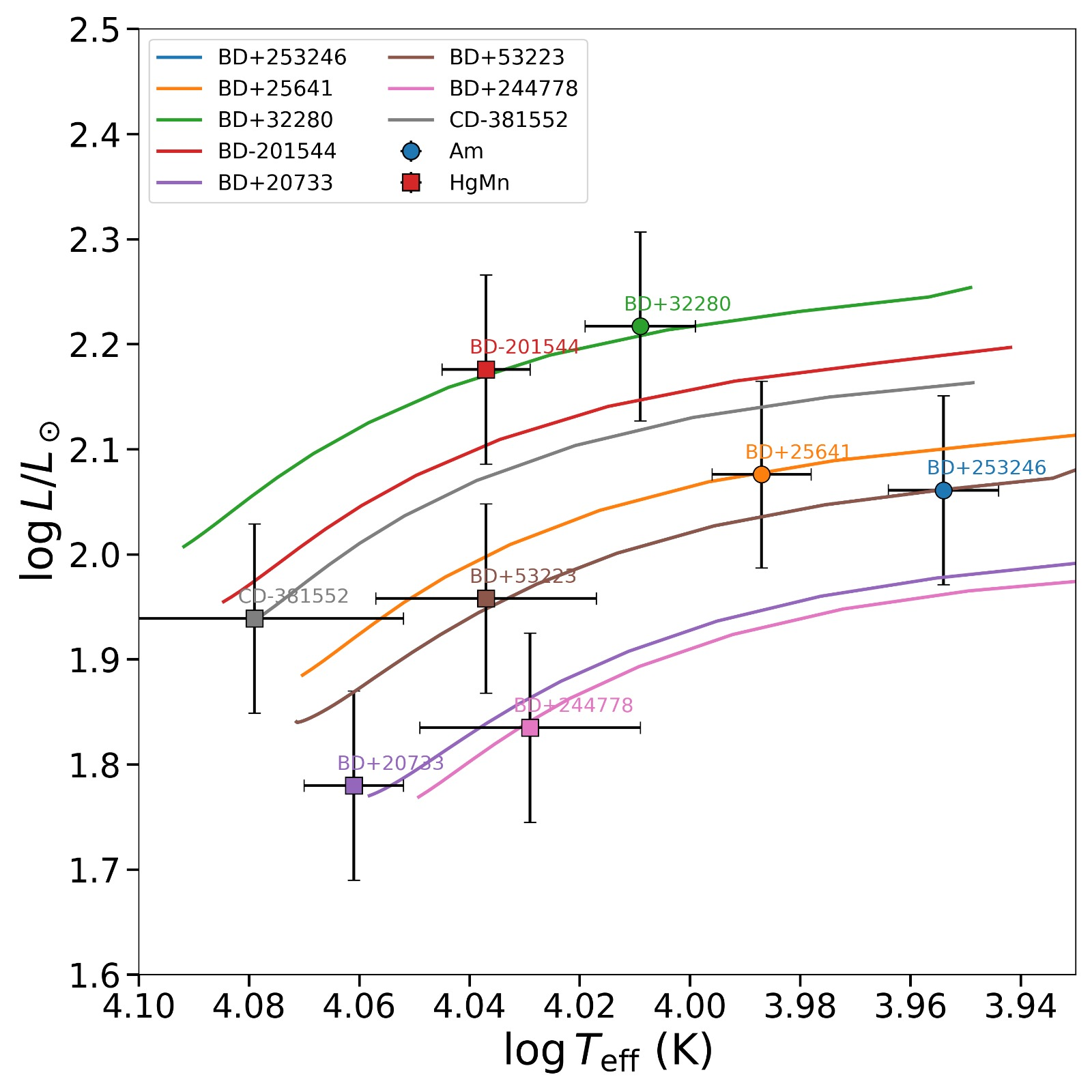}
 \end{minipage}\hfill
 \caption{HR diagram showing the positions of the cool HgMn (red squares) and hot Am (blue circles) stars proposed to represent consecutive evolutionary stages. The evolutionary tracks corresponding to the best-fitting stellar models are overplotted for each target.}
 \label{fig:evomodel2}
\end{figure}

As a result, best-fitting evolutionary tracks were determined for each of the eight candidate cool HgMn and hot Am stars, yielding estimates of their ages, masses, and $Z_{\rm initial}$ values. The resulting parameters are listed in Table~\ref{tab:evol_params}, while the comparison between the observed stellar positions and the best-fitting evolutionary tracks is presented in Fig.~\ref{fig:evomodel2}. The selected hot Am and cool HgMn stars occupy similar regions of the H--R diagram and have comparable stellar masses in the range of approximately $2.7$--$3.2~M_{\odot}$. The selected Am stars are generally cooler and, on average, older than the HgMn stars. In particular, BD$-$20~1544 and BD+3~2280 have similar masses ($3.1$--$3.2~M_{\odot}$) and comparable $Z_{\rm initial}$ values. Similarly, BD+5~3223 and BD+25~3246 have the same mass ($2.8~M_{\odot}$) and $Z_{\rm initial}=0.030$, but differ in age.

%As can be seen from Fig.~\ref{fig:evomodel}, an apparent evolutionary sequence is suggested between HgMn and Am stars, particularly in the intermediate-mass range of $2.7$--$3.3~M_{\odot}$. This behaviour supports the hypothesis proposed in the literature that these two classes of CP stars may be evolutionarily related. Based on this indication, candidate pairs of HgMn and Am stars that appear to follow similar evolutionary paths were identified. These selected systems are BD+20~527, BD+25~641, BD+3~2280, BD+69~850, BD+25~3246 Am stars and BD+20~733, BD-20~1544, BD+24~4778, CD-38~1552, BD+5~3223 HgMn stars. For each of these stars, dedicated evolutionary models were computed using the stellar parameters derived in this study (e.g., rotational velocities and atmospheric parameters). The comparison between the observed positions and the corresponding evolutionary models is presented in Fig.~\ref{fig:evomodel2}. The parameters obtained from the evolutionary modeling are listed in Table~\ref{tab:evol_params}. Based on the positions of the stars on this HR diagram and their associated error bars relative to the evolutionary tracks, the typical uncertainties of the derived parameters are estimated to be about 15--20\% in age, 5--10\% in radius, and $\pm 0.004$ in the initial metallicity $Z$. The possible connection between these stars is discussed in Sect.~\ref{discuss}.

\begin{table}
\centering
\caption{Evolutionary model parameters of the cool HgMn and hot Am stars identified as occupying similar evolutionary tracks in the HR diagram.}
\label{tab:evol_params}
\begin{tabular}{llll}
\hline
Name & Age    & $M$           & Initial $Z$  \\ %& $\dot{M}$  \\
     & (Gyr)  & ($M_{\odot}$) &  ($\pm 0.005$)           \\ % & ($M_{\odot}~\mathrm{yr}^{-1}$) \\
\hline
\multicolumn{4}{c}{\textbf{Am Stars}} \\
\hline
%BD+20 527    & $0.42 \pm 0.06$ & $2.71 \pm 0.20$ & 0.0140  & $10^{-13}$ \\
BD+25 641    & $0.34 \pm 0.05$ & $3.00\pm 0.30$ & 0.028  \\ % & $10^{-14}$ \\ %
BD+3 2280    & $0.29\pm 0.04$ & $3.20 \pm 0.30$ &  0.016   \\ %& $10^{-13}$ \\
%BD+69 850    & $0.37 \pm 0.06$ & $3.15 \pm 0.32$ & 0.0135  \\ % & $10^{-13}$ \\
BD+25 3246   & $ 0.35 \pm 0.05$ & $2.80 \pm 0.20$ & 0.030   \\ %& $10^{-13}$ \\
\hline
\multicolumn{4}{c}{\textbf{HgMn Stars}} \\
\hline
BD+20 733      & $0.02\pm 0.01$ & $2.70 \pm 0.20$ & 0.028  \\ % & $10^{-14}$ \\
BD$-$20 1544   & $0.21\pm 0.05$ & $3.10 \pm 0.20$ & 0.015  \\ % & $10^{-14}$ \\
BD+24 4778     & $0.22\pm 0.03$ & $2.80 \pm 0.20$ & 0.013   \\ %& $10^{-14}$ \\
CD$-$38 15527  & $0.06 \pm 0.01$ & $3.10 \pm 0.30$ & 0.014   \\ %& $10^{-14}$ \\
BD+5 3223      & $0.19\pm 0.04$ & $2.80 \pm 0.20$ & 0.030   \\ %& $10^{-14}$ \\
\hline
\end{tabular}
\end{table}

\section{Discussion and Conclusions} \label{discuss}

In this study, we carried out a spectroscopic and evolutionary investigation of weakly magnetic CP stars, focusing on the HgMn and Am subclasses. In particular, we investigated the long-debated hypothesis of an evolutionary connection between cool HgMn stars and hot Am stars, and examined whether some intermediate-mass cool HgMn stars may evolve into hot Am stars as they cool during their main-sequence evolution. To this end, high-resolution spectroscopic data were analyzed to derive reliable atmospheric parameters and detailed chemical abundance patterns for a selected sample of targets. These observational constraints were then compared with evolutionary models computed using the MESA code.

\subsection{Spectroscopic Results and Chemical Properties}

The observational analysis was based on a sample of 13 cool HgMn and 14 hot Am targets selected from the literature. The spectra were normalized and first analyzed using Balmer-line fitting to derive $T_{\mathrm{eff}}$ and $\log g$. Spectrum synthesis of Fe\,\textsc{i} and Fe\,\textsc{ii} lines was then used to refine these parameters and to determine $\xi$ and $v \sin i$. The adopted parameters were subsequently used in the chemical abundance analysis. The results show that all HgMn stars in the sample exhibit a strong overabundance of elements such as Hg, Mn, Sr, and Y, confirming their CP nature. Although all 14 objects in the initial sample were classified as Am stars in the literature, our detailed abundance analysis confirms Am-like chemical peculiarities in only six of them (see Table~\ref{tab:astro_params}). These stars exhibit various combinations of the abundance anomalies commonly associated with the Am phenomenon, whereas the remaining objects show abundance patterns more consistent with normal A stars. Thus, some stars classified as Am in the literature may in fact be chemically normal, and high-resolution spectroscopy is needed to confirm their classification.

The spectroscopically confirmed CP stars in our sample have projected $v \sin i$ consistent with the classical properties of both Am and HgMn stars. The confirmed Am stars have $v \sin i$ values ranging from 4 to 45 km~s$^{-1}$, while the HgMn stars range from 5 to 76~km~s$^{-1}$, with most of them rotating below $\sim$40~km~s$^{-1}$. In contrast, several targets initially classified as Am stars but later reclassified as chemically normal A-type stars show substantially higher rotational velocities (e.g. BD$+$50 1460, BD$-$8 3372, BD$-$5 3535, and BD$+$69 850), supporting the view that relatively rapid rotation may suppress the development of the classical chemical peculiarities by reducing the efficiency of atomic diffusion.

\subsection{Temperature Dependence of Chemical Abundances}

\citet{2003A&A...397..267A} demonstrated that the abundances of certain chemical elements in HgMn and Am stars show systematic correlations with $T_{\mathrm{eff}}$. In particular, a positive correlation between Mn abundance and $T_{\mathrm{eff}}$, and a negative correlation between Ni abundance and $T_{\mathrm{eff}}$, were reported. Motivated by these findings, we investigated the relationships between the abundances of Mn, Fe, and Ni and the $T_{\mathrm{eff}}$ of the stars in our sample.

\begin{figure}
 \centering
 \begin{minipage}[t]{0.44\textwidth}
  \centering
  \includegraphics[width=\linewidth,height=5cm]{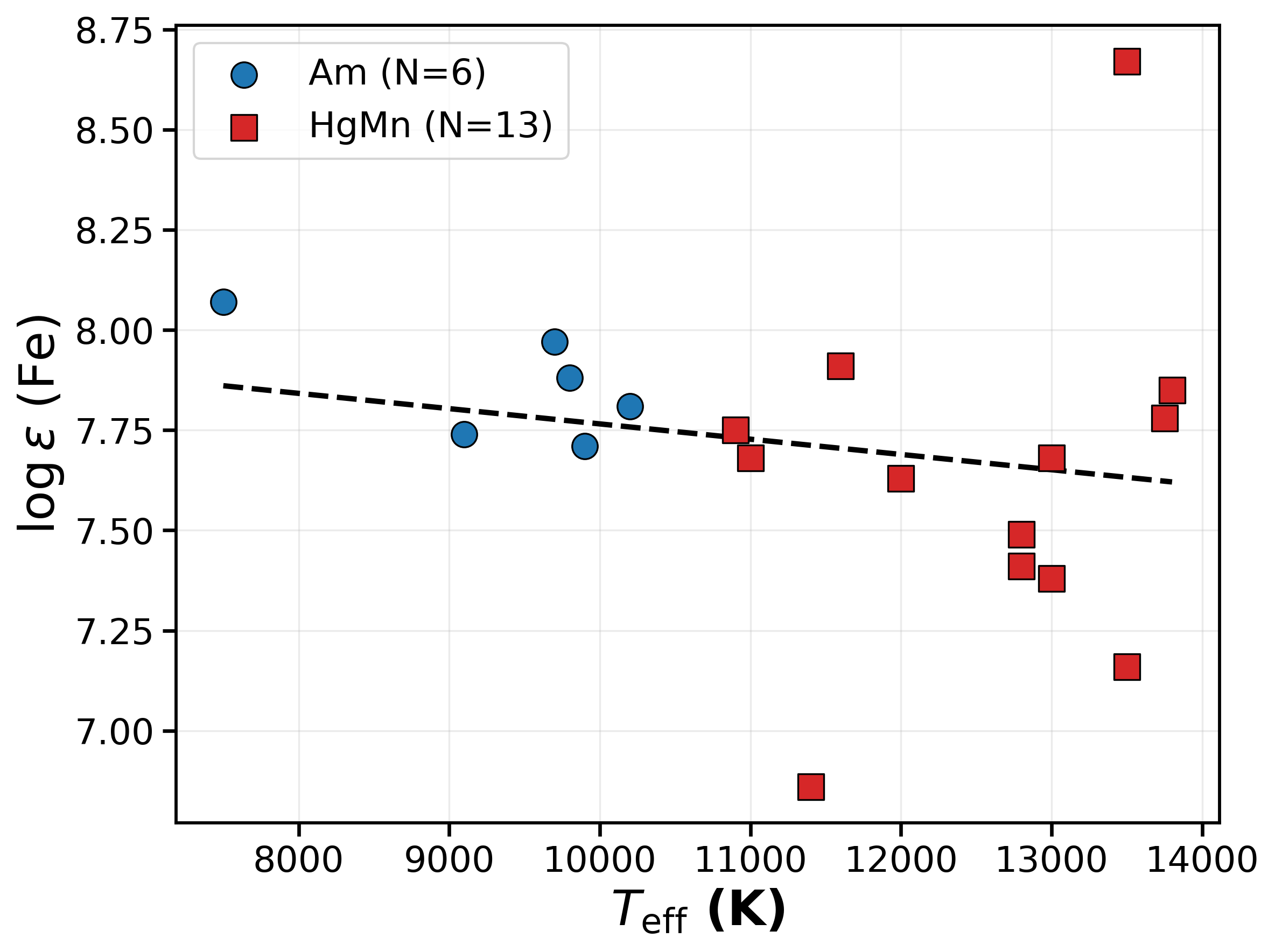}
 \end{minipage}\hfill
\begin{minipage}[t]{0.45\textwidth}
   \centering
  \includegraphics[width=\linewidth,height=5cm]{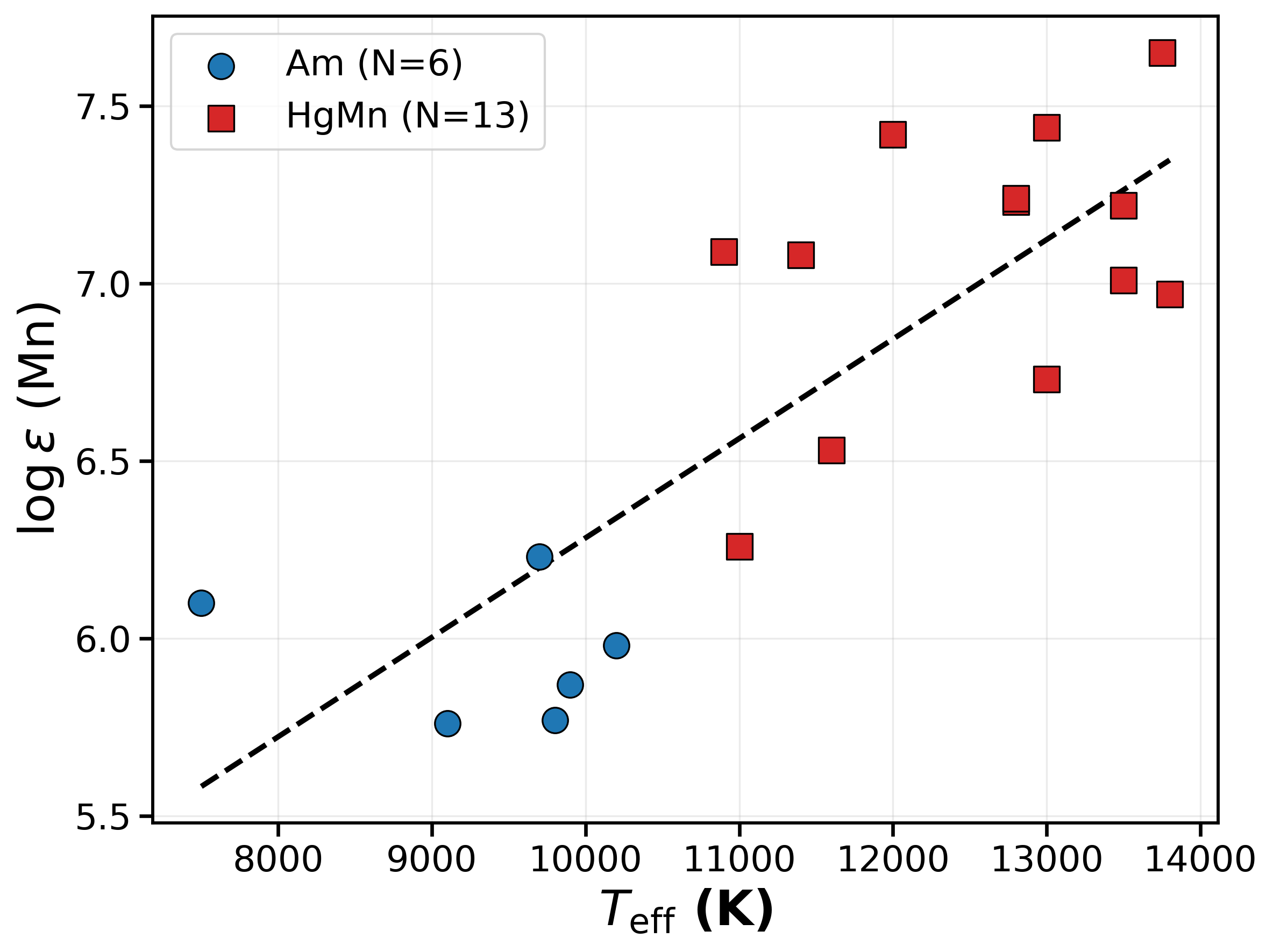}
 \end{minipage}
 \begin{minipage}[t]{0.44\textwidth}
   \centering
  \includegraphics[width=\linewidth,height=5cm]{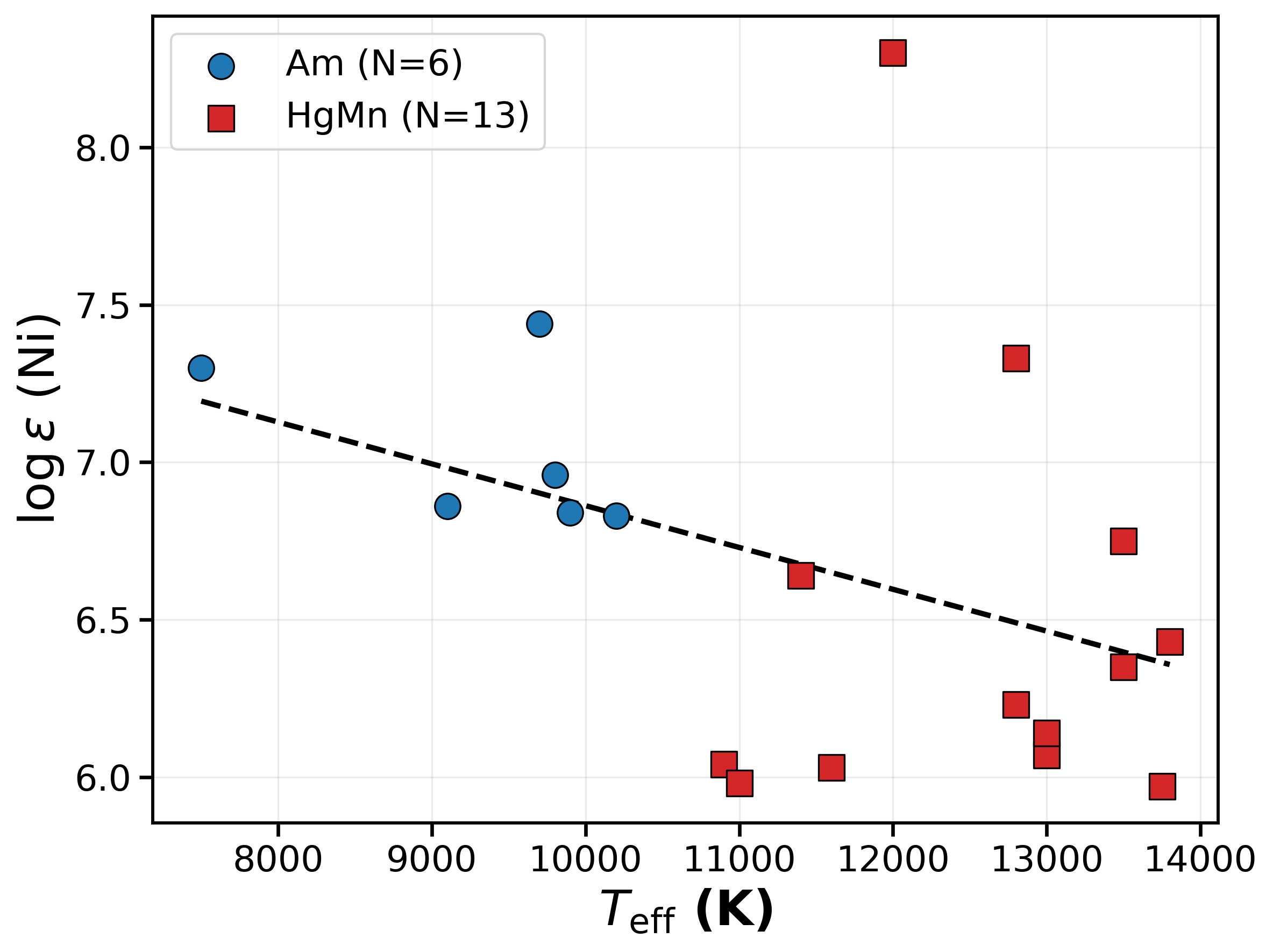}
 \end{minipage}
 \caption{Correlations between $T_{\mathrm{eff}}$ and elemental abundances for the studied stars. Upper panel shows $\log \epsilon (\mathrm{Fe})$, middle panel $\log \epsilon (\mathrm{Mn})$, and lower panel $\log \epsilon (\mathrm{Ni})$ as a function of $T_{\mathrm{eff}}$. HgMn stars are represented by red squares, while Am stars are shown as blue circles. The dashed lines indicate linear fits to the data.}
 \label{fig:abuncomp}
\end{figure}

%old version
%As shown in Fig.~\ref{fig:abuncomp}, a strong positive correlation is clearly present between Mn abundance and $T_{\mathrm{eff}}$ for both HgMn and Am stars, indicating that Mn enrichment increases with increasing temperature. In contrast, the relation between Fe abundance and $T_{\mathrm{eff}}$ appears more scattered, particularly for HgMn stars, although a weak overall negative trend is suggested. The Ni abundance, on the other hand, exhibits a clear negative correlation with $T_{\mathrm{eff}}$ for both stellar groups, consistent with the results of \cite{2003A&A...397..267A}. These findings support the presence of temperature-dependent diffusion processes that shape the observed abundance patterns in CP stars. Furthermore, the comparison of chemical abundance patterns reveals clear common trends between the two stellar groups. In particular, the increase of Mn abundance with $T_{\mathrm{eff}}$ and the decrease of Ni abundance are consistently observed in both HgMn and Am stars, indicating the efficiency of atomic diffusion processes in their atmospheres. In addition, the relatively low projected rotational velocities (typically $< 50$ km s$^{-1}$) suggest that tidal braking and/or mass-loss processes may play a significant role in their evolutionary history. These shared characteristics provide strong observational support for a possible evolutionary connection between HgMn and Am stars and point to a common physical framework governing their surface chemical peculiarities.

%new version
As shown in Fig.~\ref{fig:abuncomp}, a strong positive correlation is present between Mn abundance and $T_{\mathrm{eff}}$ for both HgMn and Am stars, indicating that Mn enrichment increases with increasing temperature. In contrast, the relation between Fe abundance and $T_{\mathrm{eff}}$ is more scattered, particularly for HgMn stars, although a weak overall negative trend is suggested. The Ni abundance exhibits a clear negative correlation with $T_{\mathrm{eff}}$ for both stellar groups, consistent with the results of \citet{2003A&A...397..267A}. These findings support the presence of temperature-dependent diffusion processes shaping the observed abundance patterns in CP stars.

\subsection{Evolutionary Connection Between Cool HgMn and Hot Am Stars}

The evolutionary models presented in Sect.~4 are broadly consistent with the possibility that some cool HgMn stars may be evolutionarily connected with hot Am stars. However, similar stellar masses, initial metallicities, and positions in the H--R diagram are not sufficient to establish a direct evolutionary sequence. The limited number of candidate pairs, together with the uncertainties inherent in the evolutionary models, prevents any firm conclusion. A larger sample, combined with homogeneous abundance analyses and evolutionary models that include atomic diffusion, will be required to test this hypothesis more rigorously.

It should also be emphasized that the proposed evolutionary connection is intended only for a subset of hot Am stars with masses and effective temperatures comparable to those of the cool HgMn stars considered in this study. The Am phenomenon spans a much broader range of $M$ and $T_{\mathrm{eff}}$, including numerous lower-mass A-type stars that are not expected to evolve from HgMn stars. In such objects, the observed chemical peculiarities are most likely produced by atomic diffusion operating in slowly rotating atmospheres, often aided by tidal braking in binary systems. Therefore, while some hot Am stars may represent evolved counterparts of cool HgMn stars, it would not be appropriate to interpret the entire Am population as descendants of the HgMn phenomenon.

Overall, the present work provides the first homogeneous spectroscopic and evolutionary framework specifically designed to examine the proposed connection between cool HgMn and hot Am stars. Although the current sample is insufficient to establish a definitive evolutionary sequence, it provides a homogeneous observational basis for future studies combining detailed abundance analyses, stellar evolution, and atomic diffusion models.

\section*{Acknowledgements}
This study has been supported by the Scientific and Technological Research Council (TUBITAK) project through 124F062. The numerical calculations reported in this paper were partially performed at TUBITAK ULAKBIM, High Performance and Grid Computing Center (TRUBA resources). The calculations have also been carried out at the Wroc{\l}aw Centre for Networking and Supercomputing (http://www.wcss.pl), under grant No.~214. Archival HARPS, UVES, FEROS, and GIRAFFE spectra obtained from the ESO Science Archive Facility were used in this work. Based on observations collected at the European Southern Observatory under ESO programme(s) 076.D-0172, 081.C-0475, 086.D-0062, 086.D-0240, 087.B-0308, 088.C-0353, 089.C-0006, 089.D-0129, 098.C-0463, 0102.C-0547, 266.D-5655, 268.D-5738, and 380.D-0161.

%%%%%%%%%%%%%%%%%%%%%%%%%%%%%%%%%%%%%%%%%%%%%%%%%%
\section*{Data Availability}

The spectroscopic data underlying this article are available from public archives, including the ESO Science Archive Facility, ELODIE, and SOPHIE databases. The processed spectra, abundance analysis results, and evolutionary models generated during this study are available from the corresponding author upon reasonable request.



%\bibliographystyle{mnras}

% Alternatively, you could enter them by hand, like this:
%\begin{thebibliography}{99}
%\end{thebibliography}

%%%%%%%%%%%%%%%%%%%%%%%%%%%%%%%%%%%%%%%%%%%%%%%%%%

%%%%%%%%%%%%%%%%% APPENDICES %%%%%%%%%%%%%%%%%%%%%

\appendix
\setcounter{table}{0}
\renewcommand{\thetable}{A\arabic{table}}

\begin{table*}
\centering
\small
\caption{Literature information for the selected targets analyzed in this study.}\label{tab:A1}
%\resizebox{\textwidth}{!}{
\begin{tabular}{lclccc}
\hline
\hline
Name & Spectroscopic Study & Reference & $T_{\rm eff}$ (K) & $\log g$ & $v \sin i$ (km~s$^{-1}$) \\
\hline
\multicolumn{6}{c}{\textbf{Am Targets}} \\
\hline
BD+20 527  & No  & --                     & --               & --             & -- \\
BD+25 641  & Yes &{\citet{2016A&A...589A..83G}}& $8900 \pm 150$  & $3.4 \pm 0.3$  & $10 \pm\ 2$ \\
BD+3 1437  & No  & --                     & --               & --             & -- \\
BD+50 1460 & No  & --                     & --               & --             & -- \\
BD-1 2074  & Yes & {\citet{2016A&A...589A..83G}}; {\citet{2021MNRAS.500.2451W}} & $9200 \pm 250$ & $4.0 \pm 0.2$ & $5.5 \pm 0.2$ \\
BD+19 2097 & Yes & {\citet{2016A&A...589A..83G}}   & $7400 \pm 150$  & $3.0 \pm 0.3$  & $25 \pm 2$ \\
BD+3 2280  & Yes & {\citet{2014A&A...562A..84R}}    & $10200 \pm 125$ & $3.9 \pm 0.2$  & $27 \pm 3$ \\
CD-49 4801 & No  & --                     & --               & --             & -- \\
BD+25 2498 & Yes & {\citet{2016A&A...589A..83G}}   & $8200 \pm 150$  & $3.4 \pm 0.3$  & $50 \pm 2$ \\
BD-8 3372  & No  & --                     & --               & --             & -- \\
BD-5 3535  & No  & --                     & --               & --             & -- \\
CD-47 7893 & No  & --                     & --               & --             & -- \\
BD+69 850  & No  & --                     & --               & --             & -- \\
BD+25 3246 & No  & --                     & --               & --             & -- \\
BD+50 2468 & Yes & {\citet{2016A&A...589A..83G}}   & $9800 \pm 200$  & $4.2 \pm 0.1$  & $120 \pm 3$ \\
\hline
\multicolumn{6}{c}{\textbf{HgMn Targets}} \\
\hline
BD+28 4     & Yes &{\citet{2019A&A...627A.138A}}& $13076 \pm 354$ & $3.5 \pm 0.2$  & -- \\
BD+15 177   & Yes &{\citet{2003A&A...402..299D}}& $13126 \pm 200$ & $4.0 \pm 0.1$  & $21 \pm 2$ \\
BD+21 535   & Yes &{\citet{2018MNRAS.480.2953G}}& $13000 \pm 1000$ & $4.0 \pm 0.4$ & $67 \pm 5$ \\
BD+20 733   & Yes &{\citet{2018MNRAS.480.2953G}}& $11572 \pm 1000$ & $4.2 \pm 0.4$ & $5 \pm 3$ \\
BD+14 787   & Yes &{\citet{2018MNRAS.480.2953G}}& $13900 \pm 1000$ & $4.1 \pm 0.4$ & $75 \pm 5$ \\
BD-16 1072  & Yes &{\citet{2018MNRAS.480.2953G}}& $12800 \pm 1000$ & $3.9 \pm 0.4$ & $16 \pm 3$ \\
BD-20 1544  & No  & --                       & --               & --             & -- \\
CD-36 14166 & No  & --                       & --               & --             & -- \\
BD+40 5068  & Yes & --                       & $12241 \pm 402$ & $4.1 \pm 0.1$  & -- \\
BD+24 4778  & Yes &{\citet{2018MNRAS.480.2953G}}& $10950 \pm 1000$ & $4.1 \pm 0.4$ & $35 \pm 5$ \\
CD-38 15527 & Yes &{\citet{2018MNRAS.480.2953G}}& $12400 \pm 1000$ & $3.9 \pm 0.4$ & $25 \pm 3$ \\
BD+15 4033  & Yes &{\citet{2018MNRAS.480.2953G}}& $13200 \pm 1000$ & $3.6 \pm 0.3$ & $8 \pm 2$ \\
BD+5 3223   & Yes &{\citet{2018MNRAS.480.2953G}}& $11000 \pm 1000$ & $3.8 \pm 0.3$ & $8 \pm 2$ \\
\hline
\hline
\end{tabular}
\end{table*}

\setcounter{table}{1}
\renewcommand{\thetable}{A\arabic{table}}
\begin{table*}
    \centering
    \caption{The result of the analysis of chemical abundances of Am targets. The numbers in brackets indicate the number of lines used in the analysis.}\label{tab:table_abundance1}
    \begin{tabular}{lccccccc}
    \hline
Element & BD+20~527              & BD+25~641            & BD+3~1437           & BD+50~1460           & BD-1~2074           & BD+19~2097            &  BD+3~2280 \\
\hline
$_{6}$C & -                       & -                     &-                     & -                     &  -                    &  -                    & - \\
$_{11}$Na & -                     & -                     & -                    & -                     &-7.94~$\pm$~0.20 (1) &-4.89~$\pm$~0.20 (2) & - \\
$_{12}$Mg& -4.74~$\pm$~0.30 (2) &-4.23~$\pm$~0.34 (1) &-4.37~$\pm$~0.11(7) &-4.67~$\pm$~0.30 (2) &-4.42~$\pm$~0.21 (11)&-3.64~$\pm$~0.20 (2) &-4.09~$\pm$~0.22 (8)\\
$_{13}$Al &-                      &-                      &-5.70~$\pm$~0.20 (1)&-                      &-5.05~$\pm$~0.20 (2) &-                    & - \\
$_{14}$Si&-                       & -                     &-4.71~$\pm$~0.11 (8)&-4.62~$\pm$~0.30 (2) &-4.49~$\pm$~0.26 (14)&-4.42~$\pm$~0.58 (15)&-4.18~$\pm$~0.24 (8) \\
$_{16}$S& -                       & -                     & -                    & -                     &-4.47~$\pm$~0.25 (11)&-4.36~$\pm$~0.42 (4) &-4.02~$\pm$~0.35 (10)\\
$_{20}$Ca & -5.34~$\pm$~0.30 (2)&-4.21~$\pm$~0.32 (2) &-5.44~$\pm$~0.12 (7)&-5.56~$\pm$~0.30 (2) &-5.55~$\pm$~0.11 (14)&-6.03~$\pm$~0.44 (13)&-5.15~$\pm$~0.28 (8)\\
$_{21}$Sc &-8.58~$\pm$~0.30 (2) &-8.15~$\pm$~0.33 (2) &-9.13~$\pm$~0.13 (4)&-8.86~$\pm$~0.31 (2) &-9.60~$\pm$~0.26 (4) &-9.64~$\pm$~1.10 (6) &-8.65~$\pm$~0.37 (7)\\
$_{22}$Ti &-7.04~$\pm$~0.35 (5) &-6.37~$\pm$~0.41 (9) &-7.03~$\pm$~0.12 (25)&-7.26~$\pm$~0.29 (6)&-6.77~$\pm$~0.15 (42)&-6.66~$\pm$~0.41 (23)&-6.70~$\pm$~0.26 (37)\\
$_{23}$V  &-7.60~$\pm$~0.32 (1) &-8.42~$\pm$~0.30 (2) &-6.94~$\pm$~0.20 (2) & -                    &-7.45~$\pm$~0.21 (5) &-6.91~$\pm$~0.14 (3) &-7.24~$\pm$~0.26 (6)\\
$_{24}$Cr &-5.83~$\pm$~0.32 (3) &-6.04~$\pm$~0.37 (5) &-6.38~$\pm$~0.15 (18)&-6.63~$\pm$~0.33 (5)&-6.02~$\pm$~0.11 (45)&-5.75~$\pm$~0.45 (24)&-5.96~$\pm$~0.20 (29)\\
$_{25}$Mn&-5.48~$\pm$~0.30 (2)  &-5.81~$\pm$~0.30 (5) &-6.48~$\pm$~0.18 (9) & -                    &-6.27~$\pm$~0.16 (13)&-5.94~$\pm$~0.88 (9) &-6.06~$\pm$~0.28 (8) \\
$_{26}$Fe&-4.31~$\pm$~0.26 (9)  &-4.07~$\pm$~0.42 (19)&-4.59~$\pm$~0.16(79) &-4.94~$\pm$~0.27(15)&-4.16~$\pm$~0.34 (442)&-3.97~$\pm$~0.47(150)&-4.23~$\pm$~0.12(113)\\
$_{27}$Co& -                      & -                     &-                      & -                    &-7.22~$\pm$~0.27 (1)  & -                     &-6.23~$\pm$~0.27 (1) \\
$_{28}$Ni&-6.81~$\pm$~0.32 (2)  &-4.60~$\pm$~0.31 (2) &-5.65~$\pm$~0.19 (8) & -                    &-5.08~$\pm$~0.19 (28) &-4.74~$\pm$~0.45 (22)&-5.21~$\pm$~0.23 (10) \\
$_{29}$Cu& -                      & -                     & -                     & -                    &-                       &-10.25~$\pm$~0.40 (1)& -\\
$_{30}$Zn& -                      & -                     & -                     & -                    &-6.65~$\pm$~0.25 (1)  & -                     & -\\
$_{38}$Sr& -                      & -                     &-9.62~$\pm$~0.20 (1) & -                    &-8.95~$\pm$~0.25 (1)  &-7.36~$\pm$~0.32 (2) &-9.17~$\pm$~0.30 (1)\\
$_{39}$Y&-8.24~$\pm$~0.32 (1)   & -                     &-9.55~$\pm$~0.20 (1) &-9.91~$\pm$~0.30 (1)&-8.87~$\pm$~0.25 (3)  &-8.71~$\pm$~0.30 (2) &-9.03~$\pm$~0.30 (2) \\
$_{40}$Zr& -                      & -                     &-9.14~$\pm$~0.23 (1) & -                    &-8.48~$\pm$~0.25 (3)  &-9.02~$\pm$~0.38 (3) & - \\
$_{56}$Ba& -                      & -                     & -                     & -                    &-                       &    -                  & - \\
\hline
Element & BD+25 2498              & BD-8 3372             & BD-5 3535            & CD-47 7893            & BD+69 850             & BD+25 3246             & BD+50 2468 \\
\hline
$_{6}$C & -                       & -                     &-                     & -                     &  -                    &-3.70~$\pm$~0.18 (20)&-3.33~$\pm$~0.30 (2) \\
$_{11}$Na&  -5.03~$\pm$~0.31 (1)& -                     & -                    &-5.83~$\pm$~0.30 (1) &-                      &-5.26~$\pm$~0.24 (3) &-4.94~$\pm$~0.30 (2) \\
$_{12}$Mg& -4.77~$\pm$~0.38 (10)&-                      &-                     &-4.31~$\pm$~0.53 (7) &-4.30~$\pm$~0.22 (5) &-4.55~$\pm$~0.19 (10)&-4.15~$\pm$~0.36 (3)\\
$_{13}$Al &-                      &-                      &-5.70~$\pm$~0.30 (1)&-6.81~$\pm$~0.30 (1) &-                      &-5.25~$\pm$~0.24 (3)& -5.13~$\pm$~0.30 (1) \\
$_{14}$Si&-4.78~$\pm$~0.29 (13) & -                     &-                     &-4.78~$\pm$~0.41 (6) &-4.58~$\pm$~0.23 (5) &-4.36~$\pm$~0.20 (23)&-4.42~$\pm$~0.34 (3) \\
$_{16}$S&-4.76~$\pm$~0.39 (6)   & -                     & -                    &-4.39~$\pm$~0.33 (5) &-4.03~$\pm$~0.30 (1) &-4.37~$\pm$~0.19 (18) &-4.52~$\pm$~0.30 (1)\\
$_{20}$Ca & -5.36~$\pm$~0.42 (6)&-4.07~$\pm$~0.20 (1) &-4.97~$\pm$~0.30 (2)&-6.11~$\pm$~0.28 (4) &-5.17~$\pm$~0.40 (5) &-5.42~$\pm$~0.21 (30)&-5.52~$\pm$~0.36 (7)\\
$_{21}$Sc &-9.54~$\pm$~0.39 (4) &-9.30~$\pm$~0.20 (1) &-                     &-9.98~$\pm$~0.30 (2) &-8.37~$\pm$~0.35 (5) &-8.70~$\pm$~0.16 (10) &-8.95~$\pm$~0.16 (3)\\
$_{22}$Ti &-6.78~$\pm$~0.19 (24)&-6.90~$\pm$~0.28 (5) &-6.53~$\pm$~0.30 (1)&-7.46~$\pm$~0.25(23) &-7.30~$\pm$~0.30 (8) &-6.79~$\pm$~0.15 (61)&-6.97~$\pm$~0.14 (14)\\
$_{23}$V  &-7.42~$\pm$~0.17 (3) &-                      &-8.77~$\pm$~0.30 (1)&-7.76~$\pm$~0.28 (3) &-7.81~$\pm$~0.30 (1) &-7.61~$\pm$~0.21 (14) &-6.98~$\pm$~0.30 (1)\\
$_{24}$Cr &-5.85~$\pm$~0.22 (45)&-6.36~$\pm$~0.44 (5) &-6.49~$\pm$~0.31 (2)&-6.76~$\pm$~0.38 (22)&-6.31~$\pm$~0.26 (7) &-6.07~$\pm$~0.14 (83)&-6.26~$\pm$~0.15 (8)\\
$_{25}$Mn&-6.17~$\pm$~0.21 (8)  &-                      &-5.79~$\pm$~0.31 (2)&-6.34~$\pm$~0.48 (4) &-6.15~$\pm$~0.30 (2) &-6.28~$\pm$~0.13 (31) &-7.35~$\pm$~0.30 (1) \\
$_{26}$Fe&-4.33~$\pm$~0.36 (135)&-5.13~$\pm$~0.25 (9) &-4.59~$\pm$~0.27 (5)&-5.13~$\pm$~0.18(47) &-4.62~$\pm$~0.21 (22)&-4.30~$\pm$0.24 (463)&-4.55~$\pm$~0.17(30)\\
$_{27}$Co& -                      & -                     &-                     &-7.29~$\pm$~0.30 (1) &-                      &-6.45~$\pm$~0.24 (2) &-\\
$_{28}$Ni&-5.20~$\pm$~0.18 (11) &-4.44~$\pm$~0.20 (1) &-                     &-5.38~$\pm$~0.30 (5) &-5.41~$\pm$~0.30 (4) &-5.18~$\pm$~0.15 (54)&-5.45~$\pm$~0.33 (5) \\
$_{29}$Cu& -                      & -                     & -                    & -                     &-                      &-7.23~$\pm$~0.24 (2)& -\\
$_{30}$Zn& -                      & -                     & -                    & -                     &-                      &-6.62~$\pm$~0.24 (2)& -\\
$_{38}$Sr&-9.33~$\pm$~0.34 (1)  &-8.53~$\pm$~0.23 (1) &-7.36~$\pm$~0.32 (1)&-9.31~$\pm$~0.73 (4) &-                      &-8.09~$\pm$~0.24 (1) &-\\
$_{39}$Y&-8.84~$\pm$~0.31 (3)   &-7.86~$\pm$~0.20 (1) &-                     &-10.06~$\pm$~0.30 (2)&-                      &-8.81~$\pm$~0.28 (9) &-10.39~$\pm$~0.31 (1) \\
$_{40}$Zr&-8.27~$\pm$~0.30 (2)  & -                     &-8.56~$\pm$~0.31 (2)&-9.58~$\pm$~0.30 (2) &-8.32~$\pm$~0.30 (2) &-8.23~$\pm$~0.28 (9) &-8.29~$\pm$~0.30 (1) \\
$_{56}$Ba& -                      &-9.69~$\pm$~0.20 (1) & -                    & -                     &-                      & -8.42~$\pm$~0.28 (2)&-9.64~$\pm$~0.30 (1) \\
\hline
    \end{tabular}
\end{table*}

\setcounter{table}{2}
\renewcommand{\thetable}{A\arabic{table}}
\begin{table*}
    \centering
    \caption{The results of the chemical abundance analysis of HgMn targets are presented below. The numbers given in parentheses indicate the number of lines used in the analysis.}\label{tab:table_abundance2}
    \begin{tabular}{lccccccc}
    \hline
Element & BD+28 4                 & BD+15 177             & BD+21 535            & BD+20 733             & BD+14 787             & BD-16 1072            &  BD-20 1544 \\
\hline
$_{6}$C &-2.56~$\pm$~0.64 (3)   &-3.62~$\pm$~0.30 (2) &-3.54~$\pm$~0.32 (2)&-3.00~$\pm$~0.30 (2) &-3.99~$\pm$~0.27 (5) &-3.56~$\pm$~0.25 (6) &-4.58~$\pm$~0.20 (2) \\
$_{7}$N &-                        &-                      &-                     &-6.26~$\pm$~0.30 (2) &-3.87~$\pm$~0.20 (2) &-4.49~$\pm$~0.20 (5) &-3.13~$\pm$~0.20 (1) \\
$_{10}$Ne &                       &-                      &-3.93~$\pm$~0.33 (1)&-                      &-3.85~$\pm$~0.25 (1) &-3.71~$\pm$~0.27 (2) & - \\
$_{11}$Na & -                     & -                     & -                    & -                     &-                      &-                      & - \\
$_{12}$Mg& -4.86~$\pm$~0.30 (2) &-4.56~$\pm$~0.44 (5) &-4.60~$\pm$~0.16 (3)&-4.86~$\pm$~0.30 (2) &-4.04~$\pm$~0.28 (4) &-4.48~$\pm$~0.40 (4) &-5.39~$\pm$~0.24 (3)\\
$_{13}$Al &-                      &-6.34~$\pm$~0.30 (1) &-                     &-                      &-6.65~$\pm$~0.20 (1) &-6.93~$\pm$~0.20 (2) & - \\
$_{14}$Si&-                       &-4.84~$\pm$~0.32 (7) &-4.59~$\pm$~0.36 (4)&-4.56~$\pm$~0.36 (7) &-4.52~$\pm$~0.13 (9) &-4.80~$\pm$~0.43 (19)&-4.73~$\pm$~0.37 (3) \\
$_{15}$P &-4.65~$\pm$~0.30 (1)  &-                      &-5.63~$\pm$~0.32 (1)&-6.41~$\pm$~0.34 (1) &-5.12~$\pm$~0.25 (5) &-5.43~$\pm$~0.27 (23)&-5.27~$\pm$~0.20 (1) \\
$_{16}$S&-4.13~$\pm$~0.71 (8)   &-4.53~$\pm$~0.45 (11)&-4.94~$\pm$~0.41(12)&-4.61~$\pm$~0.30 (1) &-5.47~$\pm$~0.39 (10)&-5.07~$\pm$~0.24 (36)&-\\
$_{17}$Cl&-                       &-                      &-                     &-                      &         -             &-5.13~$\pm$~0.20 (1) &-\\
$_{18}$Ar&-                       &-                      &-                     &-                      &-                      &-5.77~$\pm$~0.20 (1) &-\\
$_{20}$Ca & -4.88~$\pm$~0.30 (2)&-5.02~$\pm$~0.30 (2) &-4.10~$\pm$~0.32 (2)&-5.59~$\pm$~0.30 (2) &-5.02~$\pm$~0.20 (1) &-5.43~$\pm$~0.23 (4) &-5.45~$\pm$~0.20 (2)\\
$_{21}$Sc &-7.56~$\pm$~0.30 (1) &-7.50~$\pm$~0.30 (2) &-8.33~$\pm$~0.33 (2)&-7.53~$\pm$~0.31 (1) &-                      &-8.22~$\pm$~0.29 (4) &-\\
$_{22}$Ti &-6.13~$\pm$~0.42 (4) &-5.46~$\pm$~0.17 (21)&-5.92~$\pm$~0.38(15)&-6.01~$\pm$~0.20 (21)&-6.26~$\pm$~0.18 (8) &-6.38~$\pm$~0.15 (31)&-6.98~$\pm$~0.27 (4)\\
$_{23}$V  &-6.80~$\pm$~0.32 (3) &-                      &-6.29~$\pm$~0.30 (3)& -                     &-                      &-8.01~$\pm$~0.20 (1) &-\\
$_{24}$Cr &-5.98~$\pm$~0.36 (5) &-5.48~$\pm$~0.18 (20)&-5.70~$\pm$~0.26(15)&-5.86~$\pm$~0.21 (5) &-5.94~$\pm$~0.27 (27)&-5.99~$\pm$~0.16 (16)&-6.07~$\pm$~0.15 (12)\\
$_{25}$Mn&-4.81~$\pm$~0.17 (8)  &-4.82~$\pm$~0.30 (2) &-4.60~$\pm$~0.33 (4)&-4.96~$\pm$~0.30 (2) &-5.07~$\pm$~0.26 (26)&-4.80~$\pm$~0.24 (8) &-4.95~$\pm$~0.28 (1) \\
$_{26}$Fe&-4.55~$\pm$~0.30 (29) &-4.88~$\pm$~0.45 (60)&-4.36~$\pm$~0.19(55)&-5.18~$\pm$~0.28 (39)&-4.19~$\pm$~0.13 (61)&-4.63~$\pm$~0.17(155)&-4.29~$\pm$~0.15 (42)\\
$_{27}$Co& -                      & -                     &-                      & -                    &-6.34~$\pm$~0.20 (2) &-6.75~$\pm$~0.20 (2) &- \\
$_{28}$Ni&-4.71~$\pm$~0.30 (2)  &-5.29~$\pm$~0.20 (3) &-5.97~$\pm$~0.28 (4) &-5.40~$\pm$~0.36 (3)&-5.61~$\pm$~0.13 (5) &-5.81~$\pm$~0.54 (15)&-6.00~$\pm$~0.20 (1) \\
$_{29}$Cu& -                      & -                     & -                     & -                    &-                      &-                      & -\\
$_{30}$Zn& -                      & -                     & -                     & -                    &-                      & -                     & -\\
$_{38}$Sr&-8.40~$\pm$~0.30 (1)  & -                     &-                      &-9.42~$\pm$~0.30 (2)&-                      &-7.86~$\pm$~0.20 (1) &-\\
$_{39}$Y&-                        & -                     &-                      &-                     &-                      &-                      &- \\
$_{40}$Zr& -                      & -                     &-                      & -                    &-                      &-8.00~$\pm$~0.20 (1) & - \\
$_{56}$Ba& -                      & -                     & -                     &-9.69~$\pm$~0.30 (1)&-                      &    -                  & - \\
$_{80}$Hg&-6.15~$\pm$~0.24 (1)  & -                     &-6.11~$\pm$~0.28 (1) &-                     &-6.37~$\pm$~0.30 (1) & -6.22~$\pm$~0.30 (1)& - \\
\hline
Element & CD-36 14166             & BD+40 5068            & BD+24 4778           & CD-38 15527           & BD+15 4033            & BD+5 3223         &\\
\hline
$_{6}$C & -3.71~$\pm$~0.53 (6)  &-3.48~$\pm$~0.20 (2) &-2.88~$\pm$~0.21 (2)&-2.34~$\pm$~0.20 (1) &-3.20~$\pm$~0.20 (1) &-2.35~$\pm$~0.31 (2)                &   \\
$_{7}$N &-3.47~$\pm$~0.20 (2)   &-4.88~$\pm$~0.20 (1) &-3.78~$\pm$~0.20 (1)&-3.83~$\pm$~0.22 (2) &-2.88~$\pm$~0.21 (2) &-&  \\
$_{10}$Ne &-3.99~$\pm$~0.20 (2) &-                      &-                     &-                      &-3.50~$\pm$~0.22 (2) &-&  \\
$_{11}$Na& -                      &-                      & -                    &-                      &-                      &-& \\
$_{12}$Mg& -4.80~$\pm$~0.20 (2) &-4.53~$\pm$~0.27 (3) &-4.49~$\pm$~0.17 (4)&-4.72~$\pm$~0.22 (7) &-3.80~$\pm$~0.22 (1) &-5.07~$\pm$~0.21 (3) &\\
$_{13}$Al &-                      &-6.00~$\pm$~0.20 (1) &-6.51~$\pm$~0.22 (1)&-                      &-4.21~$\pm$~0.25 (1) &- & \\
$_{14}$Si&-4.65~$\pm$~0.46 (12) &-4.84~$\pm$~0.26 (9) &-4.44~$\pm$~0.19(10)&-4.80~$\pm$~0.21 (2) &-3.82~$\pm$~0.38 (4) &-4.72~$\pm$~0.21 (3) &\\
$_{15}$P &-5.00~$\pm$~0.19 (5)  &-5.18~$\pm$~0.20 (1) &-                     &-6.41~$\pm$~0.34 (1) &-2.34~$\pm$~0.49 (5) &- &\\
$_{16}$S&-5.13~$\pm$~0.29 (10)  &-5.33~$\pm$~0.21 (5) &-2.83~$\pm$~0.21 (1)&-4.31~$\pm$~0.33 (4) &-6.29~$\pm$~0.53 (10)&-3.38~$\pm$~0.22 (2) &\\
$_{17}$Cl&-                       &-                      &-                     &-                      &         -             &- &\\
$_{18}$Ar&-                       &-                      &-                     &-                      &-                      &- &\\
$_{20}$Ca &-5.17~$\pm$~0.11 (4) &-5.83~$\pm$~0.20 (1) &-5.41~$\pm$~0.17 (6)&-5.86~$\pm$~0.20 (4) &-                      &-5.67~$\pm$~0.26 (3)&\\
$_{21}$Sc &-7.53~$\pm$~0.20 (2) &-7.94~$\pm$~0.20 (3) &-9.41~$\pm$~0.21 (2)&                       &-                      &-9.66~$\pm$~0.27 (4) &\\
$_{22}$Ti &-5.96~$\pm$~0.19 (8) &-6.70~$\pm$~0.13 (11)&-6.64~$\pm$~0.24(28)&-6.69~$\pm$~0.32 (17)&-4.29~$\pm$~0.54 (4) &-6.94~$\pm$~0.24 (21)&\\
$_{23}$V  &-6.03~$\pm$~0.20  (1)&-                      &-7.41~$\pm$~0.20 (2)&-7.32~$\pm$~0.28 (2) &-4.74~$\pm$~0.20 (2) &-9.56~$\pm$~0.20 (1) &\\
$_{24}$Cr &-5.76~$\pm$~0.32 (45)&-5.35~$\pm$~0.12 (18)&-6.02~$\pm$~0.25(25)&-6.01~$\pm$~0.27 (11)&-5.93~$\pm$~0.26 (3) &-6.20~$\pm$~0.25 (22)&\\
$_{25}$Mn&-4.39~$\pm$~0.14 (12) &-5.31~$\pm$~0.15 (5) &-5.51~$\pm$~0.17 (11)&-4.62~$\pm$~0.25 (4)&-5.03~$\pm$~0.64 (5) &-5.78~$\pm$~0.23 (4)  &\\
$_{26}$Fe&-4.26~$\pm$~0.21 (84) &-4.66~$\pm$~0.19 (72) &-4.13~$\pm$~0.14(81)&-4.41~$\pm$~0.31(29)&-3.37~$\pm$~0.31 (71)&-4.36~$\pm$~0.16 (91)&\\
$_{27}$Co&-5.92~$\pm$~0.20 (1)  & -                     &-                     &                       &-                      &-6.60~$\pm$~0.21 (1)                     &\\
$_{28}$Ni&-6.07~$\pm$~0.29 (5)  &-5.90~$\pm$~0.24 (4) &-6.01~$\pm$~0.21 (2)&-3.74~$\pm$~0.20 (2) &-5.69~$\pm$~0.31 (5) &-6.06~$\pm$~0.25 (2) &\\
$_{29}$Cu& -                      & -                     & -                    & -                     &-                      &-                      &\\
$_{30}$Zn& -                      & -                     & -                    & -                     &-                      & -                     &\\
$_{38}$Sr&-8.51~$\pm$~0.20 (1)  &-                      &-7.29~$\pm$~0.22 (2)&                       &-                      &-7.19~$\pm$~0.22 (2) &\\
$_{39}$Y&-                        &-                      &-                     &-6.40~$\pm$~0.20 (2)&-                       &-7.27~$\pm$~0.20 (1) & \\
$_{40}$Zr&-                       & -                     &-7.82~$\pm$~0.18 (3)&-7.45~$\pm$~0.25 (1) &-                      &-                       &\\
$_{56}$Ba& -                      &-                      &-9.56~$\pm$~0.20 (1)& -                     &-                      &-8.24~$\pm$~0.20 (1)  &                \\
$_{80}$Hg&-6.38~$\pm$~0.31 (1)   &-                      &-                     & -                     &-                      &-                      &                \\
\hline
    \end{tabular}
\end{table*}
%%%%%%%%%%%%%%%%%%%%%%%%%%%%%%%%%%%%%%%%%%%%%%%%%%

\setcounter{figure}{0}
\renewcommand{\thefigure}{A\arabic{figure}}
\begin{figure*}
 \begin{minipage}[b]{0.33\textwidth}
  \includegraphics[height=3.2cm, width=1.0\textwidth]{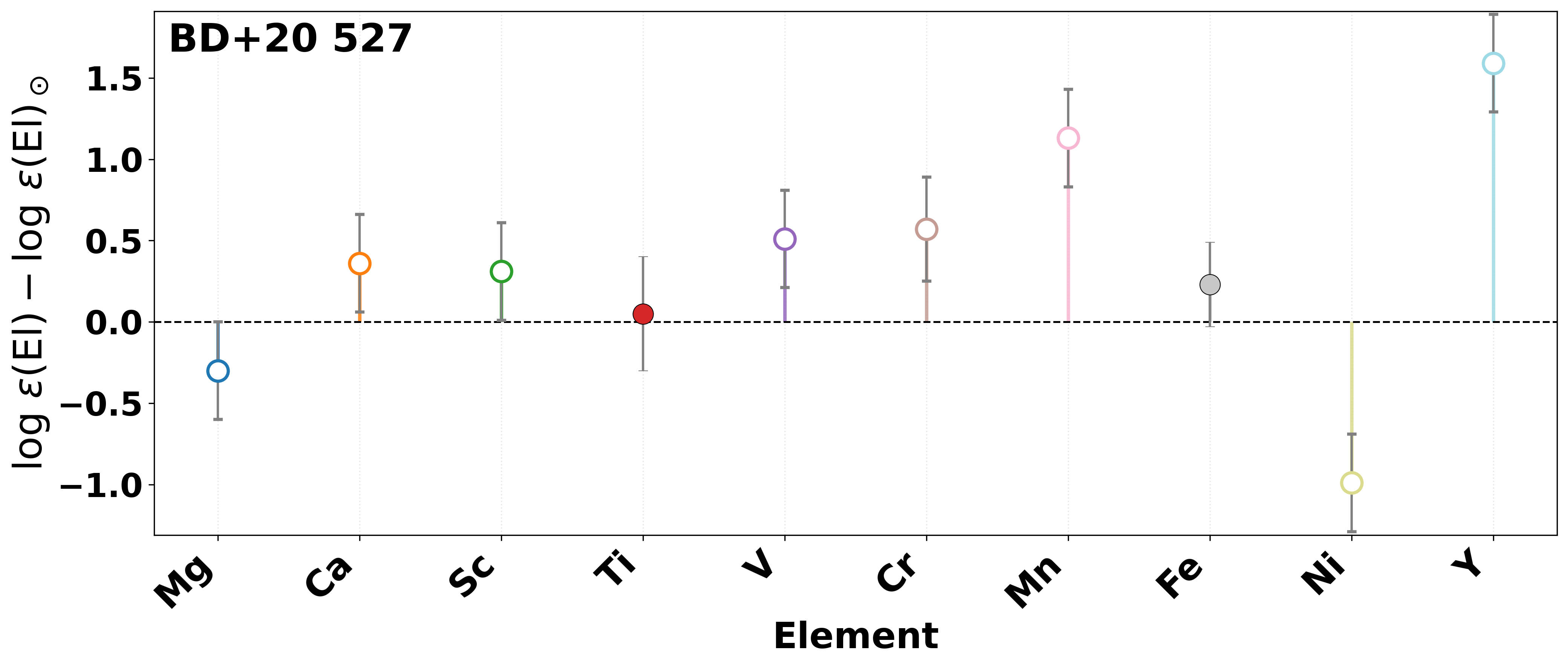}
  \end{minipage}
 \begin{minipage}[b]{0.33\textwidth}
  \includegraphics[height=3.2cm, width=1.0\textwidth]{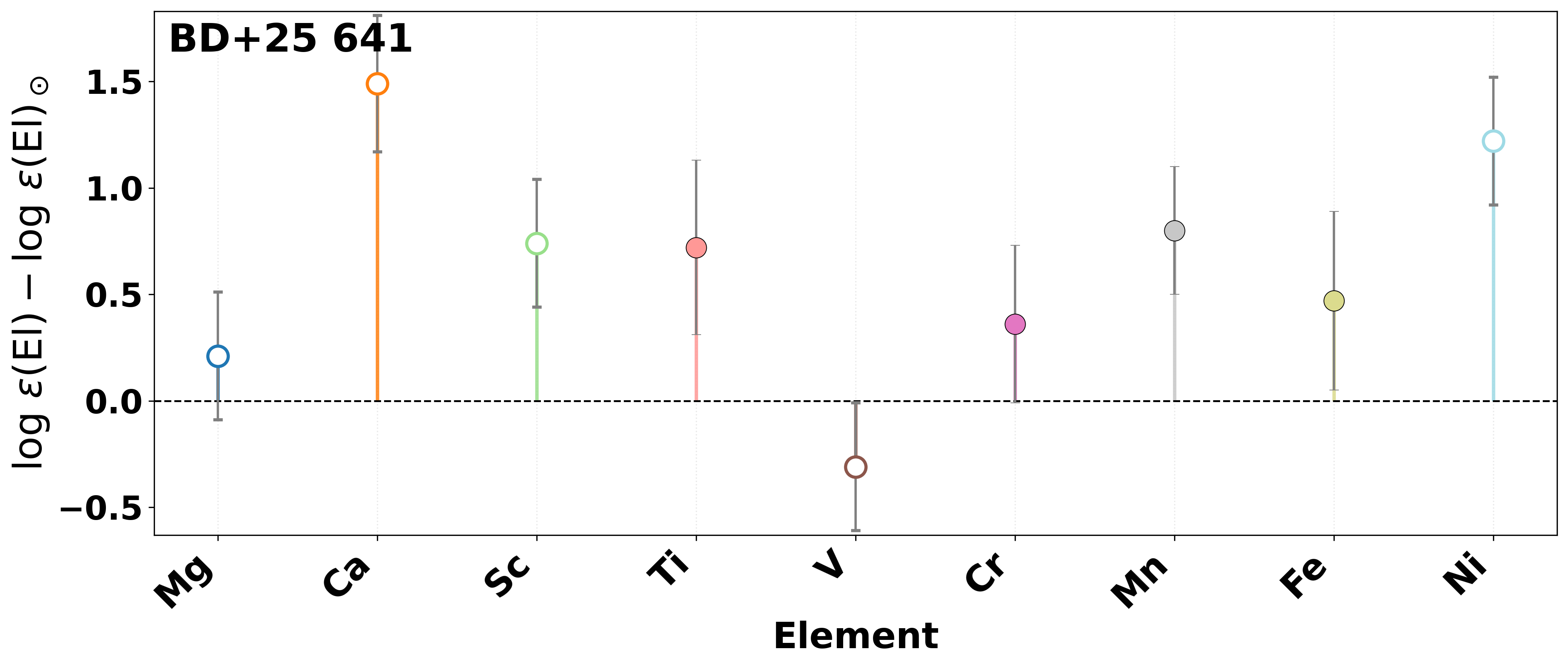}
 \end{minipage}
  \begin{minipage}[b]{0.33\textwidth}
  \includegraphics[height=3.2cm, width=1.0\textwidth]{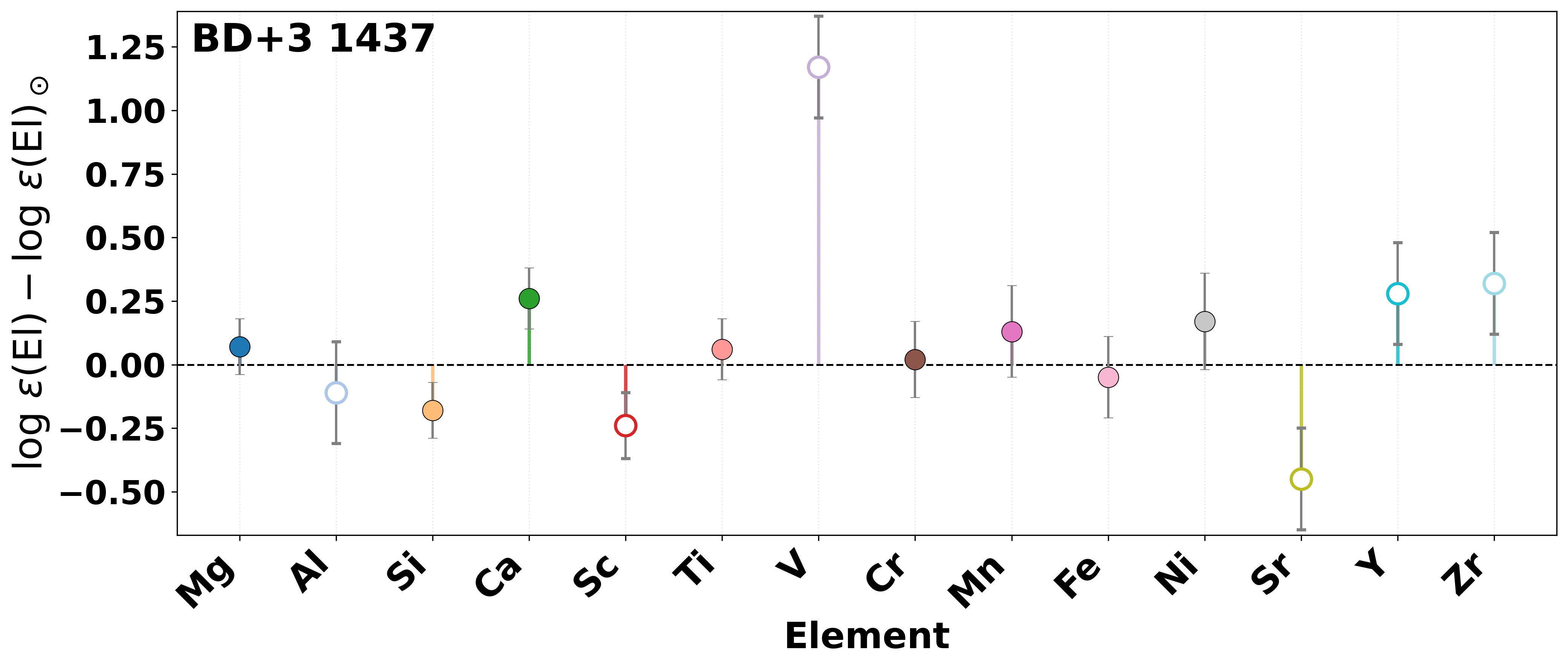}
  \end{minipage}
  \begin{minipage}[b]{0.33\textwidth}
 \includegraphics[height=3.2cm, width=1.0\textwidth]{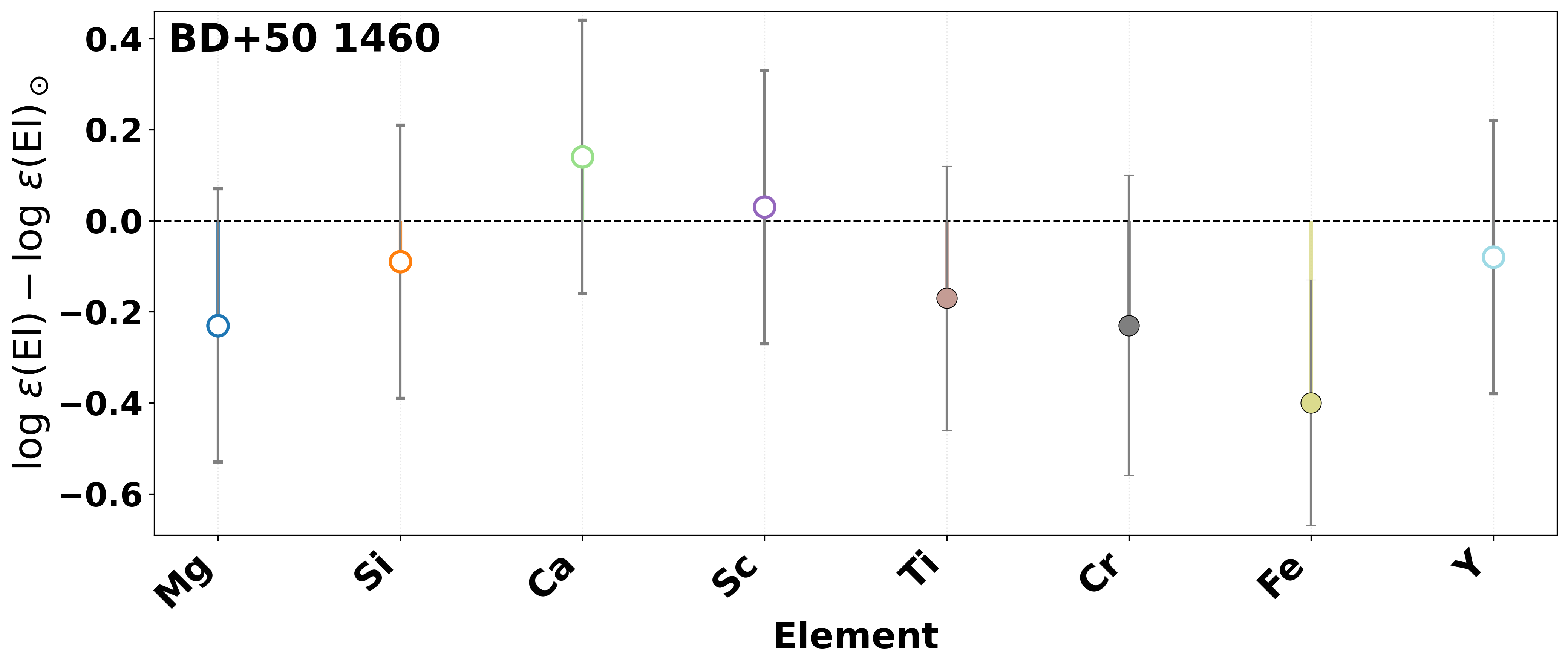}
 \end{minipage}
 \begin{minipage}[b]{0.33\textwidth}
  \includegraphics[height=3.2cm, width=1\textwidth]{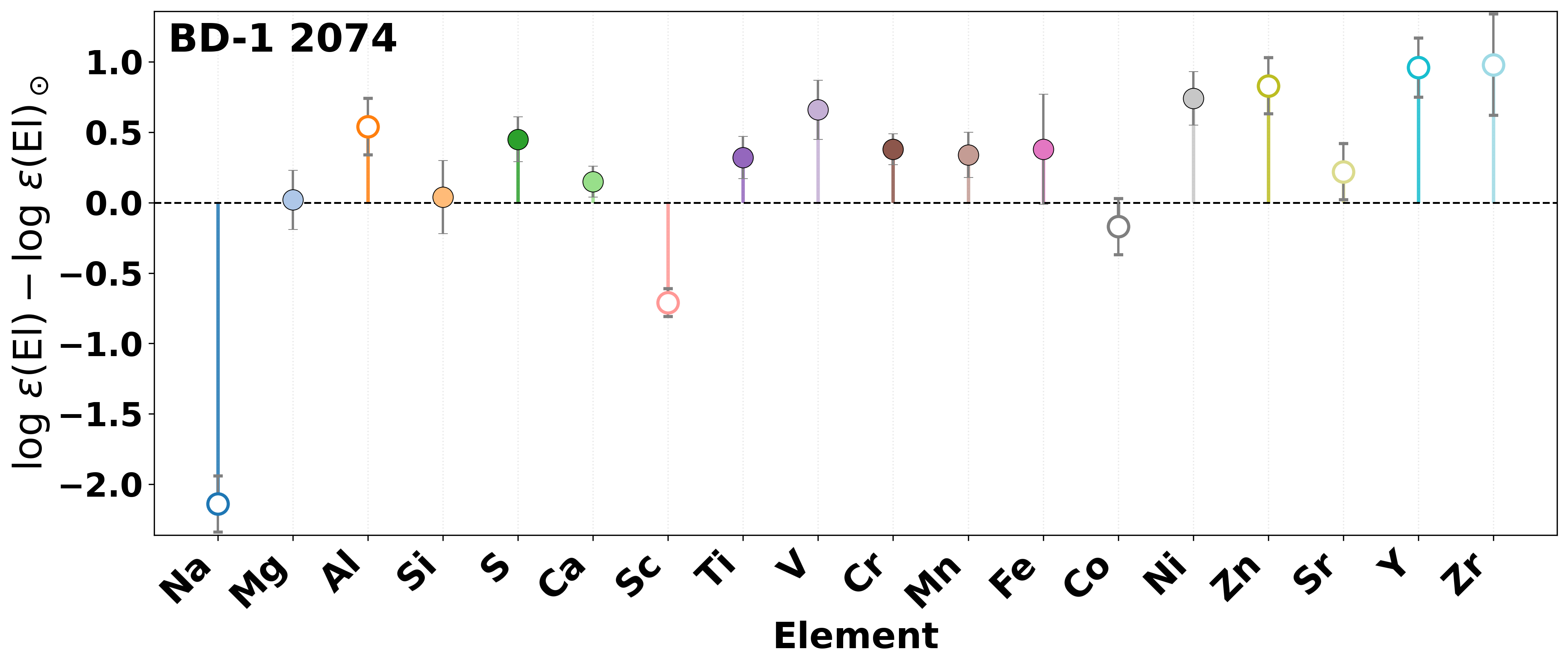}
  \end{minipage}
  \begin{minipage}[b]{0.33\textwidth}
  \includegraphics[height=3.2cm, width=1\textwidth]{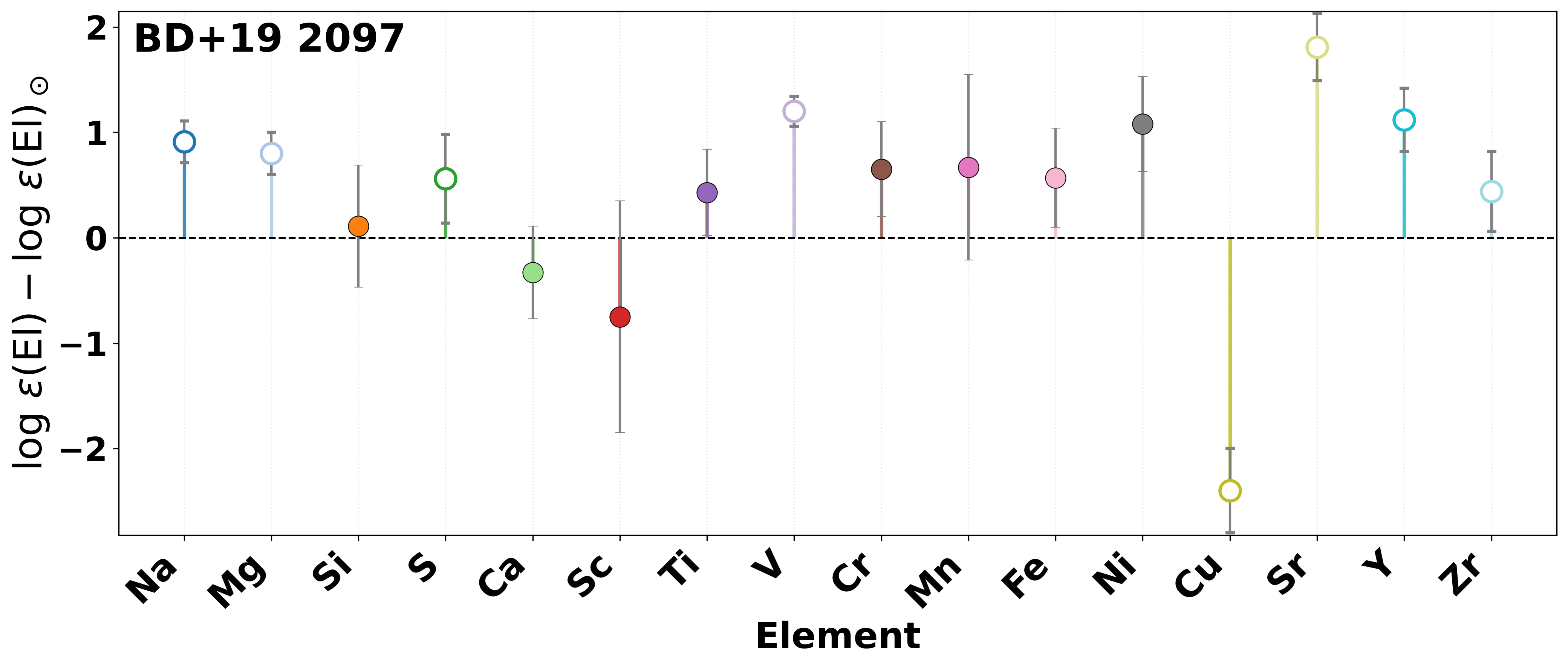}
  \end{minipage}
  \begin{minipage}[b]{0.33\textwidth}
  \includegraphics[height=3.2cm, width=1\textwidth]{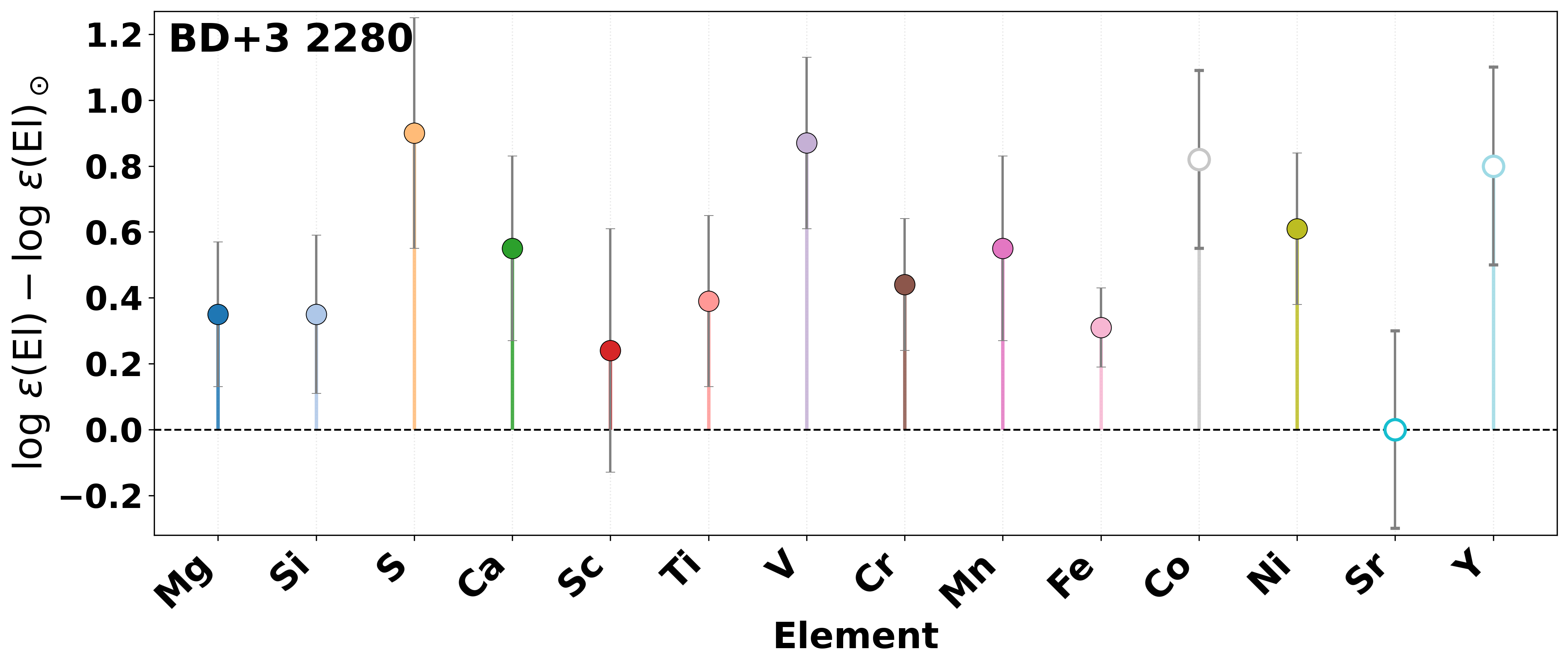}
 \end{minipage}
  \begin{minipage}[b]{0.33\textwidth}
  \includegraphics[height=3.2cm, width=1\textwidth]{BD252498_abundance_dist.png}
 \end{minipage}
 \begin{minipage}[b]{0.33\textwidth}
  \includegraphics[height=3.2cm, width=1\textwidth]{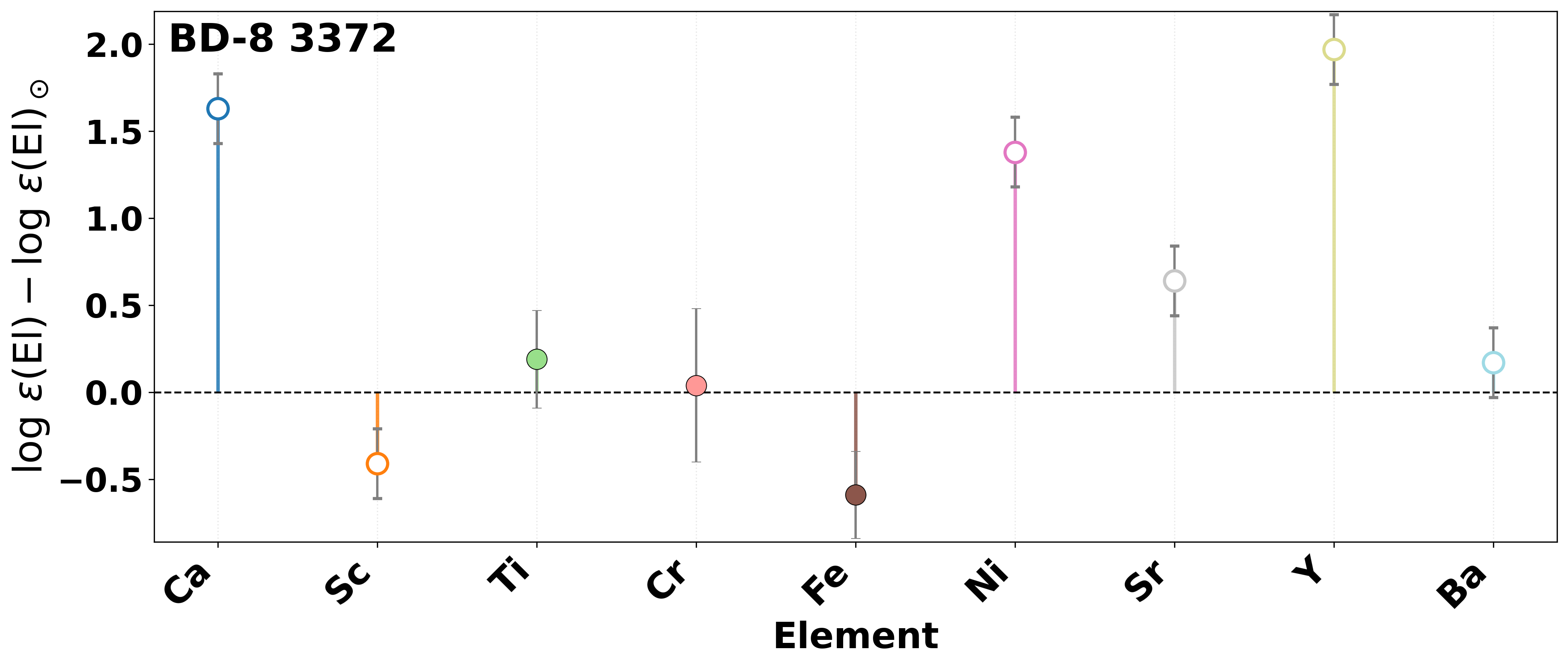}
  \end{minipage}
  \begin{minipage}[b]{0.33\textwidth}
  \includegraphics[height=3.2cm, width=1\textwidth]{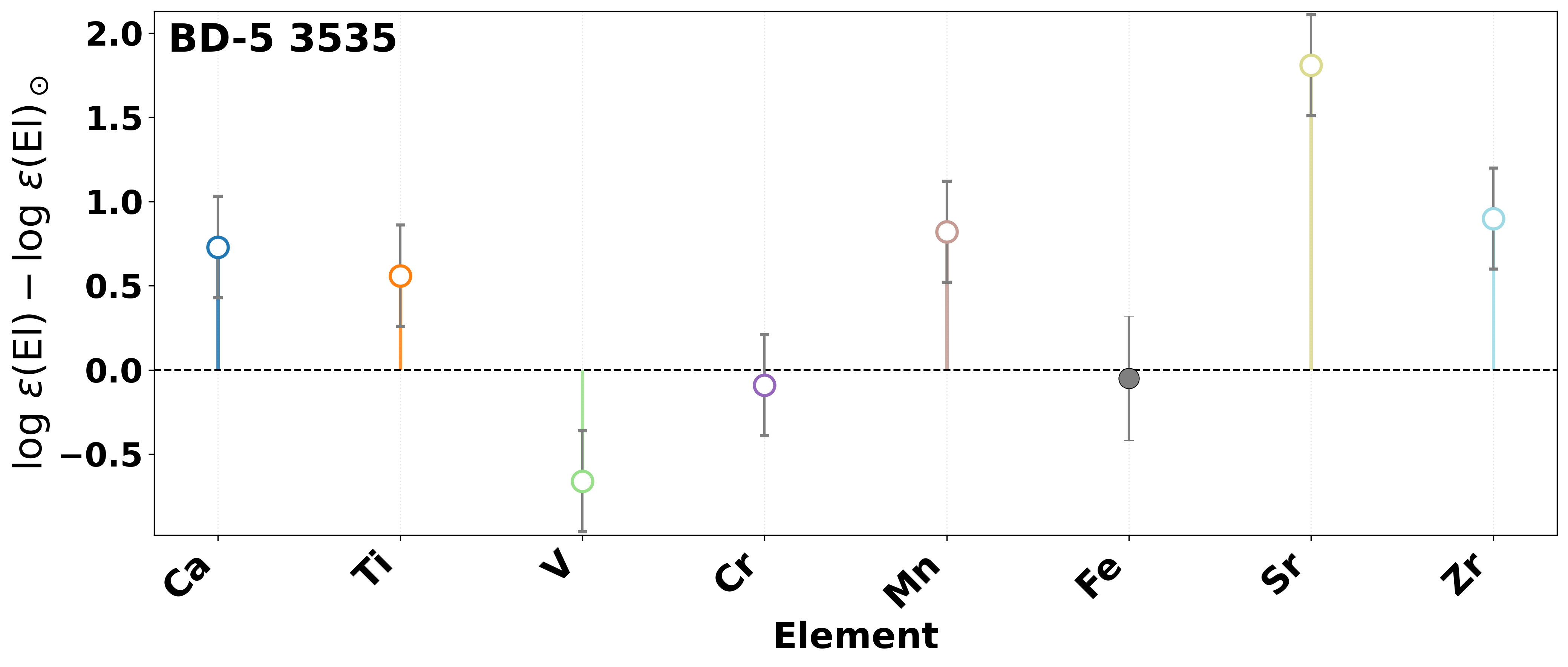}
  \end{minipage}
   \begin{minipage}[b]{0.33\textwidth}
  \includegraphics[height=3.2cm, width=1\textwidth]{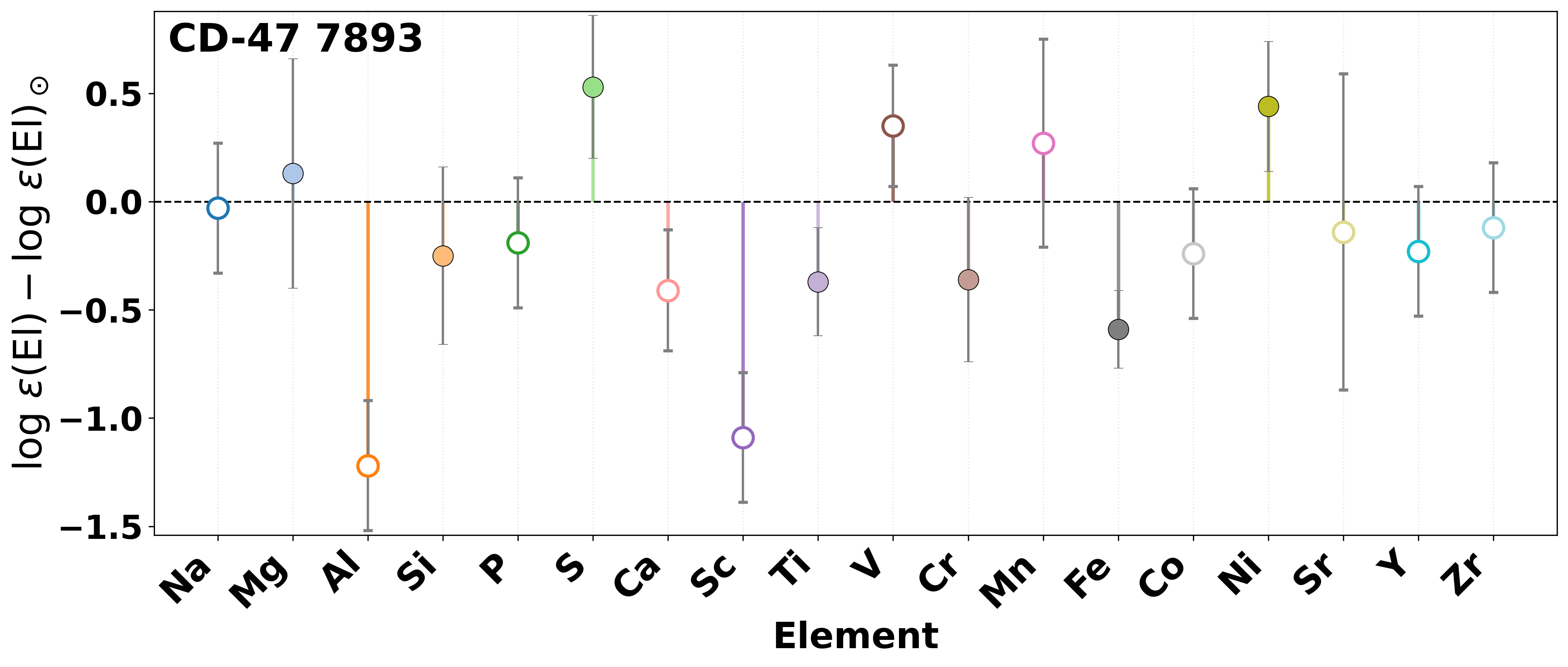}
   \end{minipage}
\begin{minipage}[b]{0.33\textwidth}
 \includegraphics[height=3.2cm, width=1\textwidth]{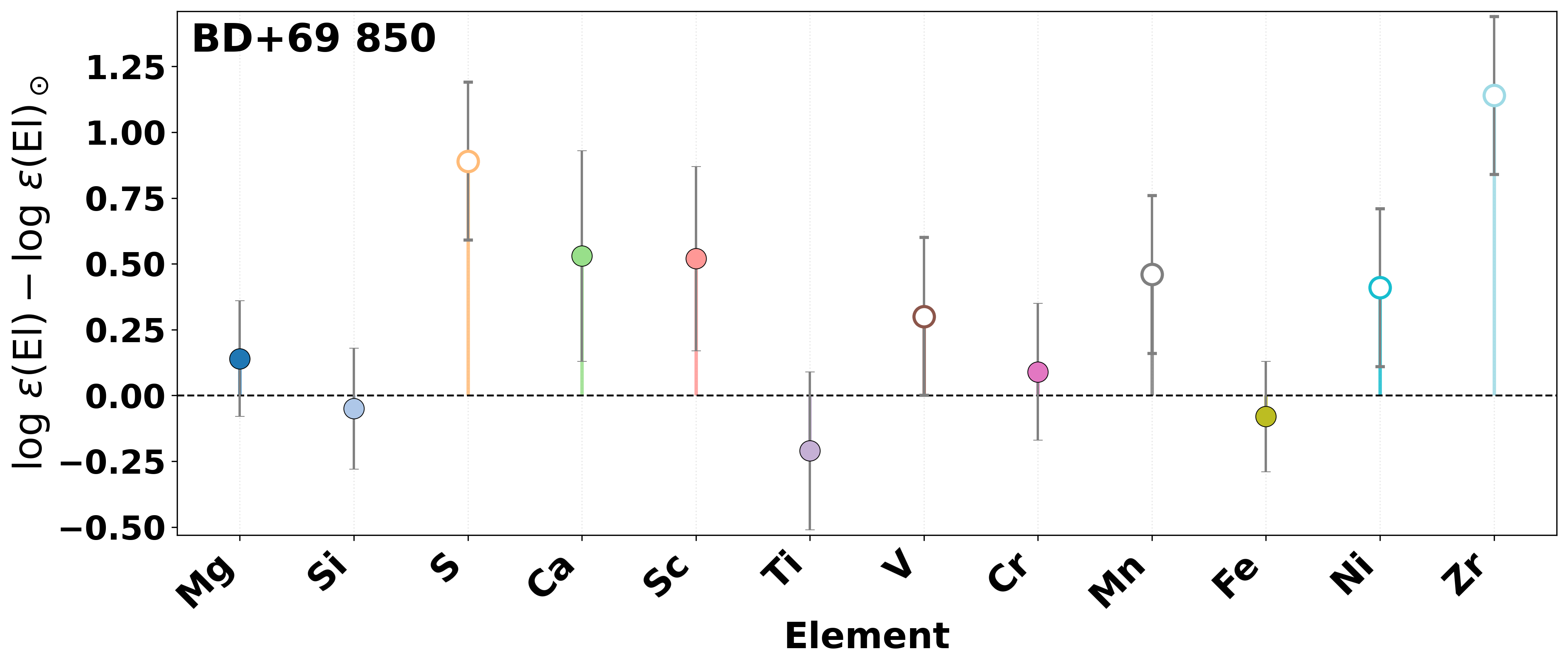}
 \end{minipage}
 \begin{minipage}[b]{0.33\textwidth}
  \includegraphics[height=3.2cm, width=1\textwidth]{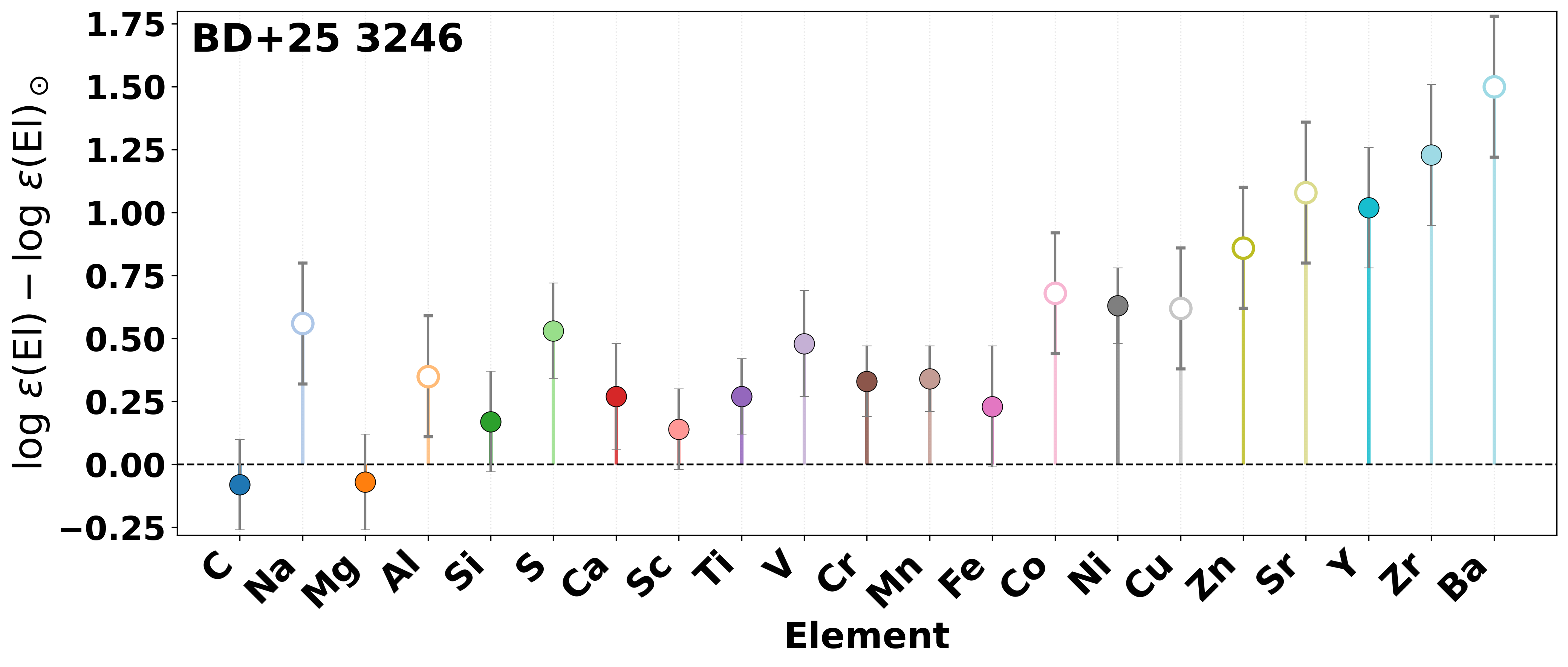}
 \end{minipage}
\begin{minipage}[b]{0.33\textwidth}
  \includegraphics[height=3.2cm, width=1\textwidth]{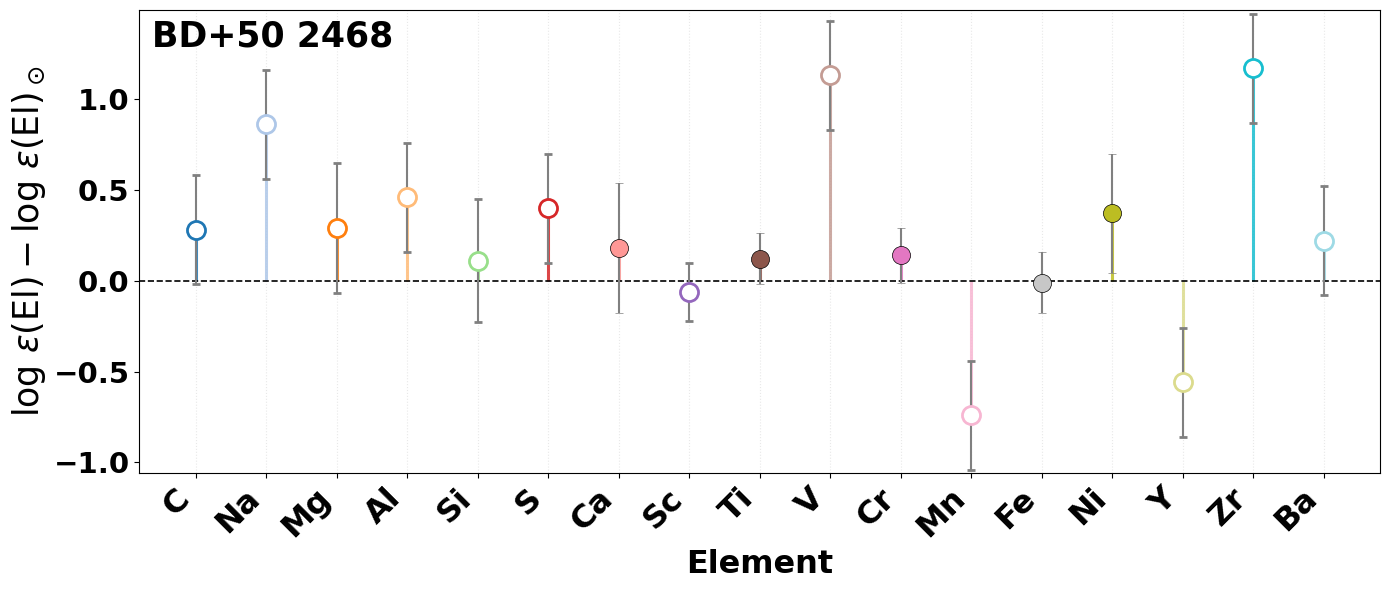}
  \end{minipage}
\caption{Elemental abundance pattern of the Am targets relative to the Sun \citep{2009ARA&A..47..481A}, expressed as $\log \epsilon(\mathrm{El}) - \log \epsilon(\mathrm{El})_{\odot}$ as a function of element. Open circles denote abundances derived from fewer than five spectral lines, while filled circles represent abundances based on five or more lines. Error bars indicate the uncertainties of the abundance determinations. The dashed line marks the solar reference level.}\label{fig:ap1}
\end{figure*}

\setcounter{figure}{1}
\renewcommand{\thefigure}{A\arabic{figure}}
\begin{figure*}
 \begin{minipage}[b]{0.33\textwidth}
  \includegraphics[height=3.2cm, width=1.0\textwidth]{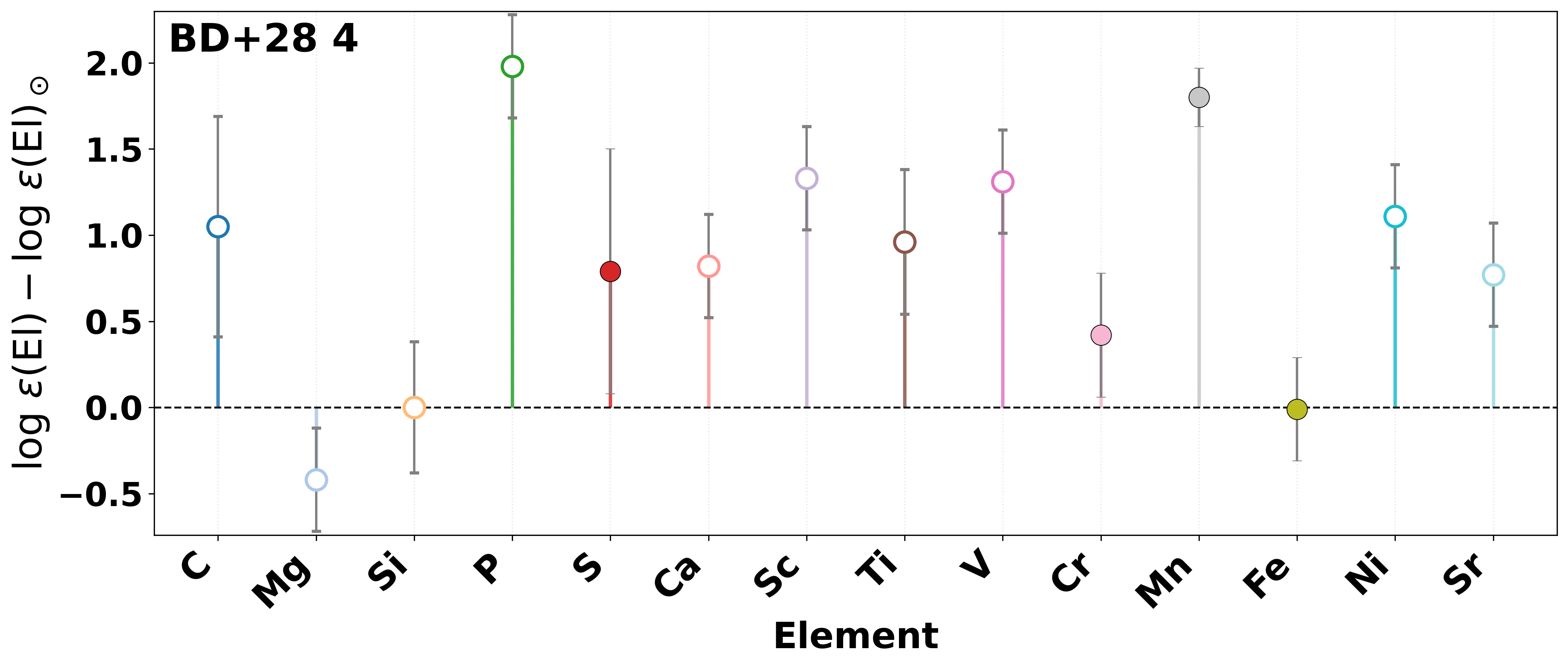}
  \end{minipage}
 \begin{minipage}[b]{0.33\textwidth}
  \includegraphics[height=3.2cm, width=1.0\textwidth]{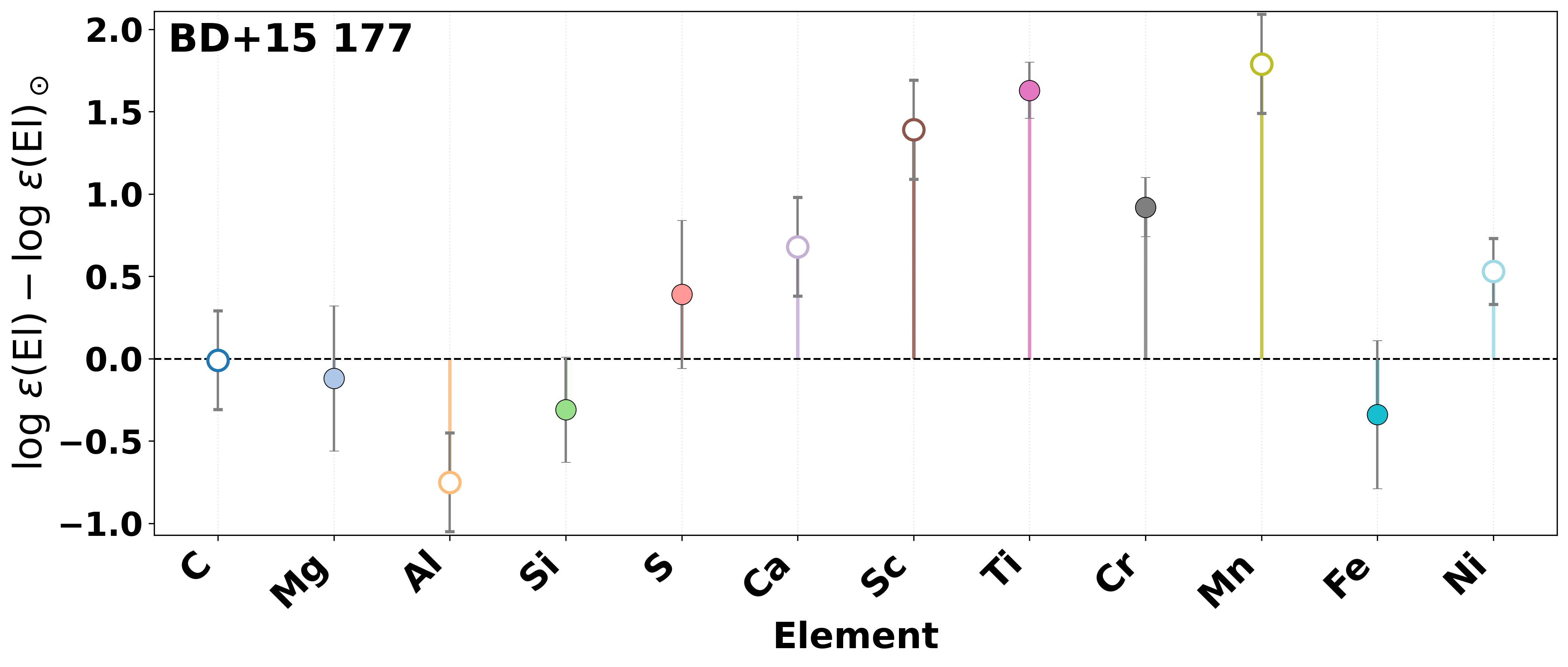}
 \end{minipage}
  \begin{minipage}[b]{0.33\textwidth}
  \includegraphics[height=3.2cm, width=1.0\textwidth]{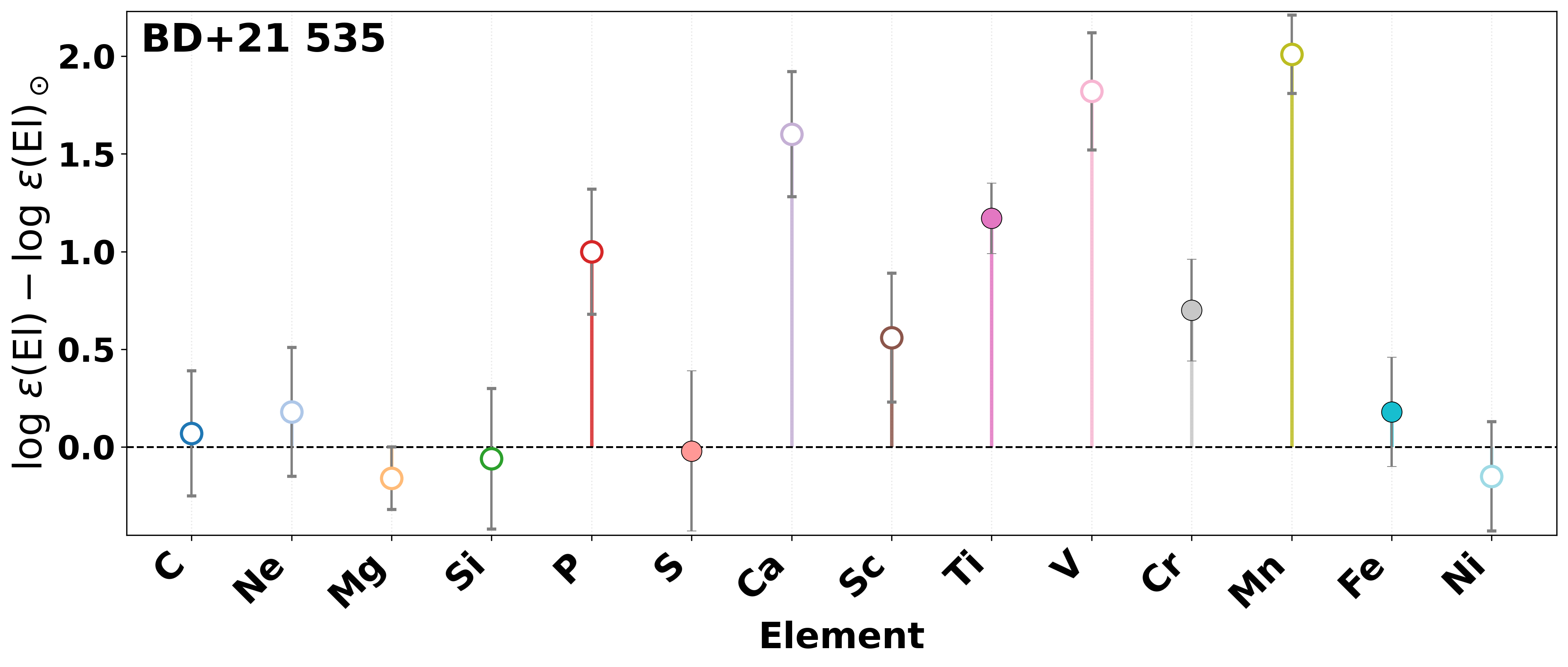}
  \end{minipage}
  \begin{minipage}[b]{0.33\textwidth}
 \includegraphics[height=3.2cm, width=1.0\textwidth]{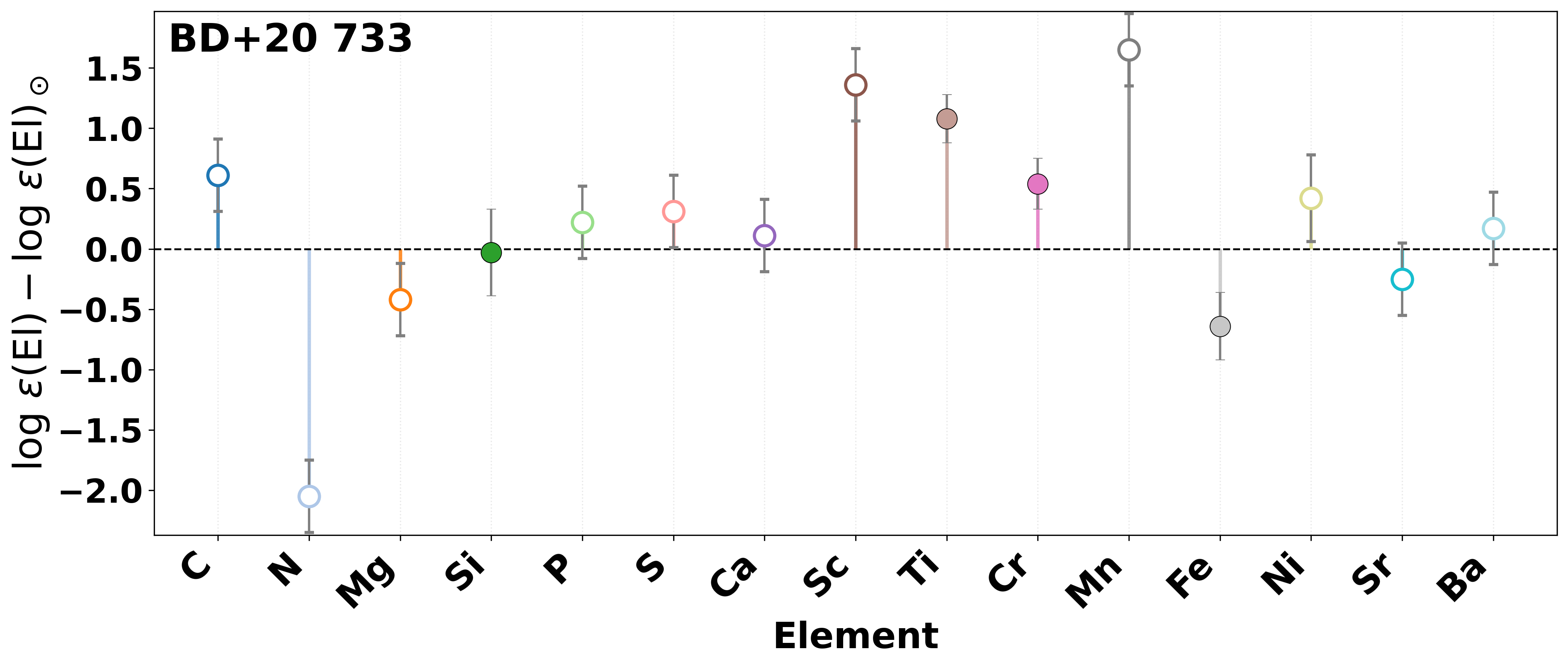}
 \end{minipage}
 \begin{minipage}[b]{0.33\textwidth}
  \includegraphics[height=3.2cm, width=1\textwidth]{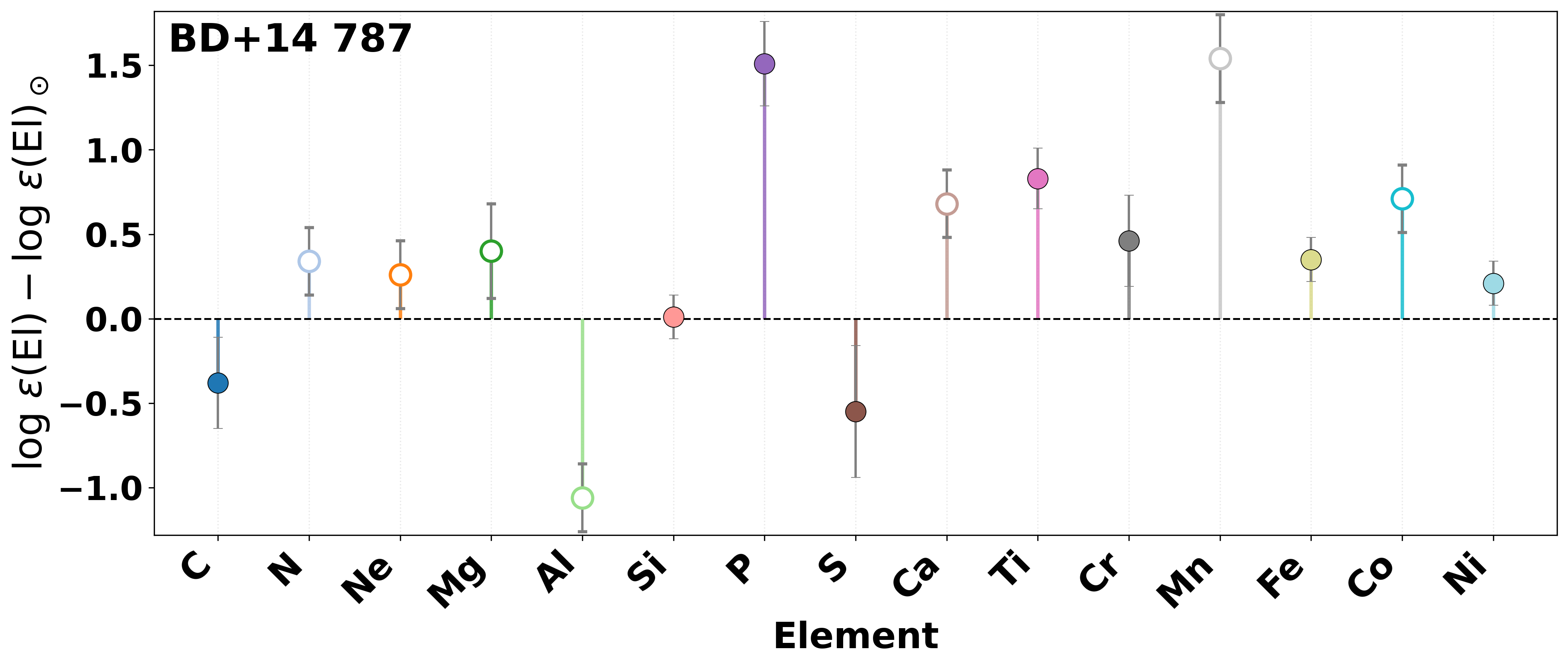}
  \end{minipage}
  \begin{minipage}[b]{0.33\textwidth}
  \includegraphics[height=3.2cm, width=1\textwidth]{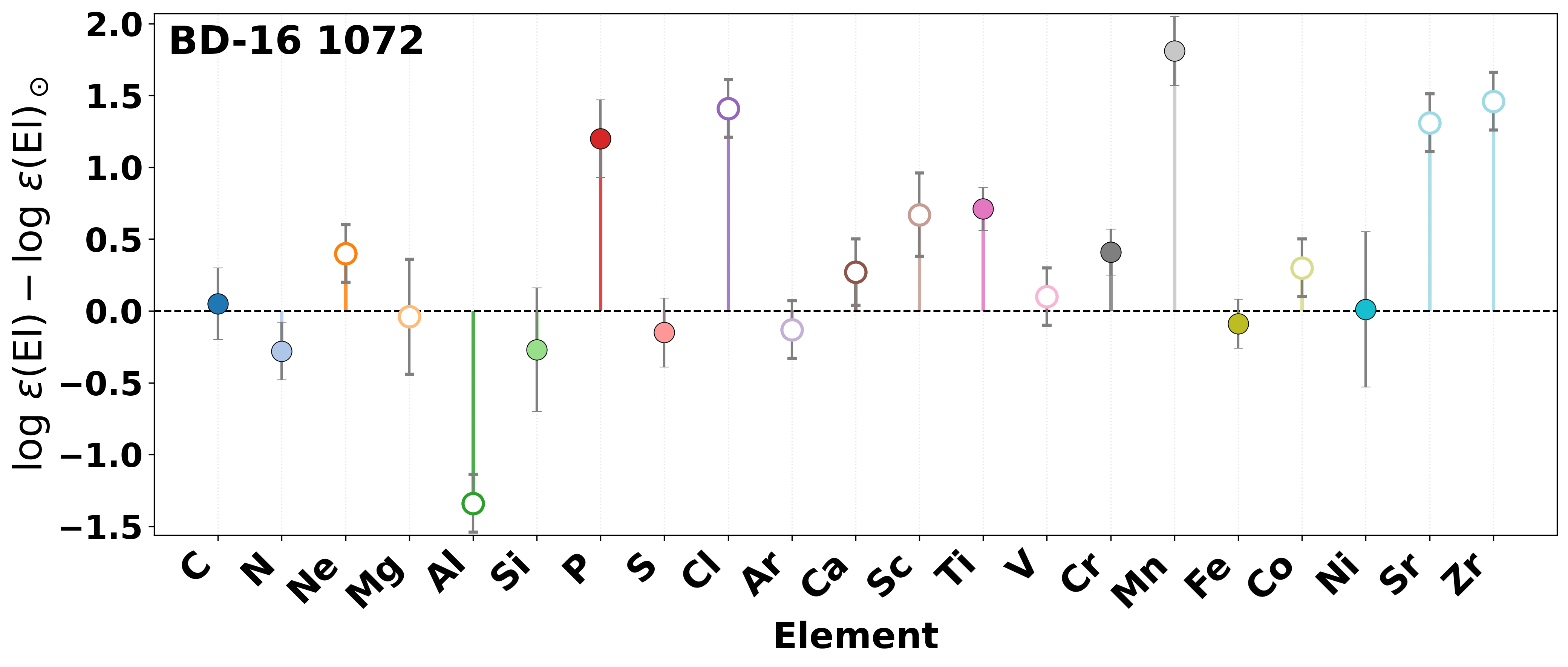}
  \end{minipage}
  \begin{minipage}[b]{0.33\textwidth}
  \includegraphics[height=3.2cm, width=1\textwidth]{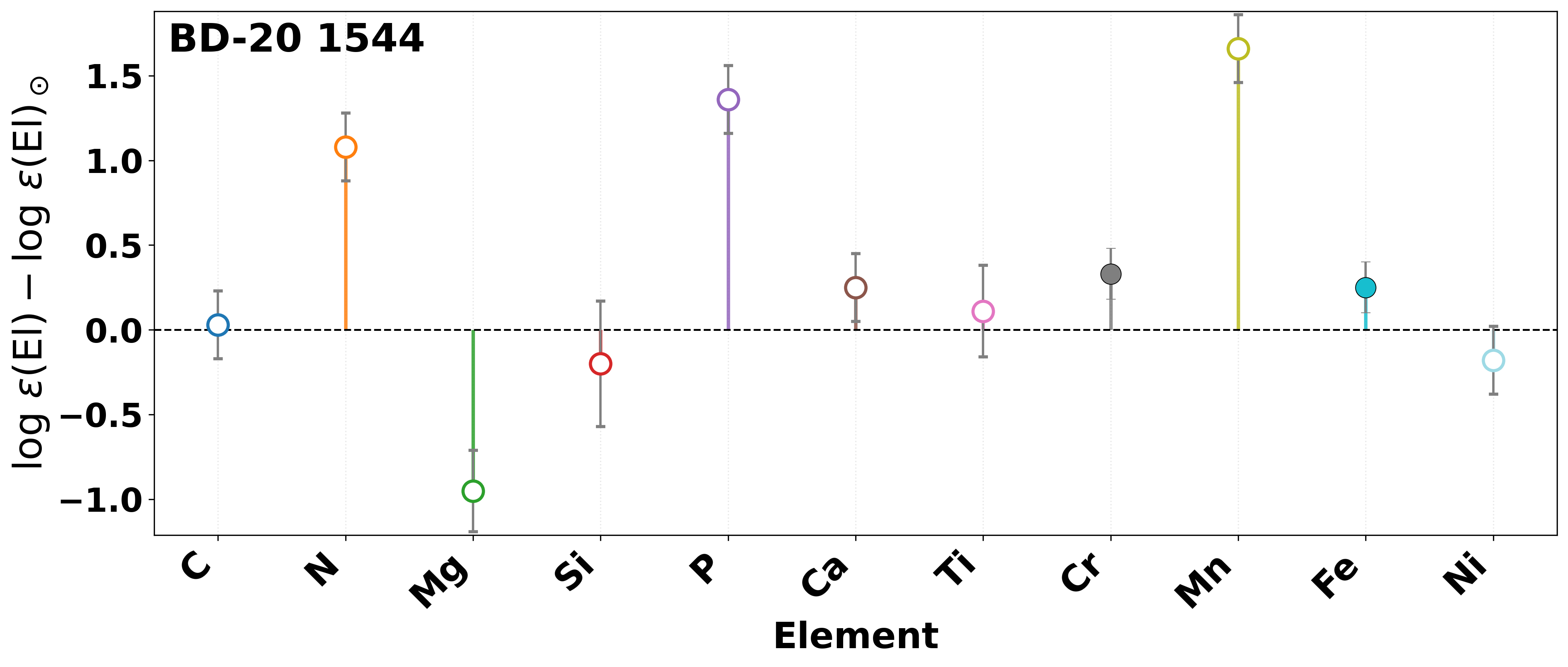}
 \end{minipage}
  \begin{minipage}[b]{0.33\textwidth}
  \includegraphics[height=3.2cm, width=1\textwidth]{CD3614166_abundance_dist.png}
 \end{minipage}
 \begin{minipage}[b]{0.33\textwidth}
  \includegraphics[height=3.2cm, width=1\textwidth]{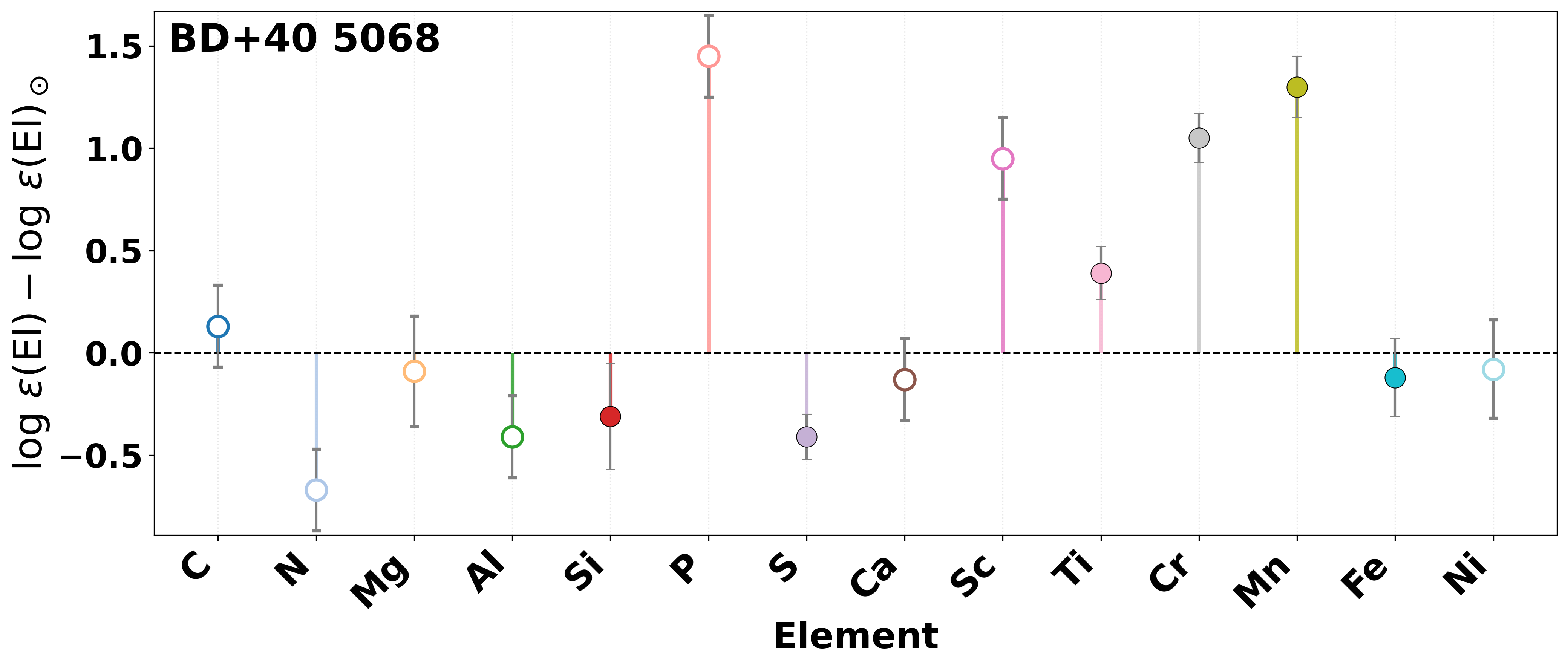}
  \end{minipage}
  \begin{minipage}[b]{0.33\textwidth}
  \includegraphics[height=3.2cm, width=1\textwidth]{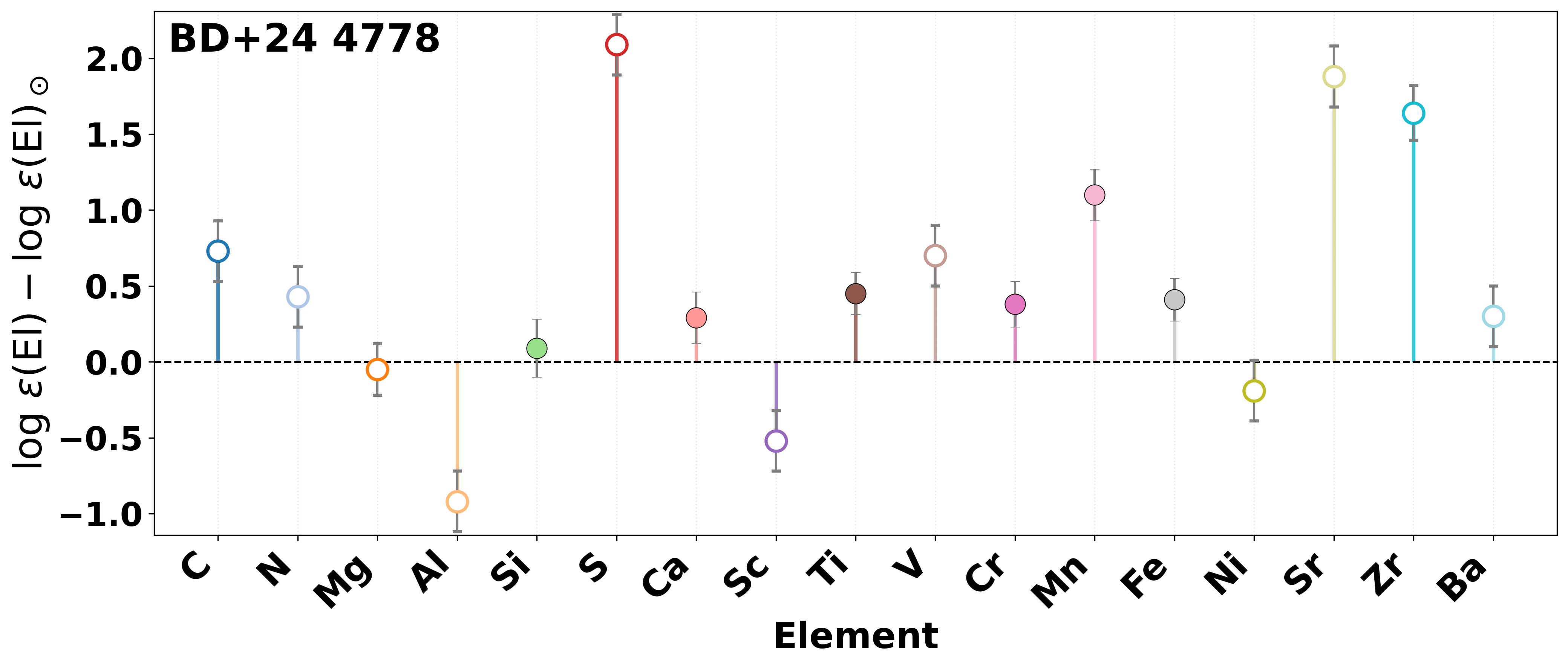}
  \end{minipage}
   \begin{minipage}[b]{0.33\textwidth}
  \includegraphics[height=3.2cm, width=1\textwidth]{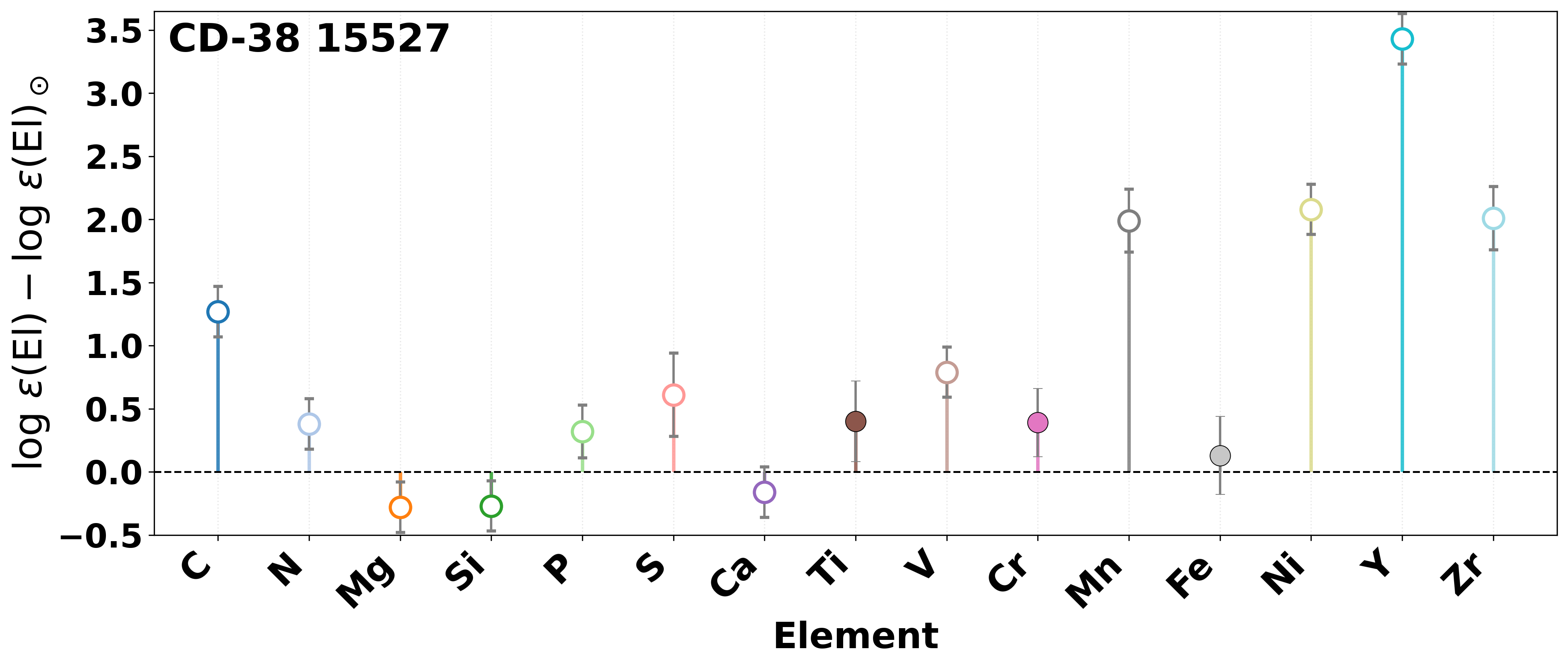}
   \end{minipage}
\begin{minipage}[b]{0.33\textwidth}
 \includegraphics[height=3.2cm, width=1\textwidth]{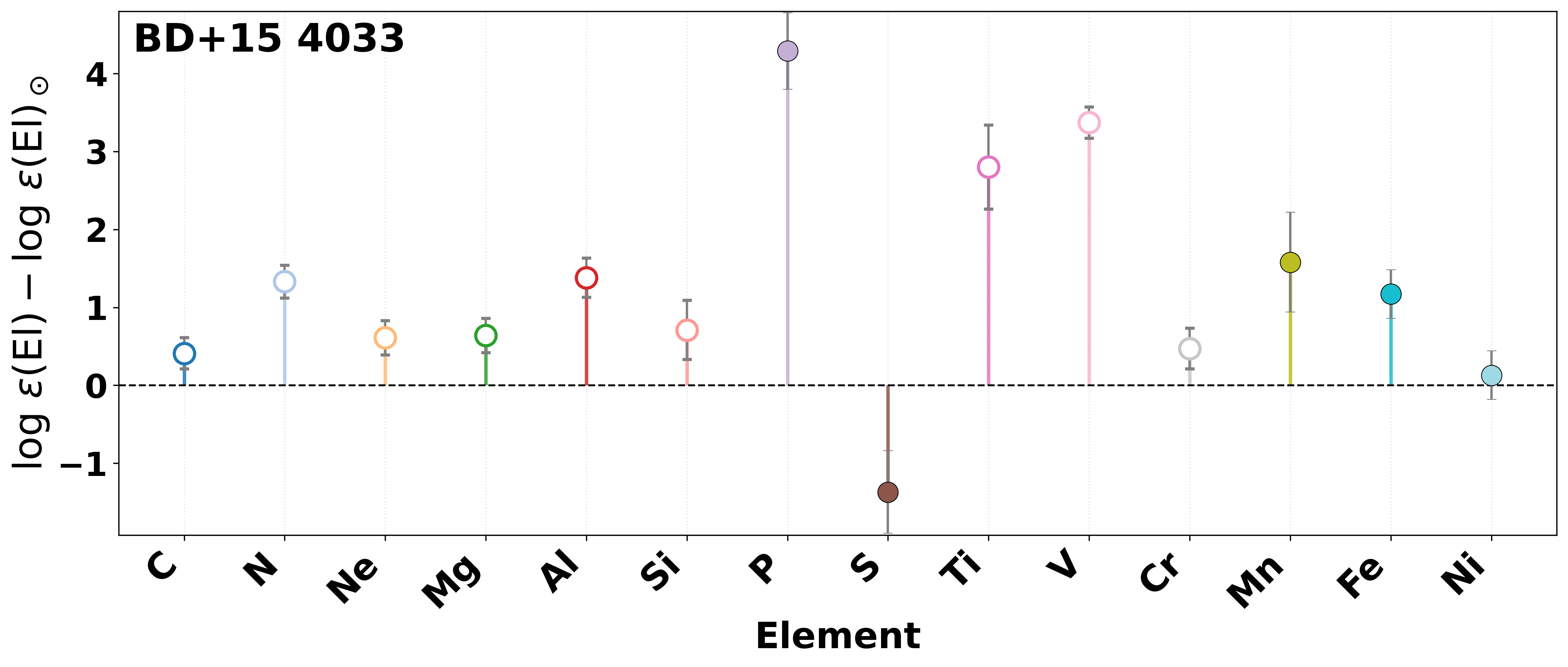}
 \end{minipage}
 \begin{minipage}[b]{0.33\textwidth}
  \includegraphics[height=3.2cm, width=1\textwidth]{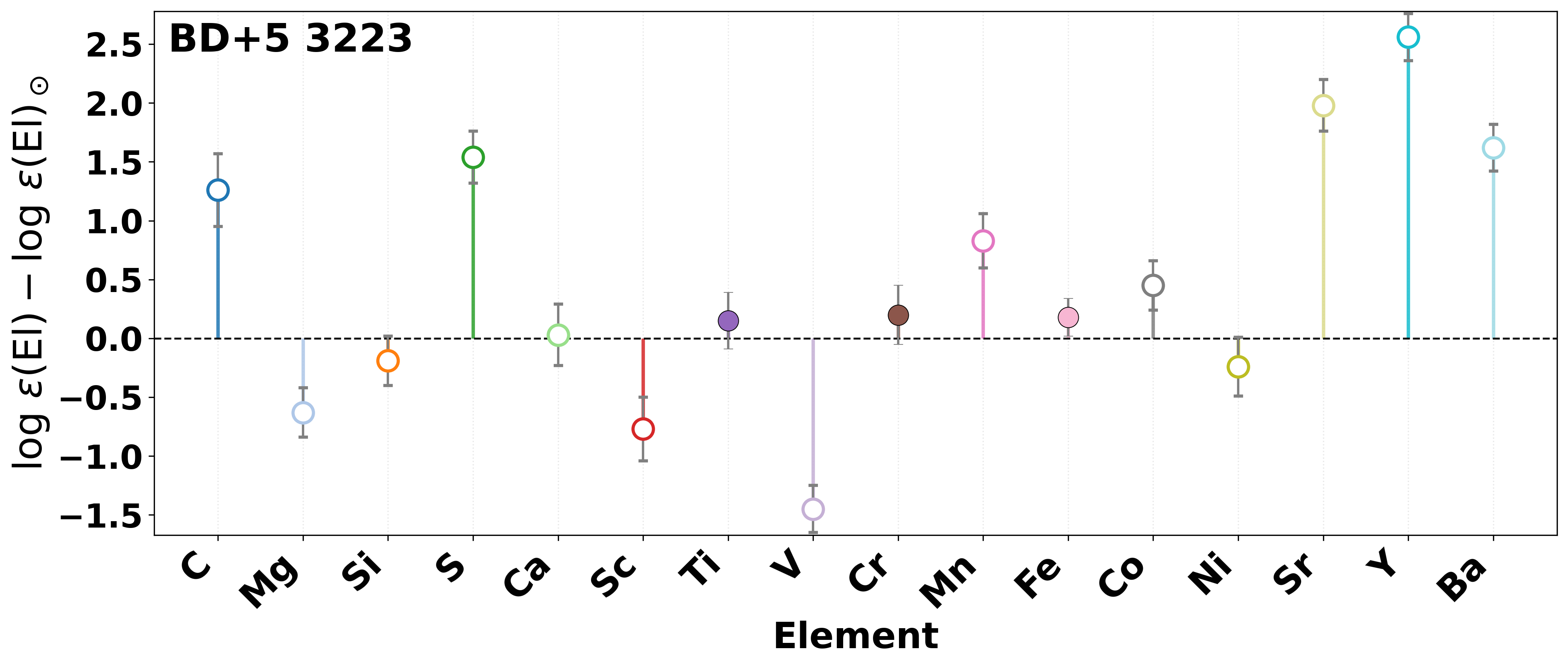}
 \end{minipage}
\caption{Elemental abundance pattern of the HgMn targets relative to the Sun \citep{2009ARA&A..47..481A}, expressed as $\log \epsilon(\mathrm{El}) - \log \epsilon(\mathrm{El})_{\odot}$ as a function of element. Open circles denote abundances derived from fewer than five spectral lines, while filled circles represent abundances based on five or more lines. Error bars indicate the uncertainties of the abundance determinations. The dashed line marks the solar reference level.}\label{fig:ap2}
\end{figure*}
% Don't change these lines
\bsp	% typesetting comment
\label{lastpage}
\end{document}